\documentclass[a4paper,11pt]{article} 

\usepackage[a4paper,left=0.99in, right=0.99in,top=1.1in, bottom=1.15in]{geometry}
\usepackage{subfig}
\usepackage{graphicx}
\usepackage{amsmath}
\usepackage{amssymb}
\usepackage{pstricks}
\usepackage{booktabs}
\usepackage{multirow}
\usepackage{cite}

\numberwithin{equation}{section}

\newcommand{\dist}{\Delta_{12}}
\newcommand{\distsq}{\Delta_{12}^2}
\newcommand{\lqcd}{\Lambda_{\rm QCD}}

\newcommand{\PU}{\text{PU}}
\newcommand{\be}{\begin{equation}}
\newcommand{\ee}{\end{equation}}
\newcommand{\bea}{\begin{eqnarray}}
\newcommand{\eea}{\end{eqnarray}}
\newcommand{\la}{\langle}
\newcommand{\ra}{\rangle}
\newcommand{\mean}[1]{\left\la\smash{#1}\right\ra}

\newcommand{\order}[1]{{\cal O}(#1)}
\newcommand{\bit}{\begin{itemize}}
\newcommand{\eit}{\end{itemize}}

\title{ 
    \vskip 2cm
    The mass area of jets
}
\author{Sebastian Sapeta and Qi Cheng Zhang\\
\\
{\it  \normalsize LPTHE, UPMC Univ.~Paris 6 and CNRS UMR 7589, Paris, France}\\
{\it  \normalsize E-mail:}
{\tt \normalsize sapeta@lpthe.jussieu.fr, cheng.zhang.5@city.ac.uk}
}
\date{}

\begin{document}
\maketitle
\thispagestyle{empty}
\begin{abstract}
We introduce a new characteristic of jets called mass area.  It is defined so as
to measure the susceptibility of the jet's mass to contamination from soft
background. The mass area is a close relative of the recently introduced
catchment area of jets. We define it also in two variants: passive and active.
As a preparatory step, we generalise the results 
for passive and active areas of two-particle jets to the case where
the two constituent particles have arbitrary transverse momenta.
As a main part of our study, we use the mass area to analyse a range of modern
jet algorithms acting on simple one and two-particle systems. 
We find a whole variety of behaviours of passive and active mass areas
depending on the algorithm, relative hardness of particles or their separation.
We also study mass areas of jets from Monte Carlo simulations as well as give an
example of how the concept of mass area can be used to correct jets for
contamination from pileup.
Our results show that the information provided by the mass area can be very
useful in a range of jet-based analyses.
\end{abstract}

\newpage

\setcounter{page}{1}

\tableofcontents

\section{Introduction}

In the present era of the LHC, as in the times of all precedent hadron
colliders, jets remain fundamental objects of interest
\cite{Ellis:2007ib,Salam:2009jx}. Their importance extends far beyond the domain
of physics of strong interactions, where they are used as representatives of
partons participating in a hard process. They play also a significant role in a
whole range of processes involving decays of heavy particles. Those include, for
example, a top quark decaying into three jets, $W/Z$, Higgs boson or a
hypothetical boson $Z'$ decaying into two jets, as well as a variety of SUSY
particles which readily decay into many-jet final states.

A considerable effort is being made to improve our control on jets. On the
theoretical side, this comprises, on one hand, improving the precision of
calculations involving the canonical set of jet observables like the transverse
momentum, mass or thrust. On the other hand new concepts are being developed
including additional characteristics, like, for example, the catchment
area of jets~\cite{Cacciari:2008gn}, or new analysis techniques  based on
subjets~\cite{Butterworth:2002tt, Butterworth:2007ke, Butterworth:2008iy,
Kaplan:2008ie, Almeida:2008yp, Plehn:2009rk, Ellis:2009su, Ellis:2009me, Krohn:2009th, Kribs:2010hp, Kribs:2009yh}.

Amongst a number of properties of a jet, its mass turns out to be important in
many physical contexts. In the legitimate approximation of massless QCD partons,
the jet mass arises due to its substructure. One source of this substructure is
of course the radiation of gluons and quarks, which leads to the well known
distribution of mass of QCD jets with a significant fraction of jets with large
masses. Consider, however, a process involving a hadronic decay of a heavy
object of mass $m$. If this object, in addition, has the transverse momentum
$p_t \gg m$, a situation not unusual at the LHC, the decay products will end up
in a single jet. The reconstructed mass of such jet will be an important emblem
pointing to its origins. Moreover, such a fat jet can be analysed further with
techniques involving study of the masses of its subjets.
The jet-based reconstruction of heavy particles has been a subject of numerous
studies devoted to decay of W \cite{Seymour:1993mx}, WW scattering
\cite{Butterworth:2002tt}, decay of top \cite{Agashe:2006hk, Kaplan:2008ie,
Ellis:2009me,Almeida:2008tp }, Higgs \cite{Butterworth:2008iy, Plehn:2009rk} as
well as SUSY searches \cite{Butterworth:2007ke, Kribs:2009yh,Butterworth:2009qa,
Kribs:2010hp}.

The success of the above techniques depends crucially on the ability of precise
determination of the mass of jets measured in experiment.
In hadron colliders, however, particles that can contribute to the jet's
substructure may also come from soft radiation unrelated to the genuine hard
process of interest.
Such radiation appears, for instance, due to independent minimum-bias
collisions that happen in the same bunch crossing, a phenomenon known by the
name of pileup~(PU). But even in the absence of pileup each hard process from
single hadron-hadron collision is accompanied by soft underlying event (UE)
which can easily modify the jet's transverse momentum by a few GeV
\cite{Dasgupta:2007wa, Cacciari:2009dp}.

A major step towards quantifying the effects of UE/PU and correct for them was
made in \cite{Cacciari:2008gn, Cacciari:2007fd}, where the concept of the
\emph{jet area} was introduced, which is a measure of how much the transverse
momentum a jet from a given clustering algorithm is prone to be
affected by soft radiation. We briefly review the corresponding results in 
section~\ref{sec:area-review}.
 
In this paper, we introduce a related characteristic of a jet, which we will
call the \emph{mass  area} and which will represent the susceptibility of a
jet's mass to a soft background like UE or PU.
In line with \cite{Cacciari:2008gn} we will introduce two types of the mass
area: passive and active.  The former will correspond to pointlike background
whereas the latter will be appropriate to measure the susceptibility of the jet
mass to the soft radiation which is diffuse and uniform. 

We will analyse the passive and active mass areas of jets from four modern
clustering algorithms: $k_t$~\cite{Catani:1993hr, Ellis:1993tq},
Cambridge/Aachen~(C/A)~\cite{Dokshitzer:1997in,Wobisch:1998wt, Wobisch:2000dk},
anti-$k_t$~\cite{Cacciari:2008gp} and SISCone~\cite{Salam:2007xv}. The first
three belong to the class of sequential recombination algorithms. They introduce
a distance $d_{ij}$ for each pair of particles and a distance $d_{iB}$ for
particle and the beam. The distances depend on the basic parameter, jet radius
$R$. The algorithms start from computing the above distances for all final state
particles. If the smallest distance involves two particles, they are recombined
and replaced in the list of particles by the product of this recombination. If
the smallest distance is that between a particle and the beam the particle is
called a jet and removed from the list of entries. The procedure is repeated
until there are no entries in the list. 
The SISCone algorithm belongs to a different class of the so called cone
algorithms. They look for stable cones of radius~$R$ and subsequently apply the
Tevatron run II procedure~\cite{Blazey:2000qt} to split or merge the overlapping
cones. All the above algorithms are infrared and collinear safe and are easily
accessible via the FastJet package~\cite{Cacciari:2005hq,FastJet}.
Further details on each of them are given in section~\ref{sec:passive-area-2p}.

In \cite{Cacciari:2008gn} the jet areas were calculated for the case of 1- and
2-particle systems. In the latter case the results were obtained in the limit of
strong ordering of transverse momenta of the two particles.  In this paper we
relax the assumption of strong ordering  and start by presenting in section
\ref{sec:area-general-2p} the corresponding general results for passive and
active areas of 2-particle jets.
Subsections~\ref{sec:passive-area-2p} and \ref{sec:active-areas-2p} are quite
technical. Though they contain very useful material, the reader interested in
the main part of our study may skip them on the first reading.

In section \ref{sec:mass-area} we introduce the concept of the mass area of a
jet and define its passive (subsection~\ref{sec:passive-mass-areas-2p}) and
active (subsection \ref{sec:active-mass-area}) variants. There, we also analyse
their properties for the system of 1- and 2-particles. In particular, we compare
results from the four algorithms and examine the dependence on the relative
hardness of the constituents of 2-particle jets. At the end of each subsection
we discuss the problem of logarithmic dependence of the mass area of QCD jets on
the jet's transverse momentum. We give it a quantitative description in terms of
the anomalous dimension and compare the results across the jet algorithms.
Throughout the paper we work in the small $R$ approximation which is justified
by the observation that the corrections from higher powers of $R$ are
accompanied by small coefficients~\cite{deFlorian:2007fv, Dasgupta:2007wa}.
 
In section~\ref{sec:mc-study}, we turn to a study of jets simulated
with Pythia. We illustrate how the features found for simple 1- and 2-particle
systems help understanding mass areas of more realistic jets
(subsection~\ref{sec:mc-study-ma}). Then, we give an example of practical
application of mass areas to correct jet mass for the contamination from pileup
(subsection~\ref{sec:pileup-cor}). Finally, we summarize our results in
section~\ref{sec:conclusions} and provide some extra details in two
appendices~\ref{app:passive-formulae} and \ref{app:active-formulae-sis}.

\section{Essential definitions, notation and brief review of jet areas}
\label{sec:area-review}

\subsection{Passive area}
\label{sec:passive-area}

Consider a set of particles $\{p_i\}$ which are clustered with an infrared safe jet algorithm into a~set of jets $\{J_i\}$. 
Suppose now that we add to the set $\{p_i\}$ a single infinitely soft particle~$g$, which hereafter we shall call the \emph{ghost},  and repeat the clustering on the new set of particles $\{p_i, g\}$.
Because we use an algorithm which is infrared safe and because our extra particle $g$ has infinitely small transverse momentum this clustering will not change the set of jets  $\{J_i\}$. The ghost particle $g$ can be either clustered with one of the real particles, in which case it ends up in one of the jets $J_i$, or it can form a new jet with $g$ being its only constituent.

The passive, scalar\footnote{The 4-vector passive area was also defined in
\cite{Cacciari:2008gn}. Though we will not use it directly in the current study,
we note as an aside that the concept of 4-vector passive area is implicitly
present in the calculations of passive mass area of
section~\ref{sec:passive-mass-areas-2p}. In particular, all the results from
that section could be alternatively obtained with a direct use of 4-vector
passive area.} area of the jet $J$ is defined \cite{Cacciari:2008gn} as the area
of the region in the $(y,\phi)$ plane in which the ghost particle $g$ is
clustered with $J$ 
\begin{equation}
\label{eq:passive-area-def}
a(J) \equiv \int dy\, d\phi\, f(g (y,\phi),J),
\hspace{30pt}
f (g, J) = 
\bigg \{
\begin{array}{cl}
1  & {\rm for \ } g {\rm \ clustered\  with\ } J       \\
0  & {\rm for \ } g {\rm \ not \ clustered\  with\ } J
\end{array}\,.
\end{equation}
Such definition provides a measure of the susceptibility of the jet to soft radiation in the limit in which this radiation is pointlike.

For a set o particles that consists only of a single particle $p_1$ the passive area of the corresponding jet $J_1$ is $a(J_1) = \pi R^2$ for all four jet clustering algorithms: $k_t$, C/A, anti-$k_t$ and SISCone.

\begin{figure}
    \centering
    \includegraphics[width=0.45\textwidth]{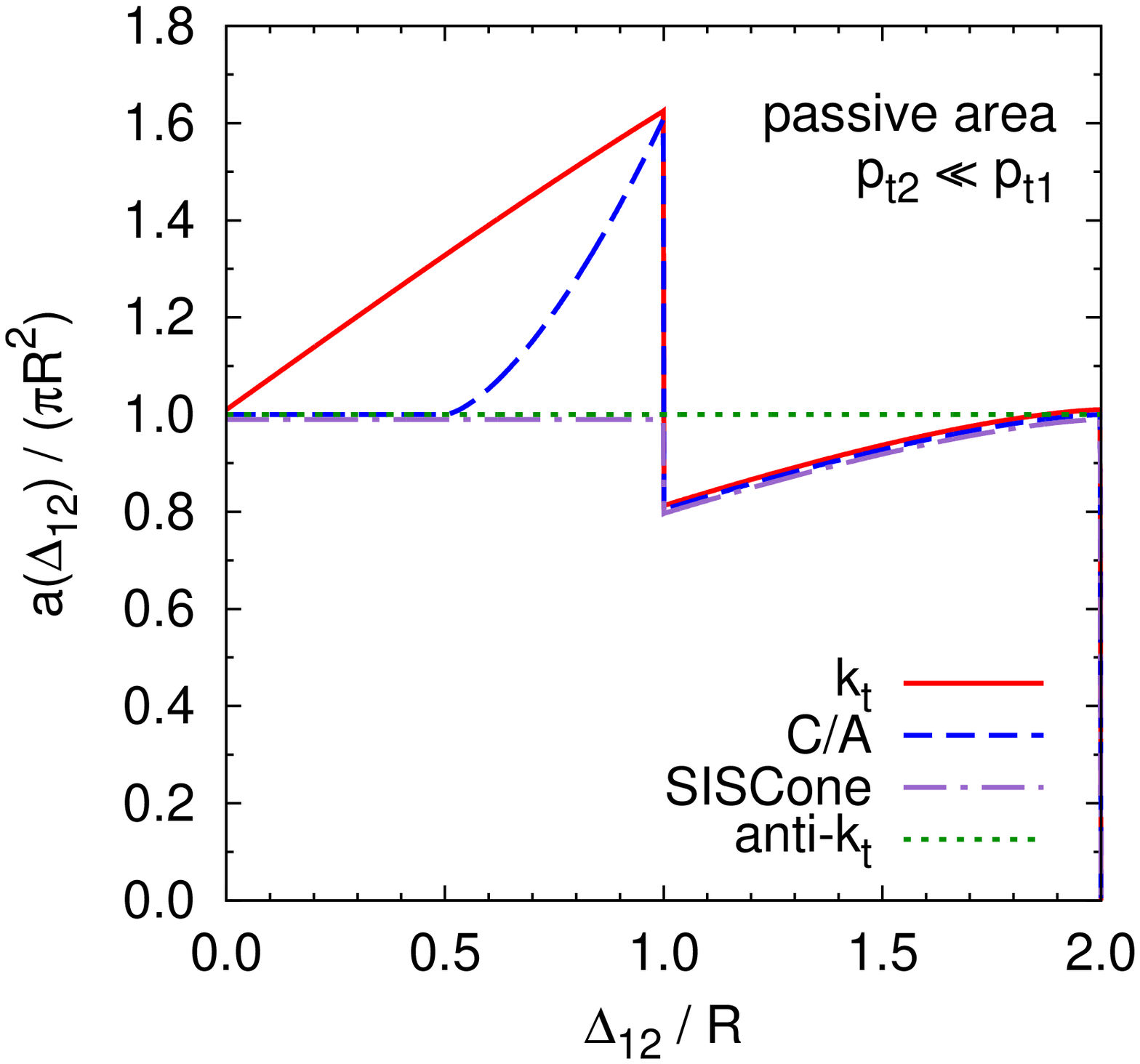}
    \hspace{20pt}
    \includegraphics[width=0.45\textwidth]{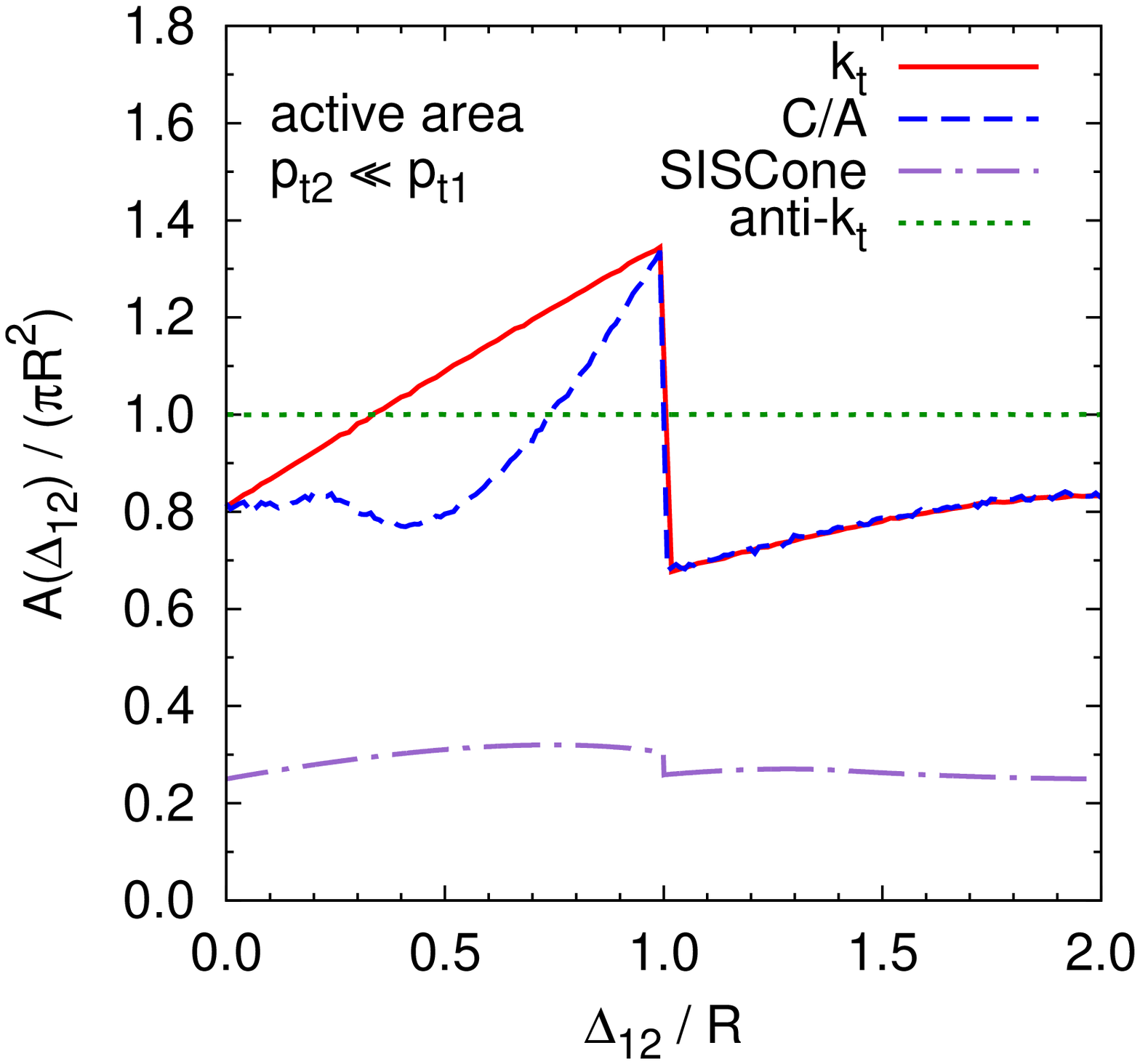}
    \caption{
        Passive (left) and active (right) area of the hardest jet in a 
        2-particle event with $p_{t2}\ll p_{t1}$ 
	and the interparticle separation $\dist$.
        All curves for passive areas as well as the anti-$k_t$ and SISCone 
        curves for active area represent the analytic formulae 
        obtained in \cite{Cacciari:2008gn,Cacciari:2008gp}. 
	The active area results for the $k_t$ and C/A algorithms were computed
	using the {\tt FastJet 2.4.2}  package \cite{Cacciari:2005hq,FastJet}.
    }
    \label{fig:passive-area2p}
\end{figure}

Adding a second particle $p_2$ leads to the result which depends on the jet
definition (i.e. jet algorithm and jet radius) and the geometrical distance
between particles $p_1$ and $p_2$ in the  $(y,\phi)$ plane $\distsq=
(y_1-y_2)^2+(\phi_1-\phi_2)^2$. 
The analytic results for $a(\dist)$ of the harder jet in the limit $p_{t1} \gg
p_{t2} \gg \lqcd \gg p_{tg}$ for all four algorithms were obtained in
\cite{Cacciari:2008gn,Cacciari:2008gp}.
In Fig.~\ref{fig:passive-area2p} (left) we show the corresponding functions,
normalised to the 1-particle passive area. We notice substantial dependence on
the algorithm especially in the region $\dist < R$ where the two particles form
a single jet. There, the areas from  the $k_t$ and C/A algorithms are notably
different from $\pi R^2$ and vary significantly with the distance between the
particles. On the contrary the areas from SISCone and anti-kt are identical with
the 1-particle area for $\dist < R$ and in the latter case also for $\dist > R$.
All results recover the correct limit of $\pi R^2$ when $\dist$ goes either to
$0$ or to~$2 R$.

\subsection{Active area}
\label{sec:active-area-2p}

Suppose that we add to the set of particles $\{p_i\}$ not a single ghost like in
the case of passive area  but a dense coverage of ghost particles randomly
distributed in the $(y,\phi)$ plane. Again, the original jets $\{J_i\}$ are not
modified, but, they can contain many ghosts which are clustered together with
real particles. In addition, now ghost may also cluster among themselves leading
to formation of jets with no physical particle -- the pure \emph{ghost jets}.

The active scalar area of a jet $J$ is defined \cite{Cacciari:2008gn} as a
number of ghosts contained in this jet per the density of ghosts per unit area 
averaged over many sets of ghosts. 
If the number of ghosts from a particular ghosts ensemble~$\{g_i\}$ clustered with the jet $J$ is ${\cal N}_{\{g_i\}}(J)$ and the number of ghosts from this ensemble per unit area is $\nu_{\{g_i\}}$ then the active scalar area is given by
\begin{equation}
    \label{eq:active-area-def}
    A(J) \equiv 
    \lim_{\nu_{g} \to \infty}
    \left\langle A(J\,|\,\{g_i\}) \right\rangle_g\,,
    \qquad
    A(J\,|\,\{g_i\}) = 
    \frac{{\cal N}_{\{g_i\}}(J)}{\nu_{\{g_i\}}}\,,
\end{equation}
where in addition to the limit of the infinite density of ghosts, the average
over many sets of ghosts is taken. The latter is necessary since the ratio
${\cal N}_{\{g_i\}}(J)/\nu_{\{g_i\}}$ depends on the particular set of ghosts
even in the limit of high $\nu_{\{g_i\}}$. Therefore, one also defines the
standard deviation of the distribution for the active area over many ghosts
ensembles
\begin{equation}
    \label{eq:sd-active-area-def}
    \Sigma^2(J) =
    \lim_{\nu_{g}\to \infty}
    \left\langle A^2(J\,|\,\{g_i\}) \right\rangle_g - A^2(J)\,.
\end{equation}
The active area is meant to measure the susceptibility of a jet to the soft
radiation which is uniform and whose density is high.

Similarly to the scalar area also the 4-vector active area may be defined as
\begin{equation}
    \label{eq:active-vec-area-def}
    A_\mu(J) \equiv 
    \lim_{\nu_{g} \to \infty}
    \left\langle A_\mu(J\,|\,\{g_i\}) \right\rangle_g\,,
    \qquad
    A_\mu(J\,|\,\{g_i\}) = 
    \frac{1}{\nu_{g}\mean{p_{tg}}} \sum_{g_i\in J} p_{\mu g}\,,
\end{equation}
where $\mean{p_{tg}}$ is the average ghost transverse momentum. The 4-vector area will prove useful in section \ref{sec:active-mass-area} were shall discuss the active mass area. 
For small jets, the scalar area and the transverse component of the 4-vector
area are virtually equal, $A(J) \simeq A_t(J)$, and $A_\mu$ is a massless vector
which points in the direction of the jet. For larger, jets the 4-vector area
becomes massive and its direction differs from that of the jet.

The active area can be studied numerically for any infrared safe jet clustering
algorithm, most easily using the FastJet package
\cite{Cacciari:2005hq,FastJet}. In addition, the analytic results can be
obtained in some cases for the anti-$k_t$ algorithm and for SISCone.

\begin{table}[t]
   \centering
   \begin{tabular}{ccccc} \toprule
   algorithm   & \multicolumn{2}{c}{$A/(\pi R^2)$}  &
   \multicolumn{2}{c}{$\Sigma/(\pi R^2)$}  \\ \midrule
               & 1-particle-jet & ghost-jet & 1-particle-jet & ghost-jet\\ \cline{2-5}
   $k_t$       &  0.812         & 0.554     &  0.277         & 0.174    \\ 
   C/A         &  0.814         & 0.551     &  0.261         & 0.176    \\
   SISCone     &  1/4           &  --       &  0             & --       \\ 
   anti-$k_t$  &  1             &  --       &  0             & --       \\ 
   \bottomrule
   \end{tabular}
   \caption{
       Summary of the results from \cite{Cacciari:2008gn, Cacciari:2008gp} for
       active areas and their fluctuations in the case of 1-particle and
       pure ghost jets.  The numbers for the $k_t$ and C/A algorithms where
       obtained from numerical study with FastJet
       \cite{Cacciari:2005hq,FastJet} whereas those for anti-$k_t$ and SISCone
       represent exact values from analytic calculations.  
       All results are normalised to $\pi R^2$.  The results for pure ghost
       jet areas are not shown for SISCone and anti-$k_t$.  In the first case
       they depend strongly on the spilt-merge parameter, $f$,  while in the
       second case the distribution has two peaks at 0 and $\pi R^2$.
    }
   \label{tab:av-active-areas}
\end{table}

Unlike the passive area, the active area of the 1-particle jet may differ significantly from the naive expectation of $\pi R^2$. Firstly, in that it is in general a rather broad distribution over many ghost ensembles and secondly in that the average value of this this distribution may lay below $\pi R^2$. 
This is illustrated in table~\ref{tab:av-active-areas}, which summarises the results for the average active scalar areas of 1-particle and pure ghost jets and the corresponding standard deviations from four clustering algorithms obtained in \cite{Cacciari:2008gn, Cacciari:2008gp}. 
We see that the average values for the $k_t$ and C/A algorithms are
significantly lower than $\pi R^2$ with pure ghost jets having smaller jet area
than the jets with 1 hard particle. Moreover, the values of standard deviations
indicate that the distribution of active jet areas is rather broad. 
The anti-$k_t$ algorithm is special in that its 1-particle-jet active area is equal to the passive area $\pi R^2$ and does not fluctuate \cite{Cacciari:2008gp}.
For the SISCone algorithm, the active area of a single-particle jet can be calculated exactly \cite{Cacciari:2008gn} and it turns out that its value is four time smaller than that of passive area.
The active areas of ghost jets for SISCone and anti-$k_t$ exhibit somewhat more
complex behaviour. For the former the results depend on the split-merge
parameter, $f$, and for the latter the distribution has two peaks at 0 and $\pi
R^2$. That is why we do not show them in table~\ref{tab:av-active-areas}.

As in the case of passive area, discussed in the preceding subsection, also
here, adding a second particle to the system has a significant effect on the
active area of the hardest jet.
This is illustrated in Fig.~\ref{fig:passive-area2p} (right) for the case of
$p_{t2} \ll p_{t1}$, which was considered in \cite{Cacciari:2008gn}.  As we see,
the behaviour depends significantly on the algorithm. The active areas from
$k_t$ and C/A exhibit similar shape to the passive areas differing with the
latter mostly by about 20\% lower normalisation. The anti-$k_t$, as expected,
gives the same result for passive and active area.  The most drastic change is
seen for the SISCone algorithm for which the active area is almost factor four
smaller than the passive one (c.f. table ~\ref{tab:av-active-areas}).

\section{Areas for general case of 2-particle system}
\label{sec:area-general-2p}

In the current study we are interested in mass of a jet and the way it is
affected by soft background. The main contribution to jet mass comes from its
substructure. This substructure may originate, e.g., from QCD splittings.
In this case, the results for the areas of jets consisting of two strongly ordered
particles, obtained in \cite{Cacciari:2008gn} and reviewed briefly in the
previous section, are adequate. 
However, if the two constituents of our simple jet come from a~decay, then 
their transverse momenta are comparable and one expects such a jet to have
different properties.

In order to be able to discuss the problems related to jet masses for the whole
spectrum of cases between those two extremes of $p_{t2} \ll p_{t1}$ (QCD jets)
and $p_{t2} \sim p_{t1}$ (jets from decay), as a preparatory step, we will
generalise the results for jet areas from \cite{Cacciari:2008gn} to 
the case of two particles with arbitrary transverse momenta.
 
It will be convenient to quantify the relative hardness of the two particles,
$p_1$ and $p_2$, in terms of the variable
\be
\label{eq:z-def}
z = \frac{\min(p_{t1}, p_{t2})}{p_{t1}+p_{t2}}\,,
\ee
which, by definition, is always in the range $0 \leq z \leq 1/2$.

The main difference with respect to the case discussed in the previous section
will be that now, when the particles $p_1$ and $p_2$ are combined, the jet
$J_{12}$ may be centred anywhere between the positions of these two particles.
Before, such jet was centred always at the harder particle. The exact
values of $(y_{J_{12}}, \phi_{J_{12}})$ will depend on the recombination scheme
used as part of a jet definition. Out of several existing schemes, we adopt for
the study presented in this paper the widespread $E$-scheme which combines
particles by simply adding their 4-momenta. Apart from being very intuitive and
preserving Lorentz symmetry, it has been also recommended in \cite{Blazey:2000qt}.

For the 2-particle system, the centre of the jet will lie on the line segment
bounded by the positions of the particles. Therefore, the results for mass areas
from the four algorithms that we are going to study can depend only on the
distance along this line, from one (any) of the particles to the centre of the
jet. We will denote the  distance from the softer particle by $\Delta_J$.  It
will depend on $\dist$ and on the asymmetry parameter~$z$. 

For convenience we also introduce the versions of $\dist$ and $\Delta_J$ normalised to the jet radius
\be
  \label{eq:x-xJ-def}
  x   \equiv \frac{\dist}{R} \,, \qquad \qquad \qquad
  x_J \equiv \frac{\Delta_J}{R} \,.
\ee
Employing the above definitions, we may write the explicit formula for $x_J$ in
the $E$-scheme valid in the small $R$ limit
\be
  \label{eq:xJ-Escheme}
  x_J (x, z) \simeq (1-z)\, x\,.
\ee
One notices that, since, according to the definition (\ref{eq:z-def}), $z \leq
1/2$, the softer particle is always further away from the centre of the jet
than the harder one.  The distance between the latter and the jet's centre being
$x-x_J$. 
In the limit of strongly ordered particle transverse momenta,  $x_J \to x$ and
the jet gets centred at the harder of the two particles.

\subsection{Passive areas}
\label{sec:passive-area-2p}

\begin{figure}[t]
    \centering
    \includegraphics[height=0.54\textwidth]{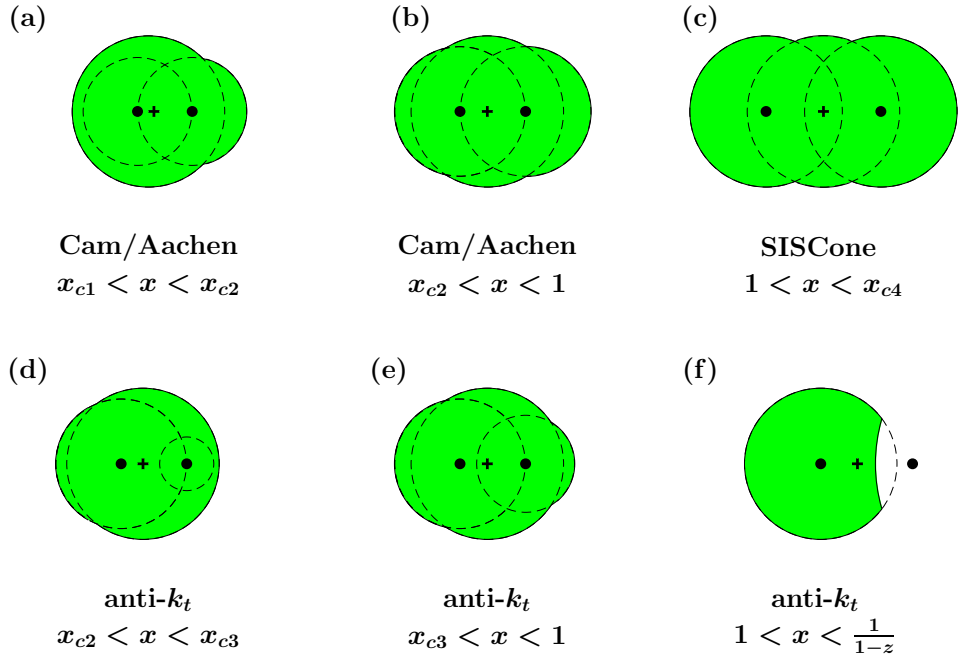}
\caption{%
    Representation of the passive area of the hardest jet in the system with two
    particles having arbitrary transverse momenta for various algorithms and
    interparticle separations. 
    The dots represent the particles and the cross centre of a jet.
    The distances $x$ and $x_J$ are defined in Eq.~(\ref{eq:x-xJ-def}) and the
    critical values of $x$ in Eqs.~(\ref{eq:xc1-solution}),
    (\ref{eq:xc2-solution}), (\ref{eq:xc4-approx}) and (\ref{eq:xc3-solution}).
    The asymmetry parameter $z$ is given in Eq.~(\ref{eq:z-def}).
}
\label{fig:2p-general-scheme}
\end{figure}

Since the system under consideration is simple and the hardest jet may consist
of at most two particles, its passive area can be calculated analytically. 
The result will depend on the order of clustering of particles $p_1$, $p_2$ and
the ghost $g$. This order is different for each algorithm hence the passive
areas will vary across them.  
As mentioned in the introduction, we work in the small $R$ approximation. One
consequence of that is that we treat the directions $y$ and $\phi$
in the $(y,\phi)$ plane on equal footing.

\paragraph{The $\boldsymbol{k_t}$ algorithm} with its 2-particle distance
measure, $d_{ij} = \min(p^2_{ti}, p^2_{tj}) \Delta_{ij}^2/R^2$, and
beam-particle measure, $d_{iB} = p_{ti}^2$, will always cluster the ghost first
with either one of the particles $p_1$, $p_2$ or the beam. In the latter case,
the contribution to the area is zero. In the former case, the ghost clusters
with the particle which is geometrically closer according to the distance
$\Delta_{ig}$ regardless of the relative hardness of particles $p_1$ and $p_2$.
Therefore, the result will be independent of $z$ and will coincide with that
found in \cite{Cacciari:2008gn} and shown already in
Fig.~\ref{fig:passive-area2p} (left) of section~\ref{sec:passive-area}. The
corresponding formula can be found in the appendix~\ref{app:passive-formulae}.

\paragraph{The Cambridge/Aachen algorithm} does not take into account the
hardness of the particles undergoing the clustering but solely the geometric
distance $\Delta_{ij}$ between them according to the measures $d_{ij} =
\Delta_{ij}^2/R^2$ and $d_{iB}=1$. 
The clustering of the system of two particles $p_1$ and $p_2$ and the ghost
proceeds as follows. If the ghost is closer than $\dist$ to either of the
perturbative particles then it is clustered first with the closer one.
Subsequently, the particles $p_1$ and $p_2$ are clustered. If, however, the
distance between the ghost and the closer particle is greater than $\dist$ then
the two perturbative particles are clustered first forming the jet $J_{12}$
centred at the point in the line segment between the positions of the particles
$p_1$ and $p_2$ at the distance $\Delta_J$ from the softer particle. 
Then, the ghost may cluster with $J_{12}$ if its distance to the jet's
centre is smaller than $R$. Therefore, the area of a 2-particle jet in the C/A
algorithm is a union of two smaller circles of radius $\dist$, centred
respectively at the particles $p_1$ and $p_2$ and the big circle with radius $R$
centred at the jet $J_{12}$.

The range $0 < x < 2$ consists of four distinct sub-ranges. The two critical
values of $x$, which we denote as $x_{c1}$ and $x_{c2}$, correspond to the
situations where one or two of the small circles start sticking out of the big
circle, as depicted in Fig.~\ref{fig:2p-general-scheme}\,(a) and (b).  
For $x$ below $x_{c1}$  or above $1$  the results will
not depend on the asymmetry parameter, $z$, and will be identical with those
found in
\cite{Cacciari:2008gn}. 

The conditions for the critical values of $x$ are given by
\bea
\label{eq:xc1}
x_{c1} + x_J     & = & 1\,, \\
\label{eq:xc2}
2\, x_{c2} - x_J & = & 1\,.
\eea
In the limit of small $R$ the approximate solutions have the following simple
forms
\bea
\label{eq:xc1-solution}
x_{c1}(z)  & \simeq & \frac{1}{2-z}\,, \\
\label{eq:xc2-solution}
x_{c2}(z)  & \simeq & \frac{1}{1+z}\,. 
\eea

\begin{figure}[t]
    \centering
    \includegraphics[width=0.45\textwidth]{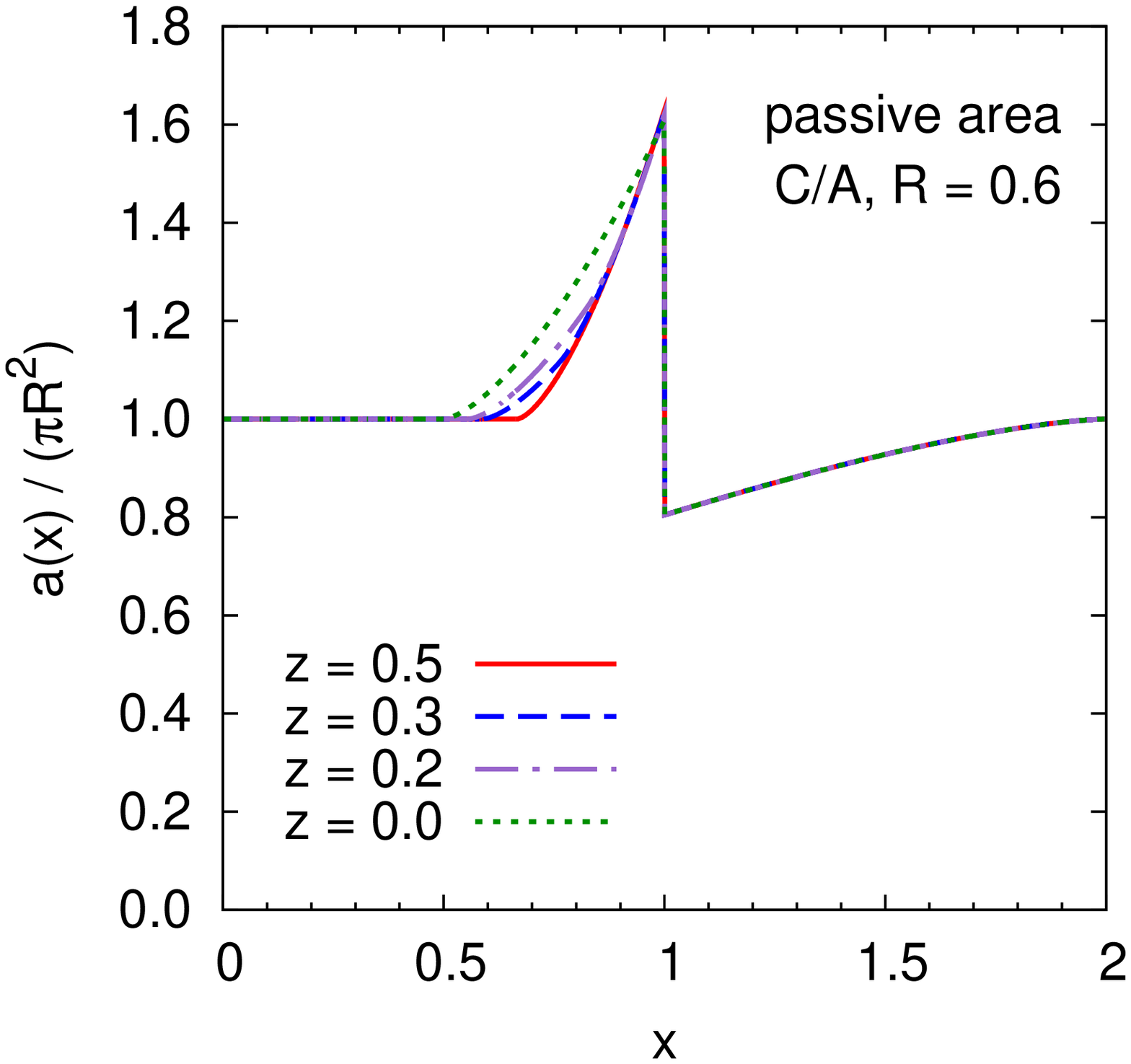}
    \hspace{20pt}
    \includegraphics[width=0.45\textwidth]{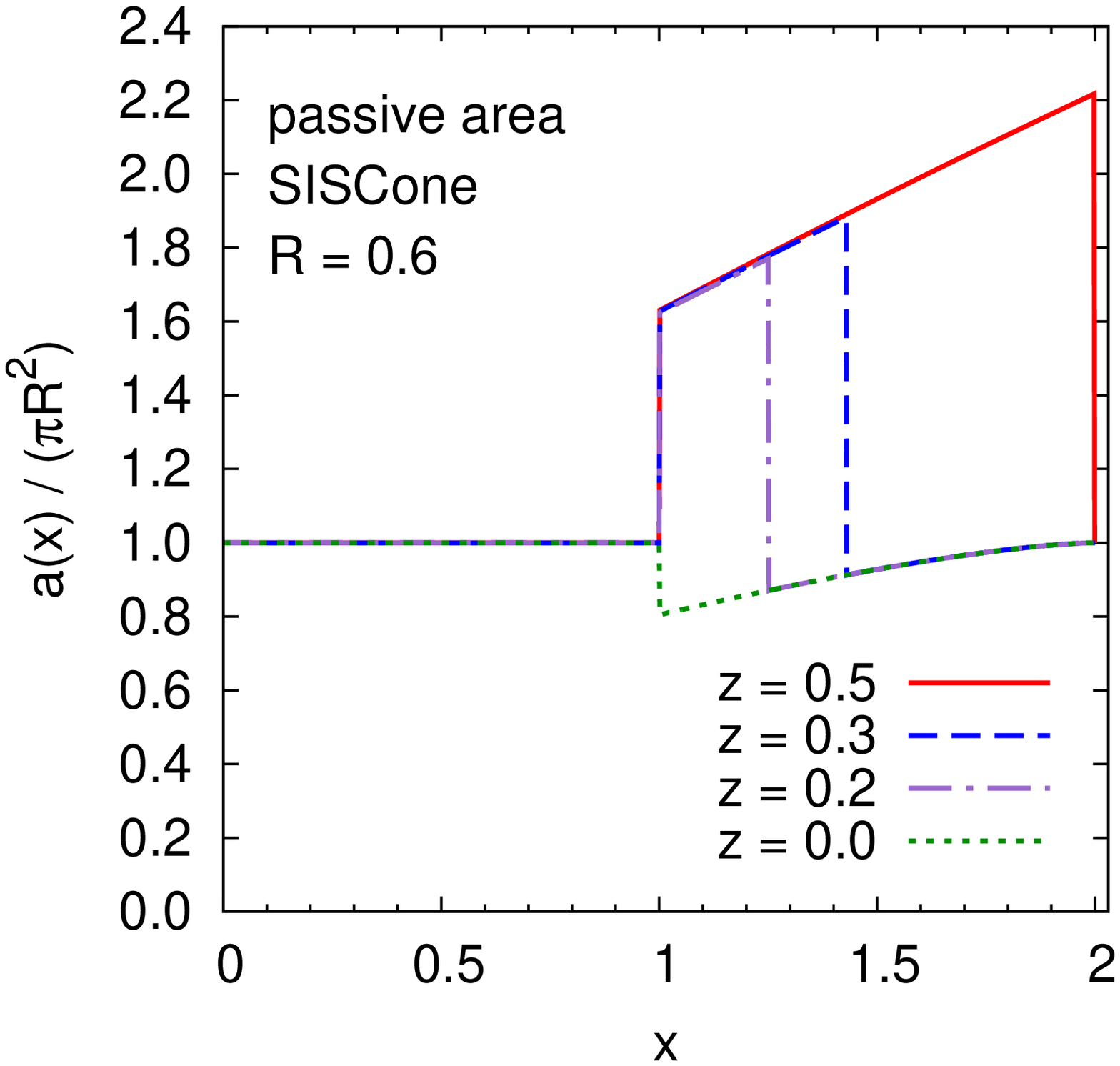}
    \caption{
    Passive areas of the hardest jet in the system of two particles with
    arbitrary ratio of transverse momenta, as functions of the interparticle
    separation, $x$, for C/A (left) and SISCone (right). The parameter $z$ is
    defined in Eq.~(\ref{eq:z-def}). The value $z=0$ corresponds to strongly
    ordered transverse momenta of the two particles and the value $z=0.5$ to the
    system of two particles of equal hardness. 
    The SISCone result does not depend on the value of $f$ parameter.
    }
    \label{fig:passive-area2p_general_sis+ca}
\end{figure}

The analytic result for the passive area from the C/A algorithm is given in
appendix~\ref{app:passive-formulae}.  The corresponding curves are shown in
Fig.~\ref{fig:passive-area2p_general_sis+ca} (left) for $R=0.6$ and several
values of the asymmetry parameters $z$. 
We note that the dependence on $z$ is mild.  In the limit $z\to 0$,
$x_{c1} \to 1/2$ and $x_{c2} \to 1$, and one recovers the
result for the system of two particles with strongly ordered transverse momenta
from \cite{Cacciari:2008gn}.  

\paragraph{The SISCone algorithm} looks for stable cones of radius $R$, which
are the cones whose direction coincides with the $E$-scheme sum of the momenta
of the particles inside. Those cones which overlap are subsequently split or
merged according to the  Tevatron run II type \cite{Blazey:2000qt} procedure.
This procedure starts from ordering stable cones according to the scalar sum of
the transverse momenta of their constituents, $\tilde p_t$.  Then, the $\tilde
p_t$ shared between the hardest jet and the next to hardest jet that overlaps
with it (with $\tilde p_{tj}$) is compared with  $f \tilde p_{tj}$, where $f$
is the overlap threshold parameter.  The cones are merged if $\tilde p_t >
f\tilde p_{tj}$ and split otherwise.

For $x<1$ only one stable cone is found, with its centre between the particles
$p_1$ and $p_2$. Any ghost within this cone belongs to the jet. Therefore the
area is identical to that of a single particle jet. 

For $1<x<2$ two stable cones are always found, centred at particles $p_1$ and
$p_2$ respectively. In addition, for $1<x<x_{c4}$ a third stable cone is found
containing both particles.  The third cone is stable as long as the distance
between the jet's centre and the softer particle is smaller than $R$.  This
gives the condition for  $x_{c4}$
\be
    x_J(x_{c4},z,R) = 1\,,
\ee
which in the limit small $R$ leads to
\be
    \label{eq:xc4-approx}
    x_{c4}(z)  \simeq \frac{1}{1-z}\,,
\ee
and since, according to the definition (\ref{eq:z-def}), $0 \leq z \leq
1/2$, the above critical value stays in the range $1 \leq x_{c4} \leq 2$.
As a next step, one has to check if the overlapping cones have a chance to be
merged.
As shown in Fig.~\ref{fig:2p-general-scheme}\,(c), all the three cones overlap
in the region $1<x<x_{c4}$. The central cone has the largest $\tilde p_t$ and
the amount of $\tilde p_t$ shared with the left jet is $ p_{t1}$ since the two
jets have only one common particle. The condition for merging the left and the
central cone is  $\tilde p_{t1} > f \tilde p_{t1}$ and it is always satisfied.
Similarly, the right cone will always be merged with the middle cone. 

Therefore, in the region $1<x<x_{c4}$ the jet area will be given by the
area of the union of the three circles, depicted in
Fig.~\ref{fig:2p-general-scheme}\,(c). 
In the region $x_{c4}<x<2$, only two stable cones are found with no common
particle so they are never merged. The two particles will end up in different
jets. The area of the harder one will be the same as for the $k_t$ and C/A
algorithms in this range of $x$.

The final formula for the passive area in the SISCone algorithm is given in 
appendix~\ref{app:passive-formulae}.
The corresponding curves are shown in
Fig.~\ref{fig:passive-area2p_general_sis+ca} (right) for $R=0.6$ and four values
of $z$.
Contrary to the $k_t$ and C/A algorithms, here the dependence on the asymmetry
parameter, $z$, is very strong for $x>1$. We note that the average 
area in this region is bigger by the factor of around two for jets consisting
of two subjets with comparable $p_t$ with respect to the jets whose constituents
are strongly ordered in transverse momenta.
As expected, in the limit $z\to 0$ one recovers the result from the
Fig.~\ref{fig:passive-area2p}~(left) since $x_{c4} \to 1$ and the third stable
cone cannot exist for any value of $x$.

\paragraph{The anti-$\boldsymbol{k_t}$ algorithm} is a sequential recombination
algorithm with hierarchy inverted with respect to the $k_t$-algorithm by using
the measures $d_{ij} = \min(p^{-2}_{ti}, p^{-2}_{tj}) \Delta_{ij}^2/R^2$ and
$d_{iB} = p_{ti}^{-2}$ .  The hardest particle in a system will cluster first
with anything within the geometric distance $\Delta < R$.
In the event with two particles of arbitrary transverse momenta and a ghost the
three competing distances are $\Delta_{1g}$, $\Delta_{2g}$ and $\dist$.  

For $0 < x < 1$, the events in which the distance between the ghost and one of
the physical particles is the smallest lead to formation of two small circles
around particles $p_1$ and $p_2$ as depicted in
Figs.~\ref{fig:2p-general-scheme}\,(d)~and~(e). If, however, $\Delta_{12} <
\Delta_{1g}, \Delta_{2g}$, then the real particles are clustered first, leading
to the jet $J_{12}$, which subsequently clusters with the ghost provided that
the distance between the two is smaller than $R$. 
Up to a certain value of $x$, which we
denote as $x_{c2}$, the two small circles are contained in the big
circle of radius $R$ and the area is simply that of 1-particle jet, i.e. $\pi
R^2$. Above $x_{c2}$ the circle centred at the harder of the two
particles protrudes and one gets the configuration shown in
Fig.~\ref{fig:2p-general-scheme}\,(d). Then,  above $x_{c3}$, the second of the
two small circles starts sticking out leading to the jet depicted in
Fig.~\ref{fig:2p-general-scheme}\,(e). 
 
The conditions for the aforementioned critical values of $x$ are given by
\bea
\label{eq:xc2-akt}
2\, x_{c2} - x_J & = & 1\,,\\
\label{eq:xc3}
\frac{z}{1-z}\, x_{c3} + x_J & = & 1\,.
\eea
The approximate solutions to each of these equations can found for $R
\lesssim 1$
\bea
\label{eq:xc2-solution-akt}
x_{c2}(z)   & \simeq & \frac{1}{1+z}\,,\\
\label{eq:xc3-solution}
x_{c3}(z) & \simeq & 
    \frac{1-z}{1-z+z^2}\,. 
\eea

For $1 < x < 2$ the particles $p_1$ and $p_2$ form two separate jets. However,
the shape of the jet centred at the harder of the two particles will not be
entirely conical.  The presence of the second particle will cause  it to be
clipped. This situation is shown in Fig.~\ref{fig:2p-general-scheme}\,(f). The
boundary~$b$ between the jets $J_1$ and $J_2$ is defined by $z\Delta_{1b} =
(1-z)\Delta_{2b}$. 
Hence, it turns out that the area of the jet $J_1$ will be reduced with respect
to $\pi R^2$ by the area of the overlap region of the circle of radius $R$
around that jet and the circle of radius $\frac{(1-z)z}{1-2z} \Delta$ and the centre
 away by $\frac{(1-z)^2}{1-2z}\Delta$ from the centre of the jet $J_1$.
Above $x=1/(1-z)$ the two circles do not overlap and the area of the harder jet
becomes perfectly conical.

\begin{figure}[t]
    \centering
    \includegraphics[width=0.45\textwidth]{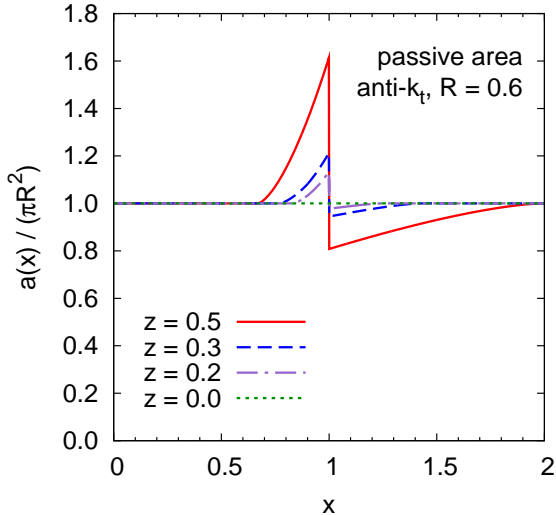}
    \caption{
    Passive jet areas for the system of two particles with arbitrary ratio of
    values of transverse momenta from the anti-$k_t$ algorithm. The asymmetry
    parameter $z$ is defined in~Eq.~(\ref{eq:z-def}) and $x$ is the distance
    between the two particles in the units of $R$.
    }
    \label{fig:passive-area2p_general-akt}
\end{figure}
In Fig.~\ref{fig:passive-area2p_general-akt}, we show the curves corresponding
to the analytic results for the passive area from the anti-$k_t$ algorithm, which
can be found in appendix~\ref{app:passive-formulae}.
One notices that, in general, the anti-$k_t$ jets are not perfectly conical. If
a jet consists of two particles of comparable hardness separated by $\dist \sim
R$ its area deviates from $\pi R^2$, the more so the closer to each other are
the transverse momenta of the two constituents.
On the other hand, if the separation between two particles is smaller than
$1/(1+z)$ or greater than $1/(1-z)$ or if their transverse momenta are strongly
ordered the resulting area of harder jet is equal to that of a single particle
jet.
For the maximally symmetric system, corresponding to, $z=0.5$, the anti-$k_t$
result for passive area coincides with that from the C/A algorithm (cf. 
formulae from appendix~\ref{app:passive-formulae}). However the two algorithms
behave very different for $z < 0.5$.

\subsection{Active areas}
\label{sec:active-areas-2p}

\begin{figure}[t]
    \centering
    \includegraphics[width=0.45\textwidth]
        {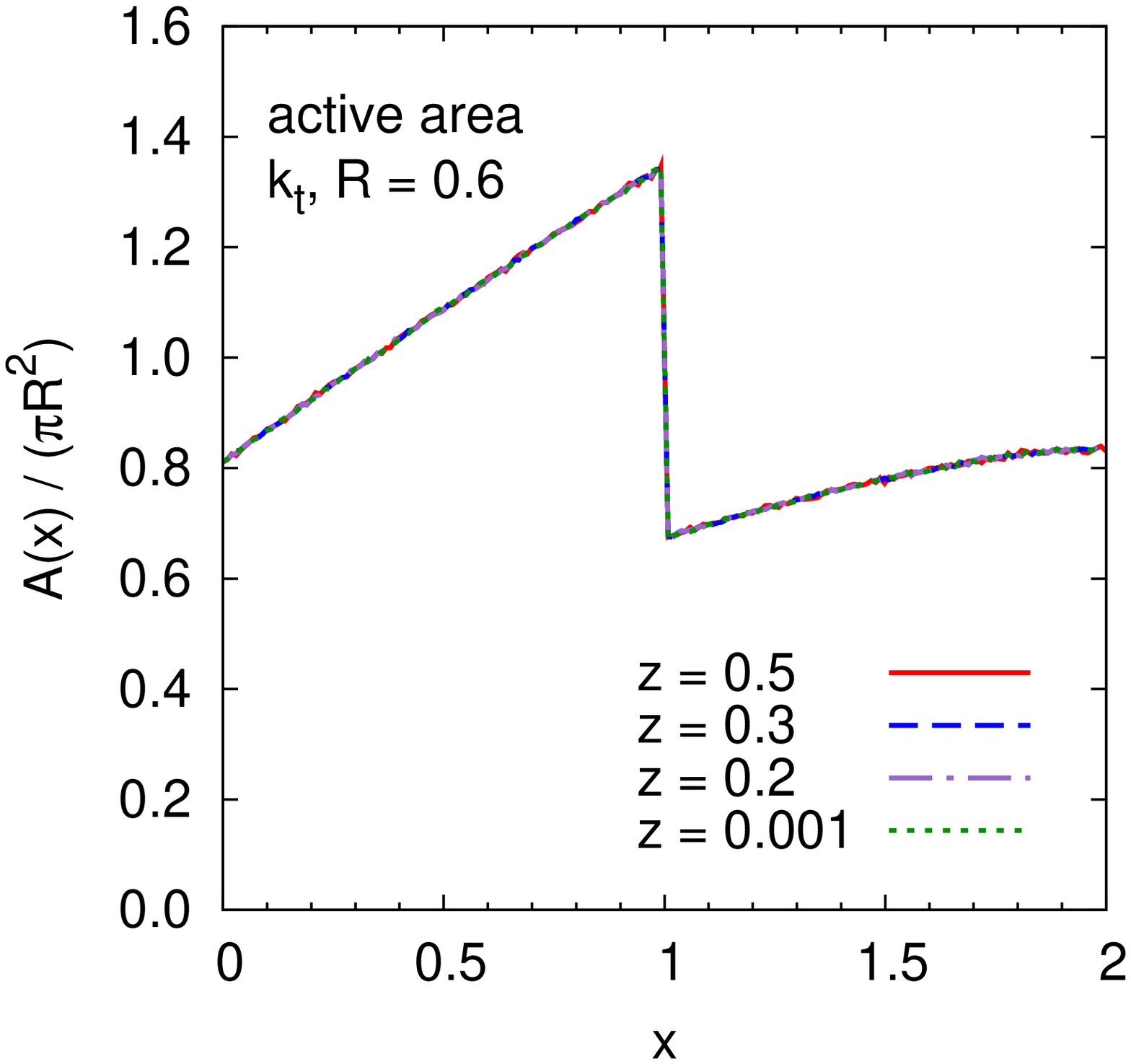}
    \hfill
    \includegraphics[width=0.45\textwidth]
        {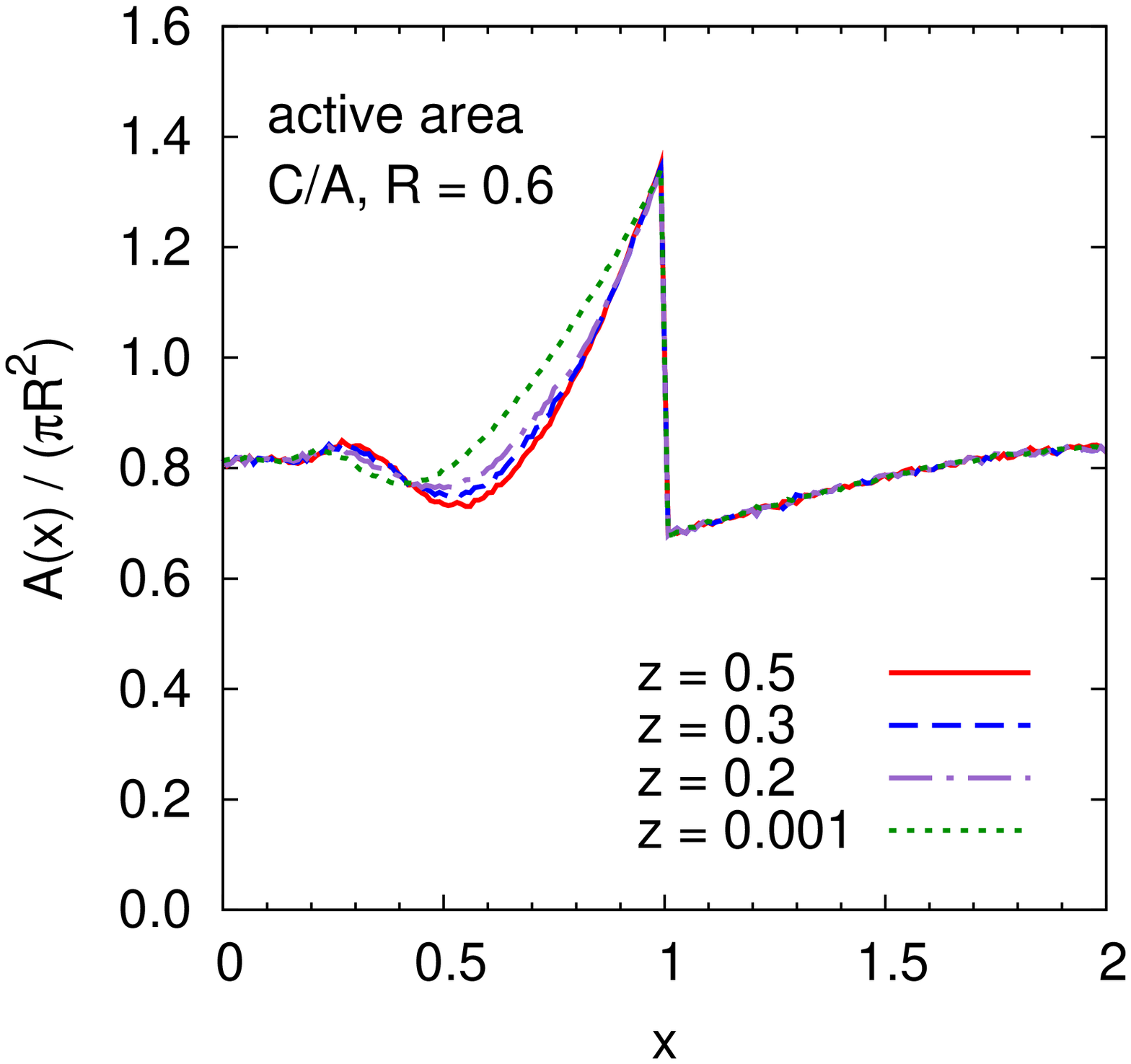}
    \vskip 5pt
    \includegraphics[width=0.45\textwidth]
        {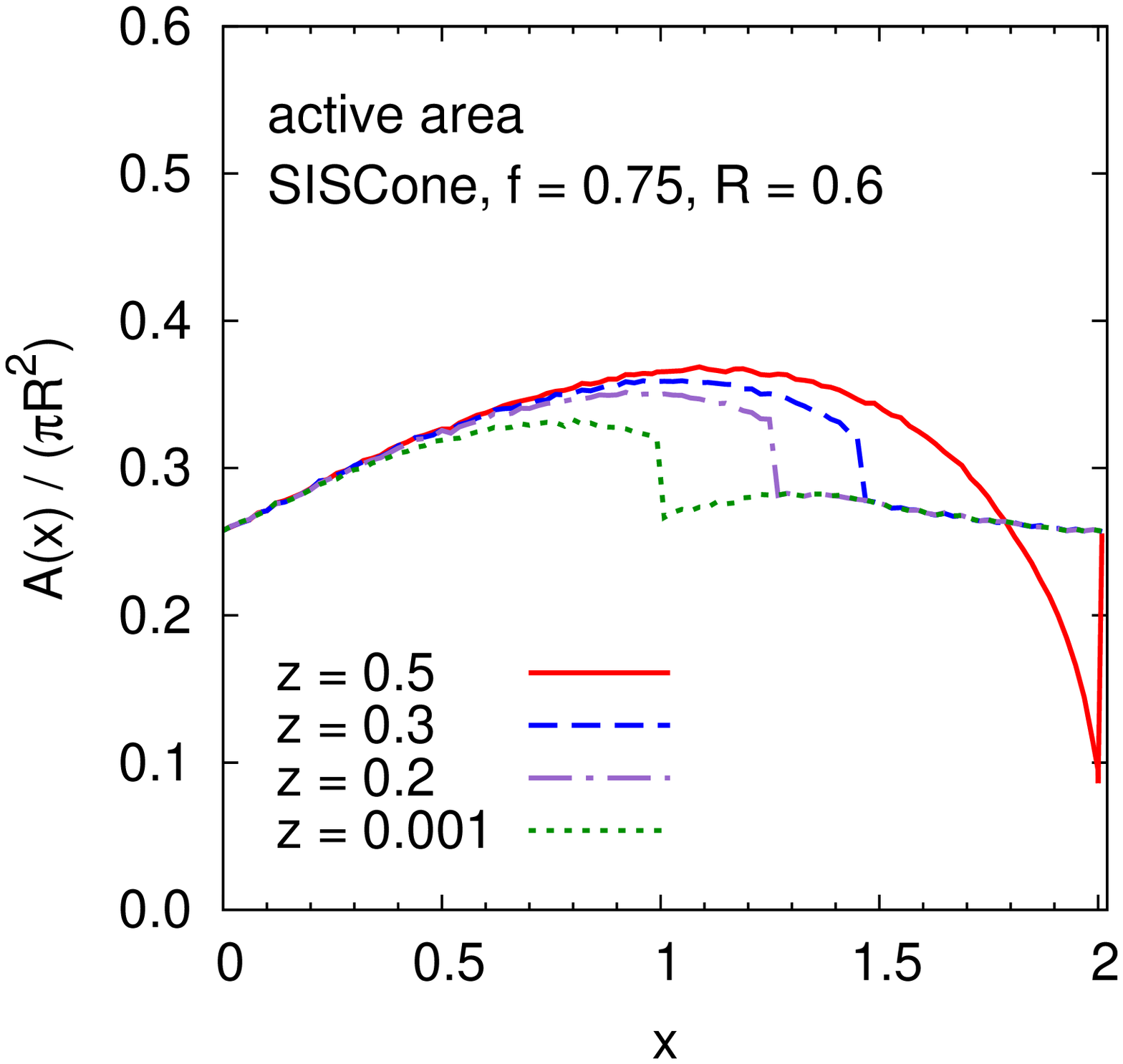}
    \hfill
    \includegraphics[width=0.45\textwidth]
        {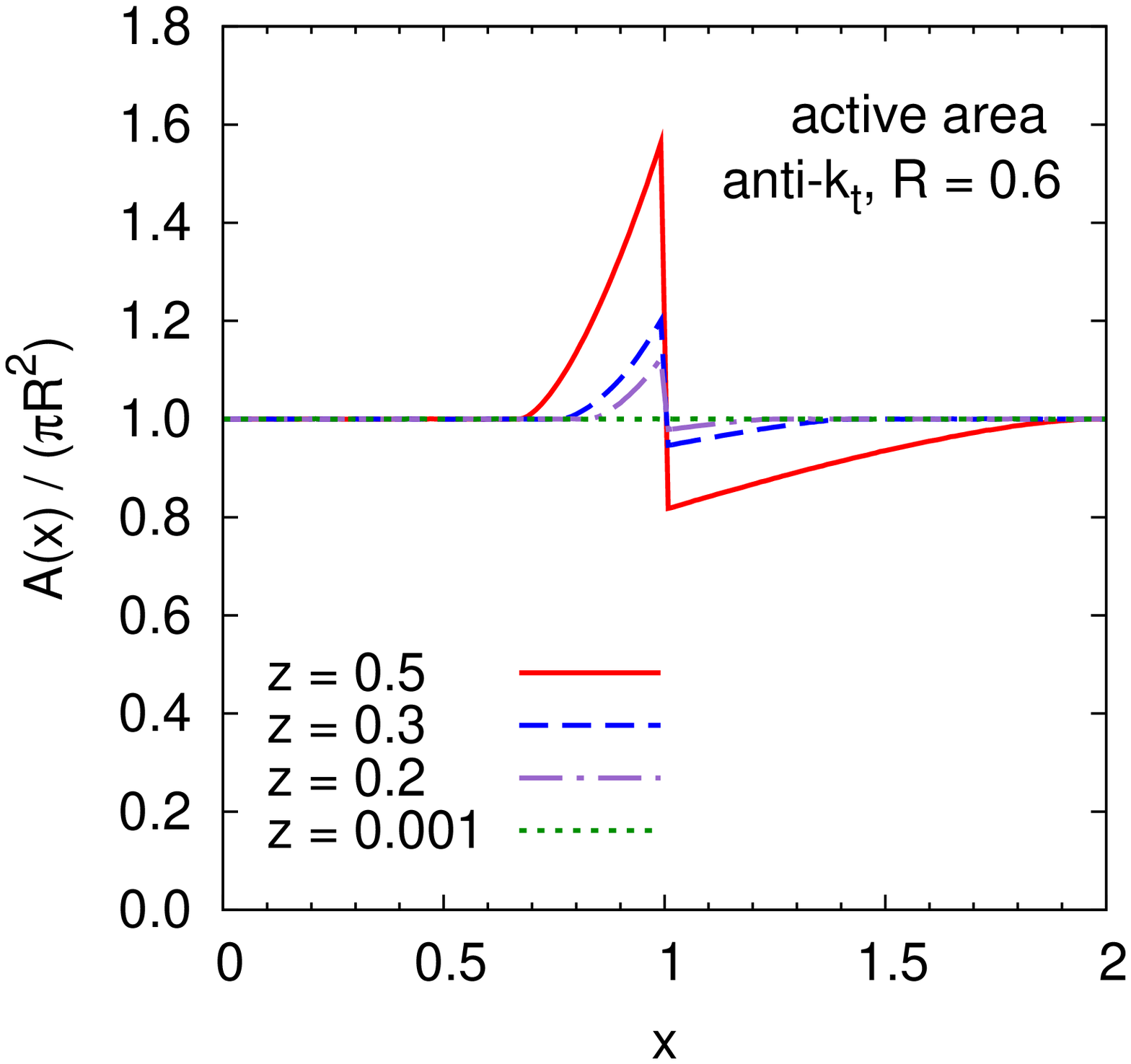}
    \caption{
    Active areas of the hardest jet for the system of two particles separated by the distance
    $x R$ in the $(y,\phi)$ plane and having arbitrary
    transverse momenta. The plots correspond to the $k_t$ (top left), C/A (top
    right), SISCone (bottom left) and anti-$k_t$ (bottom right) algorithms. 
    The asymmetry parameter, $z$, is defined in Eq.~(\ref{eq:z-def}).
    All the results obtained with FastJet~\cite{Cacciari:2005hq,FastJet}.
    }
    \label{fig:active-area2p_general_allalg}
\end{figure}

Since the computation of the active area involves clustering of a very complex
system with a large number of ghost particles, in general, one needs to rely on
the numerical analysis. 
As was the case for the passive areas also the active area results are expected
to vary between the algorithms. This is because each algorithm comes with a
different order of clustering of real particles and ghosts and it is this order
that governs the behaviour of jet areas.
 
The numerical analysis has a potential to produce slightly different results for
jets oriented along the $y$ or $\phi$ axis. This is because it operates on the
real phase space for which these directions are not equivalent. One expects,
however, that for the values of $R$ which are sufficiently small, the
corresponding differences should be largely subdominant.
In practice, all the results shown in this and the following sections correspond
to jets with the two constituent particles aligned along the rapidity axis.  We
have checked explicitly that the opposite extreme case of the particles oriented
along the $\phi$ axis gives virtually the same results for the jet areas.  The
situation for the mass areas discussed in section~\ref{sec:mass-area} is similar
except for certain configurations studied with SISCone algorithms which we will
comment on in due course.

We have studied the active areas of the hardest jet in the system of two
particles of arbitrary relative hardness. We performed analyses with the same
four jet definitions as discussed in the previous subsection, i.e. $k_t$, C/A,
SISCone and anti-$k_t$ together with the $E$-scheme for
particle recombination.
The results are presented in Fig.~\ref{fig:active-area2p_general_allalg}.
The overall picture is very similar to that from the study of passive areas
of preceding subsection.

The ``non-conical algorithms'', $k_t$ and C/A, exhibit either no dependence on
$z$, in the case of $k_t$, or only a weak $z$-dependence in the case of C/A. The
active areas from both algorithms behave very similarly to their passive area
counterparts. For $x<1$, where the two particles form a single jet, the $k_t$
active area grows steadily whereas the C/A active area stay practically constant
for low~$x$ and starts growing rapidly above certain value of $x$. The pattern
of mild $z$-dependence in the latter region is also the same as that seen for
passive area, i.e. the results are slightly smaller for more symmetric system of
two particles. For $x>1$, the hardest jet consists of a single particle and its
active area from $k_t$ and C/A does not depend on the asymmetry parameter $z$.
Apart from the similarity to the passive area results, the 2-particle active
areas are smaller by about $20\%$. This has already been observed for the
1-particle jets and the 2-particle jets with strong $p_t$-ordering 
in \cite{Cacciari:2008gn} and we have also recalled those results in
table~\ref{tab:av-active-areas} and Fig.~\ref{fig:passive-area2p}.

The SISCone algorithm gives the active area which depends quite strongly on $z$.
We see that, for $z\sim 0.5$, it stays well above the 1-particle result, $\pi
R^2/4$, also for $x>1$ and then it drops at some value, just as was the case
with passive area.  Again, this is related to the existence of the third stable
cone containing both particles $p_1$ and $p_2$. Below a critical value of $x$,
the same as that given in Eq.~(\ref{eq:xc4-approx}), this third stable cone is
being merged leading to large jets. However, there is also a difference between
the cases of passive and active areas for large $z$ and $x>1$, namely in that
the active area falls with $x$ for $1<x<x_{c4}$ whereas the passive one keeps
growing in this region.
The mechanism responsible for this effect is the same as that which leads to the
reduction of the 1-particle active SISCone area by the factor 1/4 with respect
to the passive area of 1-particle jet as explained in \cite{Cacciari:2008gn}. It
is related to additional splittings of stable cones with physical particles
which overlap with stable cones built up solely of ghosts. Such splittings
involving the central stable cone from Fig.~\ref{fig:2p-general-scheme}\,(c)
lead to narrowing the jet with increasing $x$. 
This may lead to the active area of a jet containing two particles being smaller
than the active area of a 1-particle jet. As shown in
Fig.~\ref{fig:active-area2p_general_allalg} (bottom left) such situation indeed
happens for the system with $z$ close to its maximal value 1/2 (identical
transverse momenta of the particles $p_1$ and $p_2$).  In this case, the
critical value $x_{c4}$ is reached for very high $x$ (or never in the case of
$z=1/2$) and the 2-particle active area can smoothly decrease below the
1-particle result.

The anti-$k_t$ active area results shown in
Fig.~\ref{fig:active-area2p_general_allalg} (bottom right) are identical to the
passive areas from Fig.~\ref{fig:passive-area2p_general-akt}. This comes from
the fact that the ghost particles cluster among themselves only after all
clusterings involving perturbative particles. The equivalence of the passive and
active areas from anti-$k_t$ for the 2-particle jets with strongly ordered
transverse momenta of the two constituents, corresponding to $z=0$,  has been
pointed out in ~\cite{Cacciari:2008gp}. Their equivalence for arbitrary $z$,
illustrated in Figs.~\ref{fig:passive-area2p_general-akt} and
\ref{fig:active-area2p_general_allalg} (bottom right) is also known and  has
been taken into account in the FastJet program (see the code accessible
in~\cite{FastJet}).
As in the case of SISCone, also for the anti-$k_t$ algorithm there is a
region of strong $z$-dependence. 

For all the algorithms and all the $z$ values shown in
Fig.~\ref{fig:active-area2p_general_allalg}, the 2-particle active areas tend to
the 1-particle results in the limit $x\to0$.
However, in the limit $x\to2$ the results converge to the 1-particle area only
for the ``conical algorithms'', i.e. SISCone and anti-$k_t$. For $k_t$ and C/A
the 2-particle jet areas are different from the values given in
table~\ref{tab:av-active-areas} even if the separation $x>2$. This is related to
the fact that these algorithms build up the jets starting from formation of
local structures which are subsequently merged leading to jets of very irregular
areas.

\section{Mass area}
\label{sec:mass-area}

\subsection{Jet mass}

The mass of a light quark jet arises due to its substructure. 
If a jet $J_{12}$ is obtained from clustering two subjets $J_1$ and $J_2$ 
with masses much smaller than their transverse momenta, $m_{J_{1,2}} \ll p_{t
J_{1,2}}$, then the mass of the jet $J_{12}$ in the small $R$ limit is given by
\bea
m^2_{J_{12}} & \simeq & m^2_{J_1}+m^2_{J_2}+p_{t J_1} p_{t J_2} \distsq\\[0.5em]
             &   =    & m^2_{J_1}+m^2_{J_2}+ z (1-z)\, p_{t J_{12}} \distsq\,,
\eea
with $\distsq = (y_{J_1}-y_{J_2})^2 +(\phi_{J_1}-\phi_{J_2})^2$ and $z$ 
defined in Eq.~(\ref{eq:z-def}).

Jet mass is an infrared and collinear safe quantity that can be calculated order
by order in perturbation theory.
Because of the soft and collinear singularity of the QCD matrix element for
gluon emission, the distribution of masses of the QCD jets gets strong
enhancement for low values of $m_J$. At the lowest non-trivial order (i.e. NLO
of the perturbative $\alpha_s$ expansion) the approximate result
 for the mass distributions of QCD jets is given by \cite{Ellis:2007ib,
 Almeida:2008tp, Salam:2009jx}
$\frac{d\sigma}{dm_J} =
\alpha_s(p_{tJ}) \frac{4 C}{\pi m_J} \ln \left(\frac{R\, p_{tJ}}{m_J}\right)$,
where $C$ is the colour factor of the initiating parton.
The higher order terms are enhanced by further powers of $\ln \frac{R
p_{tJ}}{m_J}$. The resummed corrections are known for jets from
$e^+e^-$~\cite{Catani:1992ua, Burby:2001uz, Dasgupta:2001sh}
and DIS~\cite{Dasgupta:2002dc}.
Contrary to the case of QCD, the distribution of jets coming from decay of a
heavy object is flat in $z$ and therefore the mass distribution of such jets is
peaked around the mass of the heavy object which originated them.

As discussed in the preceding sections, the area of a jet provides a measure of
the susceptibility of the jet's momentum to soft background. Such a measure,
combined with a method of determination of the level of this background, like
the one discussed in \cite{Cacciari:2007fd, Cacciari:2009dp}, allows one to
account for the contamination from UE/PU and correct the momentum of the jet
accordingly.

Similarly, one can define a quantity which measures how much the \emph{mass} of
a jet can be modified by the soft radiation for jets defined with a given
algorithm.
In what follows, we define such a new characteristic of jets, which we call the
\emph{mass area}, and use it to study 1- and 2-particle jets from the four
jet-clustering algorithms.

\subsection{Passive mass area}
\label{sec:passive-mass-areas-2p}

In analogy to the passive jet area from section \ref{sec:passive-area}, the passive mass area of the jet $J$ can be defined as
\be
\label{eq:passive-ma-def}
a_m(J) =\! \int dy\, d\phi\, f_m (g(\phi,y),J),
\quad
f_m (g, J) = 
\Bigg \{\!\!
\begin{array}{cl}
 \frac{1}{p_{tJ}\, p_{tg}}(m^2_{Jg}\! - m^2_J) & {\rm for \ } g {\rm \ clustered\  with\ } J       \\
 0                                  & {\rm for \ } g {\rm \ not \ clustered\  with\ } J
\end{array},
\ee
where $m_{J}$ is the mass of the jet $J$,  $m_{Jg}$ is the mass of a jet that
consists of the jet $J$ and the ghost~$g$ and $p_{tg}$ is the transverse
momentum of that ghost. The passive area defined in the above equation is
dimensionless. Its value reflects susceptibility of the mass of a jet to the
contamination from soft radiation in the limit in which this radiation is
infinitely soft and pointlike.

For a jet consisting only of a single hard particle with transverse momentum
$p_{t1}$, the passive mass area for all the four algorithms is given by
\begin{equation}
\label{eq:passive-ma-1part}
a_{m}(\text{1-particle-jet}) = \frac{\pi}{2} R^{4}\,.
\end{equation}
The above result coincides with the polar moment of inertia of a disk (or
cylinder) of radius~$R$. This correspondence is general and, in fact, the
passive mass area defined in Eq.~(\ref{eq:passive-ma-def}) is nothing but the
polar moment of inertia, i.e.  the measure of resistance of an object to
torsion. This resistance is small if the mass is distributed close to the
rotation axis (here, the jet centre) and large if the mass extends far away
from the rotation axis.

\subsubsection{Passive mass areas for general case of 2-particle system}

The calculation of the passive mass areas for the system with two particles of
arbitrary $z$ proceeds in close analogy with the calculation of passive areas
for that system which led to the results presented in section
\ref{sec:passive-area-2p}.  In particular, all the subranges of the separation
variable $x$ and the corresponding pictures from
Fig.~\ref{fig:2p-general-scheme} are valid also for passive mass areas.
However, now the integrand in the definition given of
Eq.~(\ref{eq:passive-ma-def}) is less trivial. 

The total squared mass of the system composed of two massless perturbative
particles and the ghost $g$ is given by
\begin{equation}
\label{eq:mass-12g}
m^{2}_{12g} \simeq  p_{t1} p_{t2} \Delta ^{2}_{12} 
+ p_{t1} p_{tg} \Delta ^{2}_{1g}
+ p_{t2} p_{tg} \Delta ^{2}_{2g}\,.
\end{equation}
If the two particles $p_1$ and $p_2$ come from the soft QCD splitting the last
term is negligible. If, however, they come from a decay of a heavy
object, the last two terms are commensurate. 

\begin{figure}[t]
    \centering
    \includegraphics[width=0.45\textwidth]{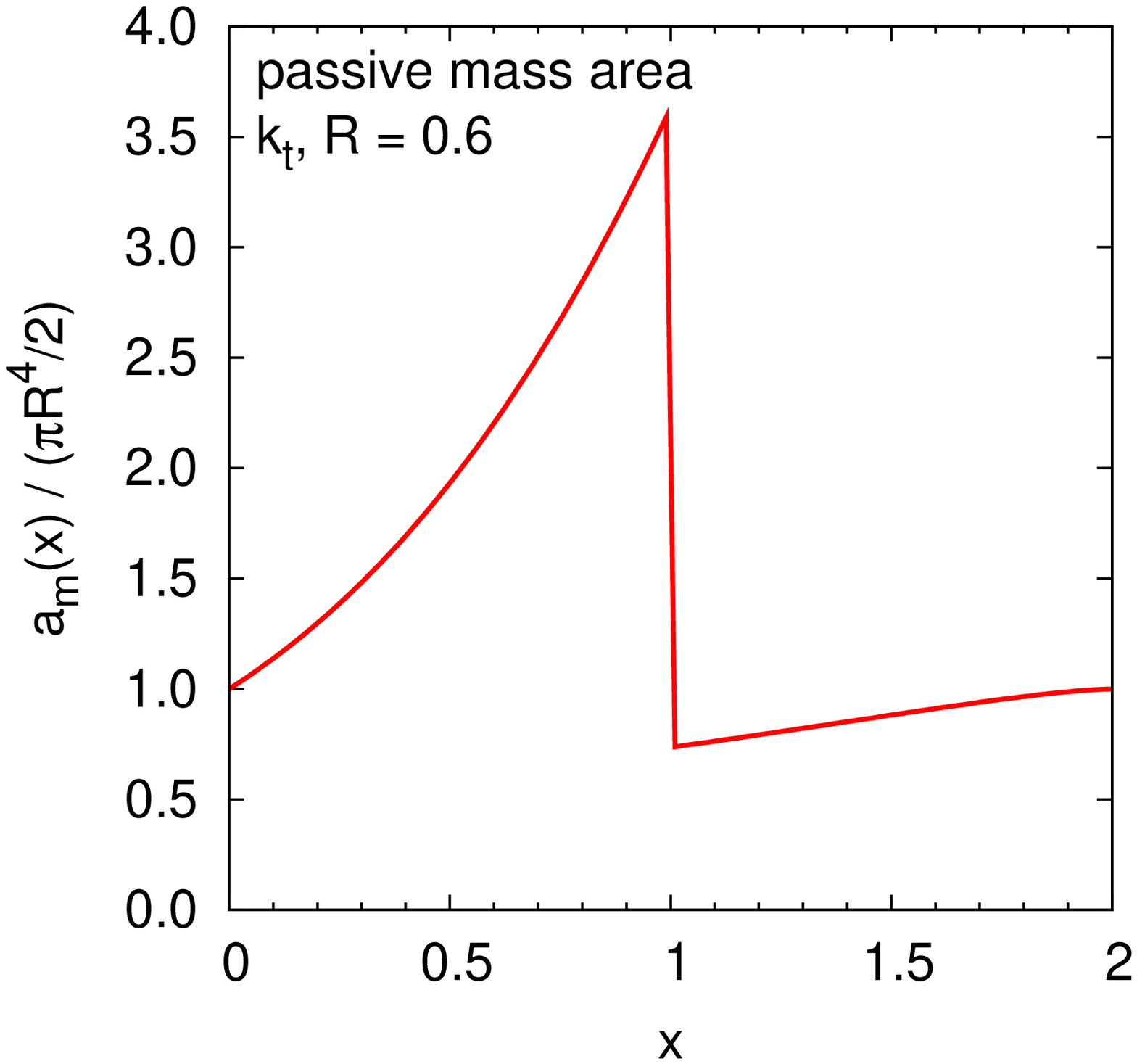}
    \hspace{20pt}
    \includegraphics[width=0.45\textwidth]{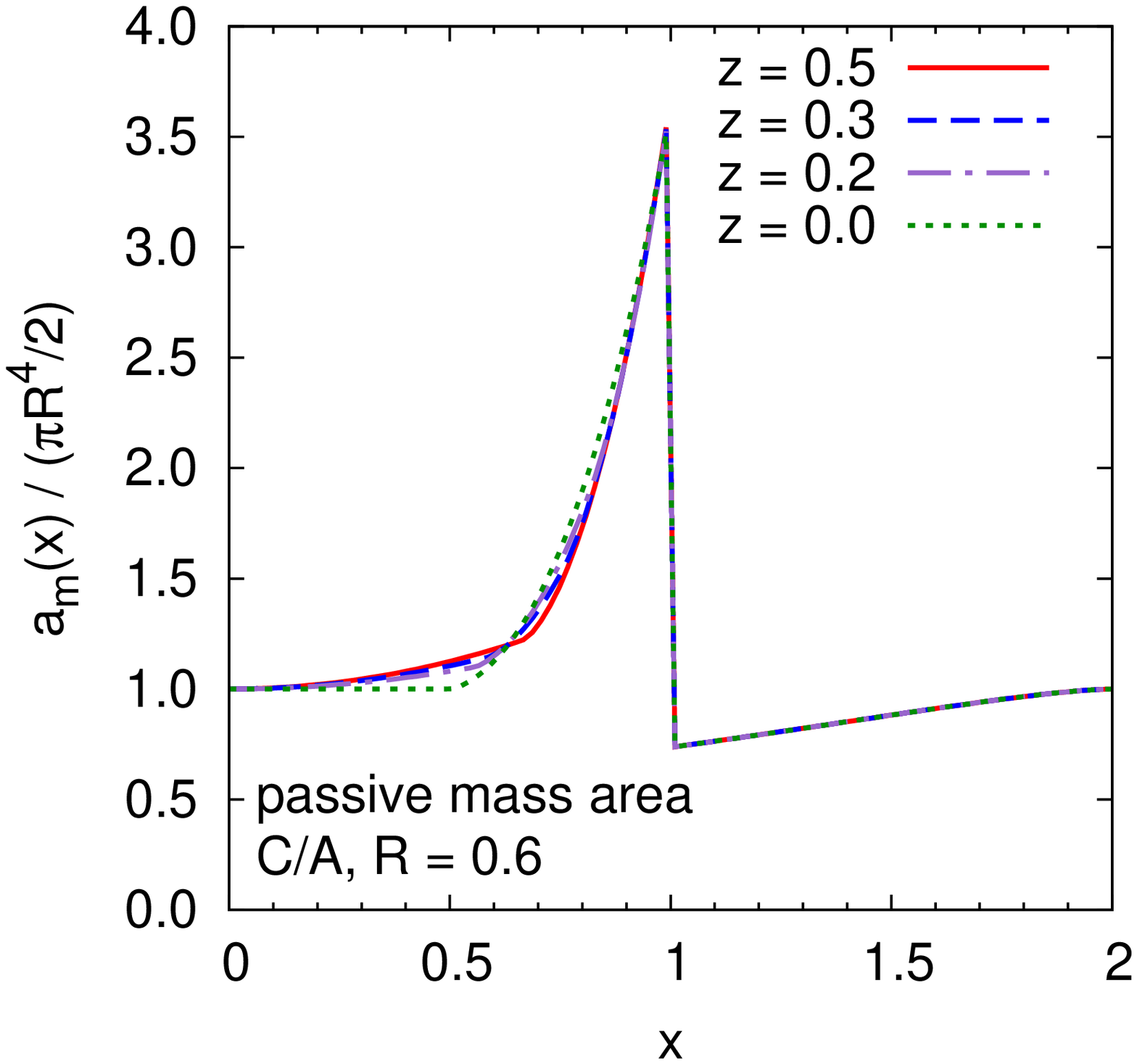}
    \caption{
    Passive mass areas of the hardest jet for the system of two particles with
    arbitrary ratio of transverse momenta in the case of the $k_t$ (left) and
    C/A (right) algorithms. The parameter $z$ is defined in Eq.~(\ref{eq:z-def})
    and $x$ is the interparticle distance in units of $R$. The value $z=0$
    corresponds to strongly ordered transverse momenta of the two particles and
    the value $z=0.5$ to the system of two particles of equal hardness.
    }
    \label{fig:passive-massarea2p_general_kt+ca}
\end{figure}

\begin{figure}[t]
    \centering
    \includegraphics[width=0.45\textwidth]{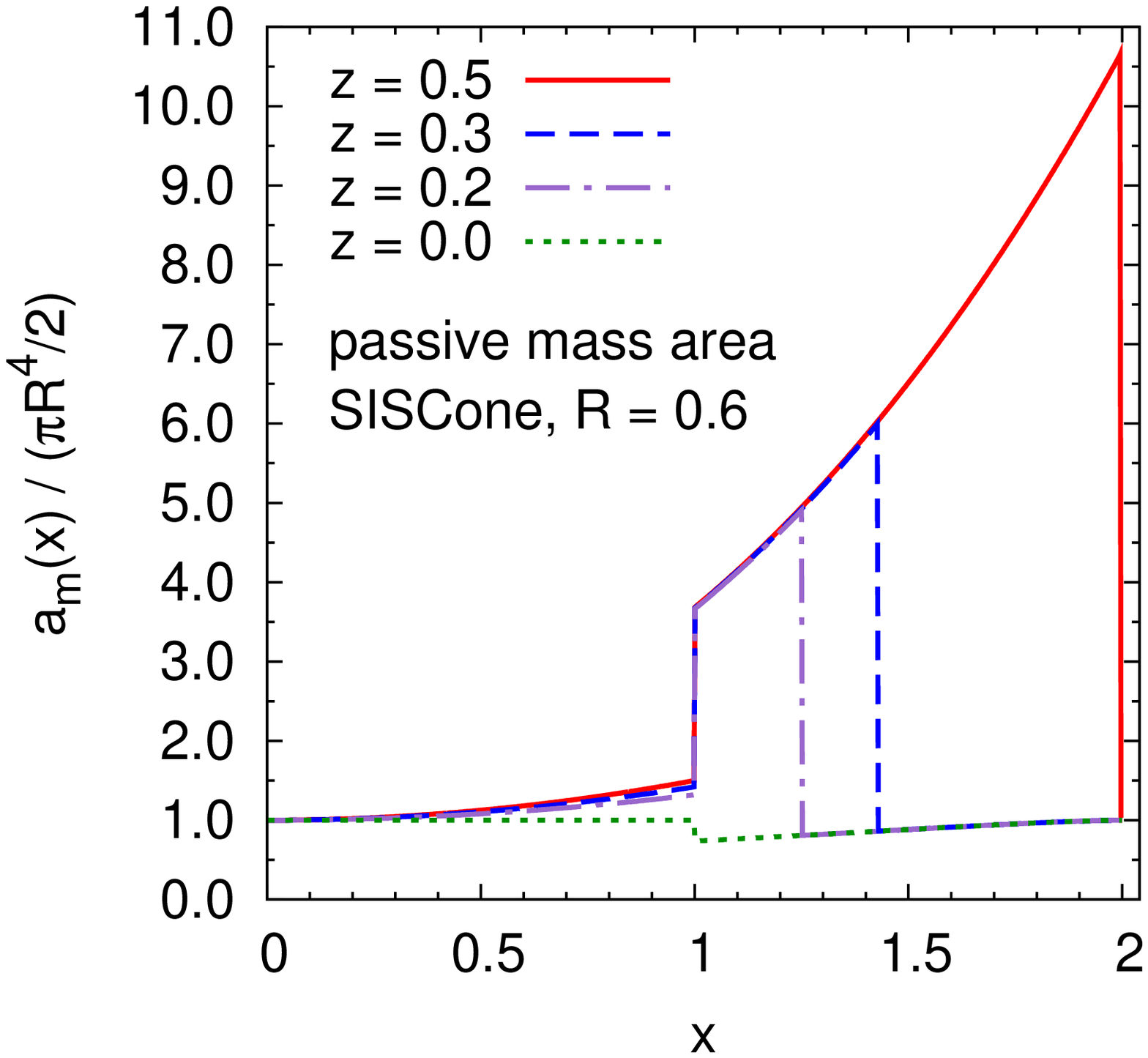}
    \hspace{20pt}
    \includegraphics[width=0.45\textwidth]{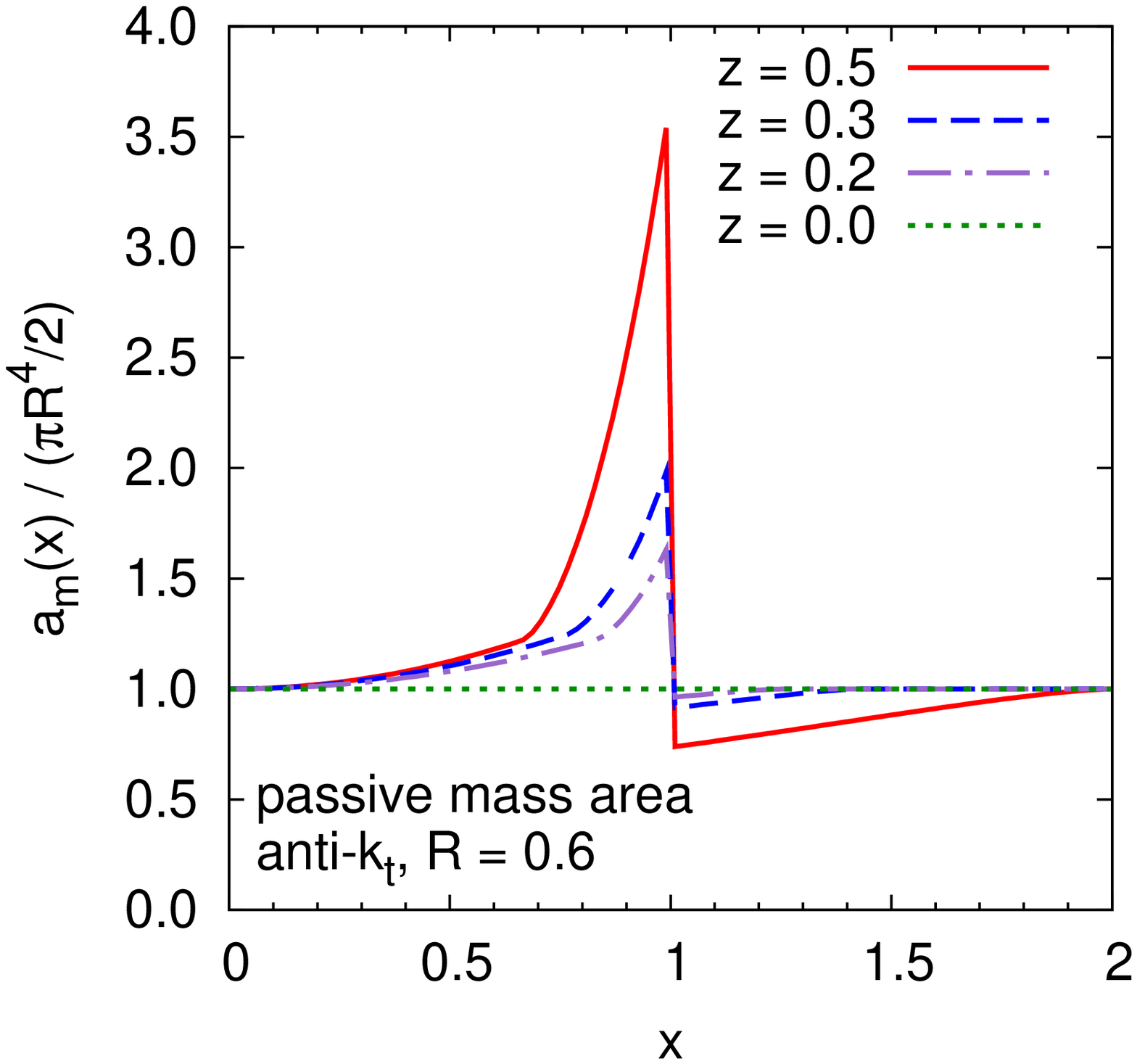}
    \caption{
    Passive mass areas of the hardest jet for the system of two particles with
    arbitrary ratio of transverse momenta in the case of SISCone (left) and
    anti-$k_t$(right). The parameter $z$ is defined in Eq.~(\ref{eq:z-def}). The
    value $z=0$ corresponds to strongly ordered transverse momenta of the two
    particles and the value $z=0.5$ to the system of two particles of equal
    hardness.  The SISCone result does not depend on the value of $f$ parameter.
    }
    \label{fig:passive-massarea2p_general_sis+akt}
\end{figure}

Plugging the above expression (\ref{eq:mass-12g}) into the definition
(\ref{eq:passive-ma-def}) allows one to obtain analytic results for passive mass
areas from all four algorithms. We give the corresponding formulae in 
appendix~\ref{app:passive-formulae}. Below, we comment on the results for each
of the four algorithms, which are shown in
Figs.~\ref{fig:passive-massarea2p_general_kt+ca} and
\ref{fig:passive-massarea2p_general_sis+akt}.

%
\paragraph{The $\boldsymbol{k_t}$ algorithm} produces jets whose passive mass
areas do not depend on the relative hardness of the two constituent particles.
This occurs in spite of the fact that the integrand in the definition
(\ref{eq:passive-ma-def}) with $m_{Jg}$ taken from Eq.~(\ref{eq:mass-12g}) does
depend on $z$. However, because the $k_t$ algorithm always clusters the ghost
first with one of the particles $p_1$ and $p_2$, the shape of the jet in the
$(y,\phi)$ plane has an additional reflection symmetry.  This, in turn, implies
that the integrated contributions from each particle differ only by the
multiplicative factor, $1-z$ for $p_1$ and $z$ for $p_2$.  Therefore, the
$z$-dependence cancels in the sum.  As shown in
Fig.~\ref{fig:passive-massarea2p_general_kt+ca} (left), the qualitative
behaviour of the mass area is the same as that of the area from
Fig.~\ref{fig:passive-area2p} (left). As long as the separation between the
particles $x<1$ the passive mass area grows fast with increasing~$x$. 
Quantitatively, however, the change of the passive mass area of the 2-particle
jet with respect to the 1-particle jet is much bigger than the corresponding
change for the passive jet area. As we see by comparing the results from
Figs.~\ref{fig:passive-area2p} (left) and
\ref{fig:passive-massarea2p_general_kt+ca} (left), the former changes by the
factor $\sim 3.6$ in the range $x < 0 < 1$ while the latter only by the factor
$\sim 1.6$.
For $x>1$ the hardest jet consists solely of a single particle. However, the
presence of the second jet in the neighbourhood causes its mass area to be
slightly smaller than $\pi R^4/2$. The value of a single particle mass area is
being slowly approached as we go to $x=2$.

\paragraph{The Cambridge/Aachen algorithm} gives jets with mass areas weakly
dependent on the asymmetry parameter $z$ as is depicted in
Fig.~\ref{fig:passive-massarea2p_general_kt+ca} (right). As in the $k_t$
algorithm,  also here the overall shape of the mass area as a function of the
distance between constituent particles $p_1$ and $p_2$ is very similar to the
shape found for the passive area (cf.
Fig.~\ref{fig:passive-area2p_general_sis+ca}).  
The quantitative change is, however, again much bigger for the mass area whose
1-particle value (\ref{eq:passive-ma-1part}) can be modified up to the factor
$\sim 3.6$ by the presence of the second particle (comparing to the
corresponding factor of $\sim 1.6$ for the area).
As can be found by inspecting Fig.~\ref{fig:passive-massarea2p_general_kt+ca}
(right) or the corresponding formulae from the
appendix~\ref{app:passive-formulae}, there is also a small qualitative
difference between passive area and passive mass area in the behaviour for
$0<x<x_{c1}$.  The mass area starts growing with $x$ right from the beginning
contrary to the area which is constant for $x<x_{c1}$. 

\paragraph{The SISCone algorithm} returns jets with mass area strongly dependent
on the separation~$x$ between two constituent particles. For $x<1$ the mass area
differs from the 1-particle result only mildly, growing slightly with $x$
(unlike the area which is constant in this region, cf.
Fig.~\ref{fig:passive-area2p_general_sis+ca}). For $x>1$, however, the mass area
of 2-particle SISCone jets jumps by the factor of four and continues growing
very fast with $x$ reaching the value $\sim 11$ for the 2-particle system with
$z=1/2$ and the separation $x=2$. We have seen already a similar behaviour for
the areas of SISCone jets shown in Fig.~\ref{fig:passive-area2p_general_sis+ca}
(right) but it is much bigger in quantitative terms for the mass area.
The cause of the big change of the mass area at $x=1$ is the same here, namely
for two particles with comparable transverse momenta there is a region of $x$
where there are three stable cones which all get merged leading to a gigantic
jet.
One can exploit this property in two different ways. If one is interested just
in measuring the jet mass with an algorithm which is as little sensitive to the
soft pointlike radiation as possible than, clearly, the result from
Fig.~\ref{fig:passive-massarea2p_general_sis+akt}~(left) strongly disfavours
SISCone.  This is especially true if the jet comes from decay of a
heavy object in which case its subjects have similar hardness and the
separation $x$ may be easily greater than 1. The result from
Fig.~\ref{fig:passive-massarea2p_general_sis+akt}~(left) could alternatively be
regarded as a useful additional characteristic of a jet.  It could be used to
devise some discriminating variable which would help separating QCD jets, which
have small mass area, from the jets coming from a heavy object decay, which
exhibit significantly larger passive mass area.

\paragraph{The anti-$\boldsymbol{k_t}$ algorithm} produces jets with mass area
growing slowly with $x$ up to the critical value $x_{c2}$ from
Eq.~(\ref{eq:xc2-solution-akt}). Between $x_{c2}$ and 1 the growth becomes much
faster and the more so the closer to each other are the values of transverse
momenta of the two constituent particles. For $x>1$ the hardest jet consist of a
single particle and its mass area slowly approaches $\pi R^4/2$ with increasing
$x$. Hence, qualitatively, the behaviour is not very different from that seen in
Fig.~\ref{fig:passive-area2p_general-akt} for the passive area except the region
below $x_{c2}$, where there was no growth in the latter case.  But as for
the three algorithms discussed above, also for anti-$k_t$, the quantitative
effect of adding a second particle is much bigger for the mass area than for the
area of a jet.
Overall, the passive mass area from the anti-$k_t$ algorithm may be substantial
especially for symmetric configurations ($z \sim 0.5$) with interparticle
separation $x\sim 1$. One notices also that the anti-$k_t$ result for $z=0.5$
coincides with that from the C/A algorithm for the same $z$ value. This is
reflected as well in the exact formulae given in
appendix~\ref{app:passive-formulae}.  However, as we go away from $z=0.5$ the
two algorithms behave very different as seen from
Figs.~\ref{fig:passive-massarea2p_general_kt+ca}~(right) and
\ref{fig:passive-massarea2p_general_sis+akt}~(right).

\subsubsection{Scaling violation of passive mass area of QCD jets}
\label{sec:scaling-violation-passive}

Mass area is sensitive to substructure of a jet.  For the QCD jets this
substructure arises due to radiative emissions of gluons.  Therefore, we expect
that the average mass area of a QCD jet will acquire logarithmic dependence on
jet's transverse momentum. The coefficient in front of this logarithm, which we
will call anomalous dimension, can be easily found in the small R approximation.
The results for jet areas were obtained in \cite{Cacciari:2008gn}.  Here, we
will determine their passive mass area counterparts. 

The mean mass area at the order $\alpha_s$ for a given jet algorithm and with a
given $R$ value can be written as 
\begin{equation}
  \label{eq:mean-mass-area}
  \langle a_{m}\rangle 
  = a_m(0) + \langle \Delta a_m\rangle
  = \frac{\pi}{2} R^4 + \langle \Delta a_m\rangle\,.
\end{equation}
The $\order{\alpha_s}$ correction in the limit of strongly ordered transverse
momenta of the particles, $p_{t2} \ll p_{t1}$, adequate for QCD jets, is given by
\begin{equation}
  \label{eq:delta-mean-mass-area}
  \langle \Delta  a_{m}\rangle \simeq \int_0^{2R} d\Delta_{12} 
  \int_{{Q_0}/\Delta_{12}}^{p_{t1}} d p_{t2} 
  \frac{dP}{dp_{t2} \, d\Delta_{12}} (a_{m}(\Delta_{12}) - a_{m}(0))\,,
\end{equation}
with $\frac{dP}{dp_{t2} \, d\Delta_{12}}$ being the probability for emitting a
gluon with transverse momentum $p_{t2}$ at relative angular distance $\dist$ and
the second term in the bracket accounting for virtual corrections. The lower
limit of the integration over $p_{t2}$ contains a cut-off $Q_0$ for the relative
transverse momentum of the particle $p_2$ with respect to particle $p_1$. The
need for such a cut-off comes from the fact that the mass area, just like the
area of jets, is not an infrared safe quantity and its value depends on
non-perturbative effects.\footnote{As argued in \cite{Cacciari:2008gn}, events
with pile-up provide a natural infra-red cut-off which replaces $Q_0$.} The
convergence of the integral over $\dist$ is guaranteed by the property that the
passive mass area of the hardest jet in the 2-particles system tends to the
1-particle result both when $\dist\to 0$ and $\dist\to 2 R$. Taking the QCD
matrix element in the soft and collinear approximation
\begin{equation}
  \label{eq:softcoll}
  \frac{dP}{dp_{t2} \, d\Delta_{12}} = 
  \frac{2 C_i }{\pi} 
  \frac{\alpha_s(p_{t2} \Delta_{12})}{\Delta_{12}\, p_{t2}}\,,
\end{equation}
and performing the integration in Eq.~(\ref{eq:delta-mean-mass-area}), one finds 
\begin{equation}
  \label{eq:delta-mass-area}
  \langle \Delta a_{m}\rangle = d_{m} \frac{2\alpha_s C_i}{\pi} \ln
  \frac{R p_{t1}}{{Q_0}}\,,
  \qquad
  \qquad
  \langle \Delta a_{m}\rangle = 
  d_{m} 
  \frac{C_i}{\pi b_0}
   \ln \frac{\alpha_s({Q_0})}{\alpha_s(R p_{t1})}\,,
\end{equation}
in the fixed and in the running coupling approximation, respectively. In the
latter case $\dist$ was replaced by $R$ in the argument of the coupling which
affects only the terms not enhanced by the logarithm of $R$. $C_i$ is a colour
factor corresponding to the parent particle and $b_0 = (11 C_A-2 n_f)/(12\pi)$.

The coefficient $d_m$, which depends on jet definition, is the aforementioned
anomalous dimension and it is given by
\begin{equation}
\label{eq:dm-def}
d_{m} = \int ^{2R} _{0} \frac{d\theta}{\theta} (a_{m}(\theta) - \frac{\pi}{2}
R^{4} )\,,
\end{equation}
In a similar manner, one can compute fluctuations of mass areas defined as
\begin{equation}
  \label{eq:passive-fluct}
  \langle \sigma^2_m \rangle = \langle a_m^2\rangle
  - \langle a_m\rangle^2 = \sigma^2_m(0) + 
  \langle \Delta a_m^2\rangle -
  \langle \Delta a_m\rangle^2 \simeq \langle \Delta a_m^2\rangle \,,
\end{equation}
where we have dropped $\sigma^2_m(0)$, which is identically zero, and $\langle
\Delta a_m\rangle^2$ as it gives higher order corrections in $\alpha_s$. A
calculation similar to the above leads to the results identical to those
given in Eq.~(\ref{eq:delta-mass-area}) with just $d_m$ replaced by $s^2_m$
where the latter is defined as
\begin{equation}
s^{2}_{m} = \int ^{2R} _{0} \frac{d\theta}{\theta} ( a_{m}(\theta) -
\frac{\pi}{2} R^{4} )^{2}\,.
\end{equation}

The analytic results for the coefficients $d_m$ and $s^2_m$, normalised to
$R^4$, for all four algorithms are given in table~\ref{tab:scalval-ma-passive}.
There, we also quote their approximate numerical values normalised to the
1-particle passive mass area.
\begin{table}[t]
\centering
\begin{tabular}{c@{\qquad}c@{\qquad}c@{\qquad}c@{\qquad}c}
  \toprule   
  algorithm 
      & \small $\displaystyle \frac{d_{m}}{R^4}$ 
      & \small $\displaystyle \frac{2\, d_{m}}{\pi R^4}$
      & \small $\displaystyle \frac{s^2_{m}}{R^4}$ 
      & \small $\displaystyle \frac{2\, s_m}{\pi R^4}$ \\[5pt]
  \midrule
  $k_t$ &
  \small $\displaystyle 
  \frac{37 \sqrt{3}}{64} 
  +\frac{\pi }{2}
  +\frac{\xi }{2}
  $
  & $ 1.799$ & 
  \small $\displaystyle 
  \frac{5629}{18432}
  + \frac{45\sqrt{3} \pi }{64}
  + \frac{5 \pi^2}{36}
  -\frac{15 \zeta (3)}{32}
  + \frac{\pi \xi }{2}
  $
  & $1.525$  \\[10pt]
  C/A &
  \small $\displaystyle 
  \frac{13 \sqrt{3}}{64}
  +\frac{\pi }{4}
  -\xi 
  $
  & $0.401$ & 
  \small $\displaystyle 
  \frac{597}{2048} 
  +\frac{\pi }{2\sqrt{3}}
  +\frac{\pi ^2}{24}
  -\frac{13 \zeta (3)}{48}
  +\frac{\pi  \xi }{3}
  $
  & $0.858$  \\[10pt]
  SISCone &
  \small $\displaystyle 
  -\frac{\sqrt{3}}{64} 
  +\frac{\pi }{24}
  -\frac{\xi }{2}
  $
  & $-0.095$  &
  \small $\displaystyle 
  -\frac{39}{2048}
  +\frac{\pi }{64 \sqrt{3}}
  -\frac{\pi ^2}{144}
  -\frac{13 \zeta (3)}{96}
  +\frac{\pi  \xi }{6}
  $
  & $0.133$  \\[10pt]
  anti-$k_t$ &
  $0$
  & $0$ & 
  $0$
  & $0$  \\  
  \bottomrule
\end{tabular} 
\caption{%
  Coefficients governing the logarithmic scaling violation of passive mass areas
  with transverse momentum of a jet for 2-particle QCD jets. The analytic
  results are normalised to $R^4$. We use the shortcut notation for
  $\xi \equiv
  \left(\psi'(1/6)+\psi'(1/3)-\psi'(2/3)-\psi'(5/6)\right)/(48\sqrt{3})
  \,\simeq\, 0.507471$ where $\psi'(x)$ is the trigamma function. In the results
  for $s^2_m$, $\zeta(3) \simeq 1.202$  is a special value of the Riemann zeta
  function.
  The numerical results are normalised to the passive mass area of a 1-particle
  jet.
}
\label{tab:scalval-ma-passive}
\end{table}
One notices that the coefficients $d_m$  depend strongly on jet algorithm. The
largest  value is found for the $k_t$ algorithm. The
next in the hierarchy is the C/A algorithm with its $d_m$ coefficient already
more than factor four smaller of that from $k_t$. SISCone produce fairly small
and negative result whereas anti-$k_t$ yields identically zero. The observed
hierarchy is consistent with the behaviour of passive mass areas of strongly
ordered system (i.e. $z=0$) from
Figs.~\ref{fig:passive-massarea2p_general_kt+ca} and
\ref{fig:passive-massarea2p_general_sis+akt}. The large coefficient for the
$k_t$ algorithm comes about due to strong rise of the passive mass area in the
region of small interparticle separations enhanced in the
integral~(\ref{eq:dm-def}). The smaller $d_m$ from C/A is related to the fact
that the mass area in this algorithm becomes significantly different from the
1-particle result at $x>1/2$ hence in the range which is less favoured
by~(\ref{eq:dm-def}). Similarly, the small and negative $d_m$ from SISCone comes
from the fact that the mass area in this algorithm deviates from $\pi R^4/2$
only for $x>1$ where it becomes lower than the mass area of a 1-particle jet.
One practical conclusion from table~\ref{tab:scalval-ma-passive} is that the
passive mass areas of jets from $k_t$ algorithm will depend much more strongly
on those jets' $p_t$ than will the passive mass areas of other algorithms. This
is a similar conclusion to that found in \cite{Cacciari:2008gn} for areas of
jets.
The values of $s_{m}$ coefficients from table~\ref{tab:scalval-ma-passive}
suggest significant fluctuations of the passive mass areas of QCD jets. Here the
pattern essentially follows that of $d_{m}$ coefficients with, however, a
somewhat smaller difference between $k_t$ and C/A algorithms.

\subsection{Active mass area}
\label{sec:active-mass-area}

We define the active mass area as follows
\begin{equation}
    \label{eq:active-ma-def}
    A_m(J) \equiv 
    \lim_{\nu_{\{g_i\}} \to \infty}
    \left\langle 
      \frac{m^2_{J{\{g_i\}}}-m^2_{J}}{\nu_{\{g_i\}}\mean{p_{tg}}p_{tJ\{g_i\}}} 
    \right\rangle_g\,,
\end{equation}
where $m_{J}$ is a mass of the pure jet $J$ and $m_{J{\{g_i\}}}$ is a mass of
the jet consisting of $J$ and a dense coverage of ghosts from some random
ensemble  $\{g_i\}$. 
Similarly, $p_{tJ\{g_i\}}$ is a transverse momentum of the whole jet with real
and ghost particles.
The ghosts have density $\nu_{\{g_i\}}$ and the
infinitesimally small average transverse momentum $\mean{p_{tg}}$.  The limit of
infinite density of ghosts is taken and, in addition, the result is averaged
over many sets of ghosts. The standard deviation of the distribution across
these ghost ensembles is given by
\be
    \label{eq:active-ma-sd-def}
    \Sigma^2_m (J) =
    \lim_{\nu_{\{g_i\}} \to \infty}
    \left\langle 
        A_m^2(J | \{g_i\}) 
    \right\rangle_g
    - A^2_m(J)\,.
\ee

Consider the system with one or more particles whose transverse momenta are
well above the ghost scale $\mean{p_{tg}}$.
In the case in which such particles are massless
\be
    \label{eq:mJ-mJg-diff}
    m^2_{J{\{g_i\}}}-m^2_{J}  = 
    2 \nu_{\{g_i\}} \mean{p_{tg}} p^\mu_{J\{g_i\}} A_{\mu}(J | \{g_i\})  \ - \
    \nu_{\{g_i\}}^2\langle p_{tg}\rangle^2 
    A^\mu(J | \{g_i\}) A_{\mu}(J | \{g_i\})\,,
\ee
where we used the definition of 4-vector active area $A_{\mu}(J | \{g_i\})$ from Eq.~(\ref{eq:active-vec-area-def}).
Note also that $p^\mu_{J\{g_i\}}$ is a 4-momentum of the whole jet 
consisting of physical and ghost particles.
%
%
The two terms on the right hand side are of two fundamentally different scales.
The second term is itself an interesting characteristic of a jet and, as we
shall see in Section \ref{sec:pileup-cor}, there are cases in which it is useful
to know it.
However, because of an extra power of an arbitrary small ghost transverse
momentum, $\langle p_{tg}\rangle$, the contribution of this second term to the
mass area, as defined in Eq.~(\ref{eq:active-ma-def}) is negligible and that is
why we drop it here.  
This, together with combining Eqs.~(\ref{eq:active-ma-def}) and
(\ref{eq:mJ-mJg-diff}), leads to the following formula for the active mass area
\begin{equation}
    \label{eq:active-ma-def2}
    A_m(\text{physical jet } J) =
    \frac{2}{p_{tJ}}\, p^\mu_{J} A_{\mu}(J)\,,
\end{equation}
which is particularly convenient to work with. 
In what follows, we will be computing active mass areas of jets using the above
equation with the 4-vector area $A_\mu (J)$ calculated with FastJet. For
definition of the latter quantity we refer to Section~\ref{sec:active-area-2p},
the original paper~\cite{Cacciari:2008gn} or FastJet
documentation~\cite{FastJet}.

\subsubsection{Active mass area for 1-particle jet}
\label{sec:ama-1p}

\begin{figure}[t]
    \centering
    \includegraphics[height=0.42\textwidth]{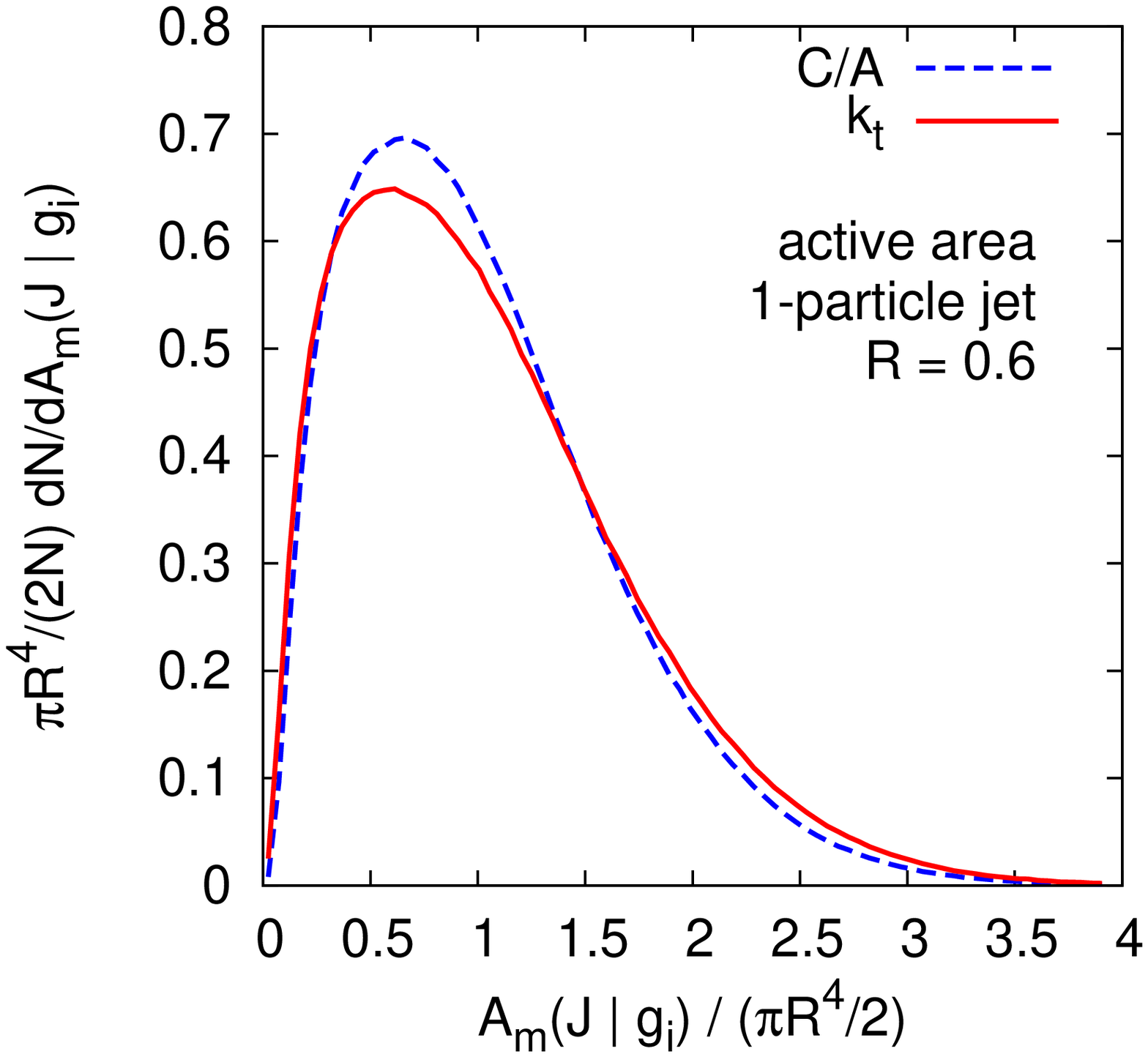}
    \hspace{20pt}
    \includegraphics[height=0.42\textwidth]{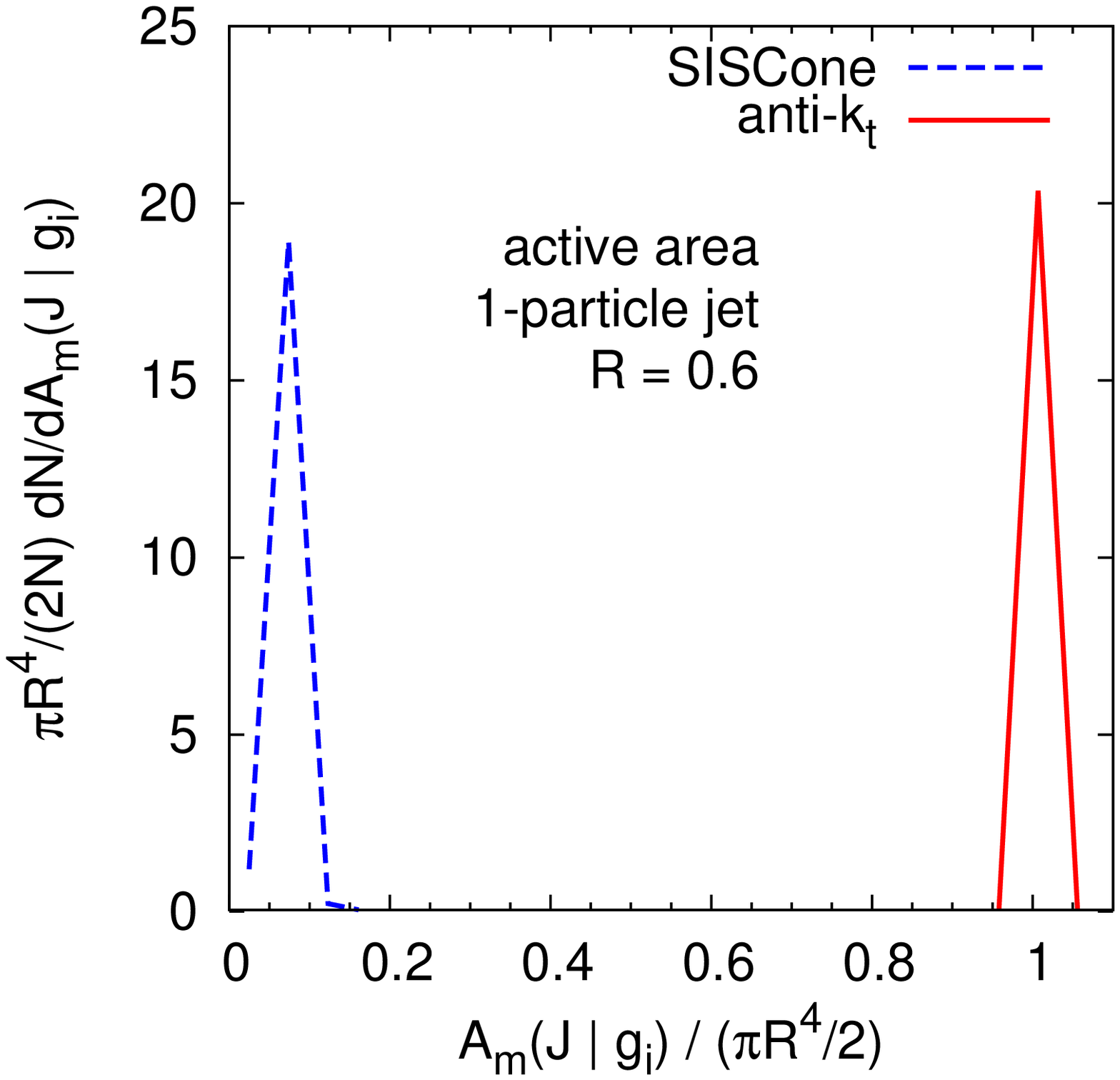}
	\caption{Distribution of active mass area $A_{m}$ 
	of 1-particle jets from the $k_t$ and C/A algorithm (left)
	and from SISCone and anti-$k_t$ (right).
	The curves correspond to numerical results obtained from FastJet.
	The width of SISCone and anti-$k_t$ distributions arises solely due to
	finite binning.
	}
 	\label{fig:dist-ma-1p}
\end{figure}
\begin{table}[t]
   \centering
   \begin{tabular}{ccc} \toprule
   algorithm   &     \multicolumn{2}{c}{1-particle-jet}       \\ \midrule
               & $A_m/(\pi R^4/2)$  & $\Sigma_m/(\pi R^4/2)$  \\ \cline{2-3}
   $k_t$       &  1.05              &  0.65                   \\ 
   C/A         &  1.02              &  0.61                   \\
   SISCone     &  1/16              &  0                      \\ 
   anti-$k_t$  &  1                 &  0                      \\ \bottomrule
   \end{tabular}
   \caption{
   Average active mass areas for 1-particle jet together with corresponding
   standard deviations for four algorithms. The numbers for $k_t$ and C/A
   correspond to the distributions from Fig.~\ref{fig:dist-ma-1p}
   (left) whereas for SISCone and anti-$k_t$ analytic results area given. The
   latter are confirmed by the numerical study as shown in
   Fig.~\ref{fig:dist-ma-1p} (right).
   }
   \label{tab:av-active-massareas}
\end{table}

\paragraph{The $\boldsymbol{k_t}$ and Cambridge/Aachen algorithms} allow
only for numerical study of active mass areas. 
The formula~(\ref{eq:active-ma-def2}) can be applied directly for 
1-particle jet. The distributions of active mass areas from the two algorithms,
normalised to $\pi R^4/2$, are shown in Fig.~\ref{fig:dist-ma-1p}
(left).  
The results from $k_t$ and C/A are very close to each other.  Similarly to the
case of the jet areas \cite{Cacciari:2008gn}, the maxima of the distributions
lie significantly below 1. 
The corresponding results for the average mass areas and their standard
derivations are given in table~\ref{tab:av-active-massareas}. The 1-particle
active mass area is very close to the 1-particle passive mass area but if
fluctuates significantly across the ghost ensembles. 
This is partly different from the jet area case where the values of active areas
were consistently $20\%$ below those of passive areas for both algorithms
\cite{Cacciari:2008gn} (cf. table~\ref{tab:av-active-areas}).
Qualitatively, however, the results shown in Fig.~\ref{fig:dist-ma-1p} (left)
and in the first two rows of table~\ref{tab:av-active-massareas} are similar 
to those found in~\cite{Cacciari:2008gn} for jet areas.

\paragraph{The SISCone and anti-$\boldsymbol{k_t}$ algorithms}  allow for
analytic study of the mass areas of 1-particle jets. 
As pointed out in~\cite{Cacciari:2008gn}, the split-merge procedure used in
SISCone always results in the split between two stable cones both if one of them
does or does not contain a hard particle. This, in turn, reduces the radius of
the hard jet by the factor 1/2 and therefore, from~(\ref{eq:passive-ma-1part}),
the active mass area by the factor 1/16. This result does not depend on the
ghost ensemble, assuming that the coverage of ghosts is sufficiently dense, hence the fluctuations vanish.

The anti-$k_t$ algorithm leads to 1-particle jets of a circular shape with
radius $R$. Therefore, the active mass area of such jets coincides with the
passive mass area result (\ref{eq:passive-ma-1part}) and the fluctuations of
the active mass area are identically zero. 

These analytic results are summarised in table~\ref{tab:av-active-massareas}.
The corresponding distributions from numerical study are shown in in
Fig.~\ref{fig:dist-ma-1p} (right). We see that both algorithms give
distribution of active mass areas for 1-particle jets which are close to
$\delta$-function. The width comes solely from finite binning. 

\subsubsection{Active mass areas for general case of 2-particle system}
\label{sec:active-mass-area-2p}

\begin{figure}[t]
    \centering
    \includegraphics[width=0.45\textwidth]
        {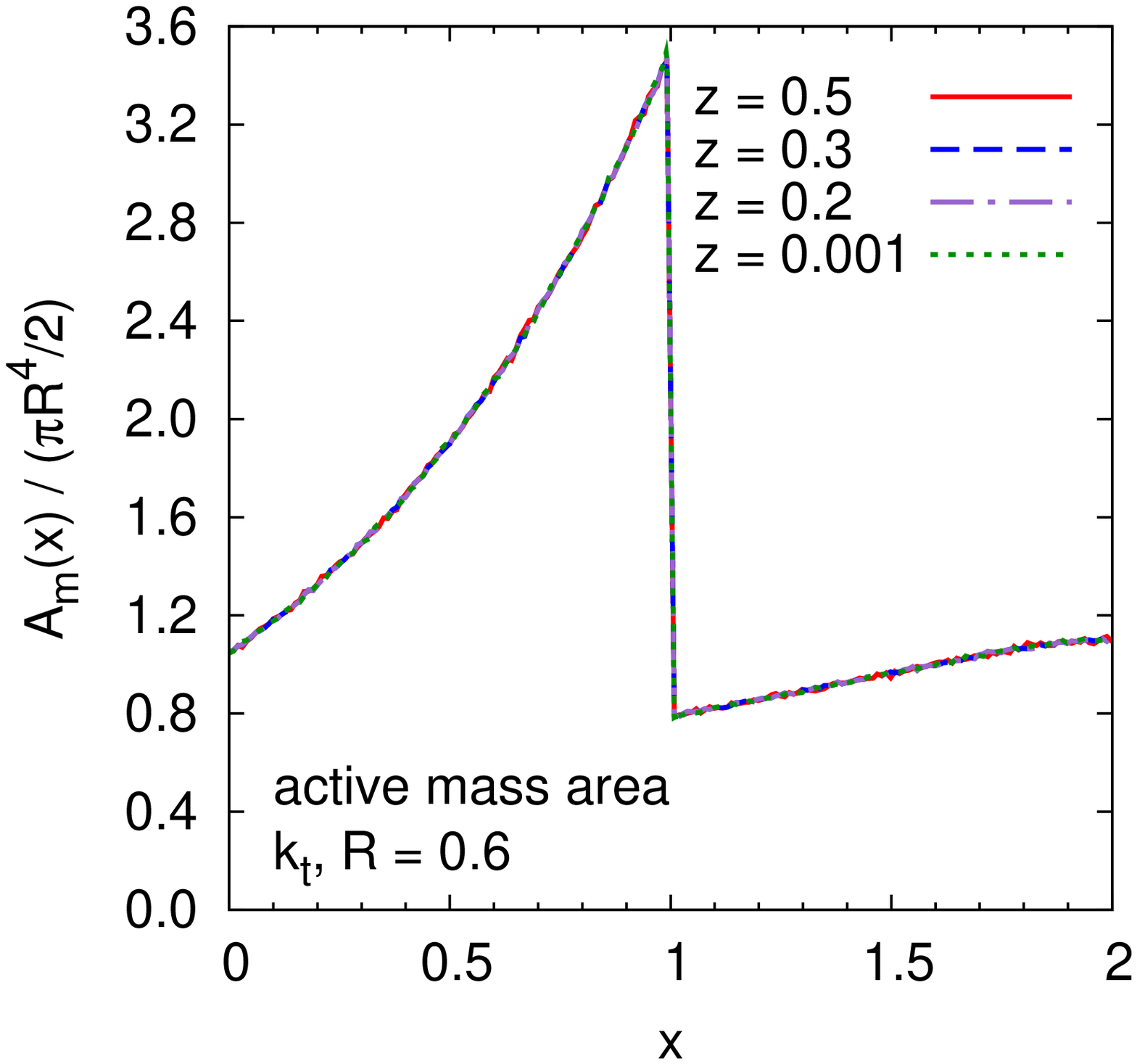}
    \hfill
    \includegraphics[width=0.45\textwidth]
        {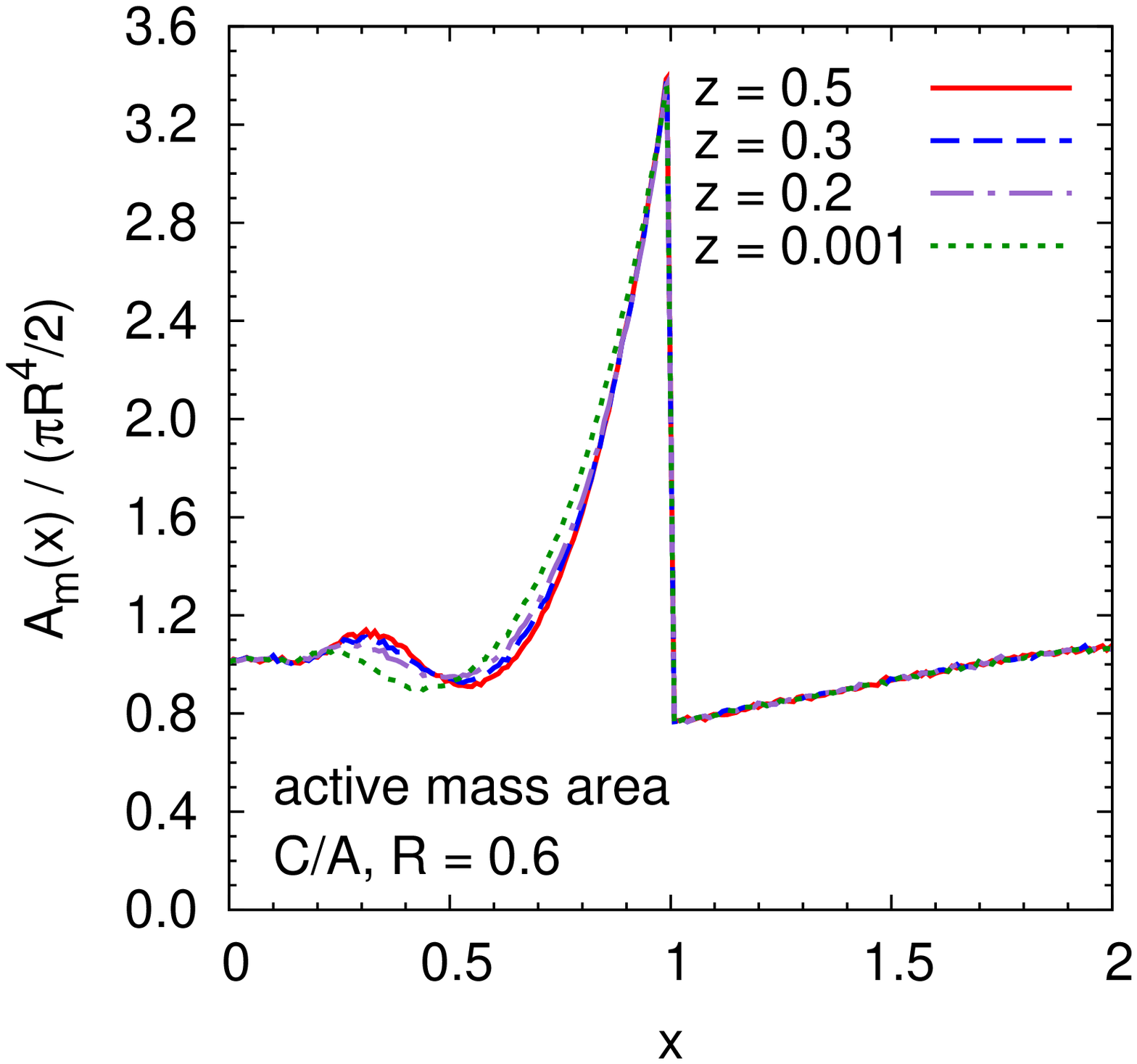}
    \vskip 5pt
    \includegraphics[width=0.45\textwidth]
        {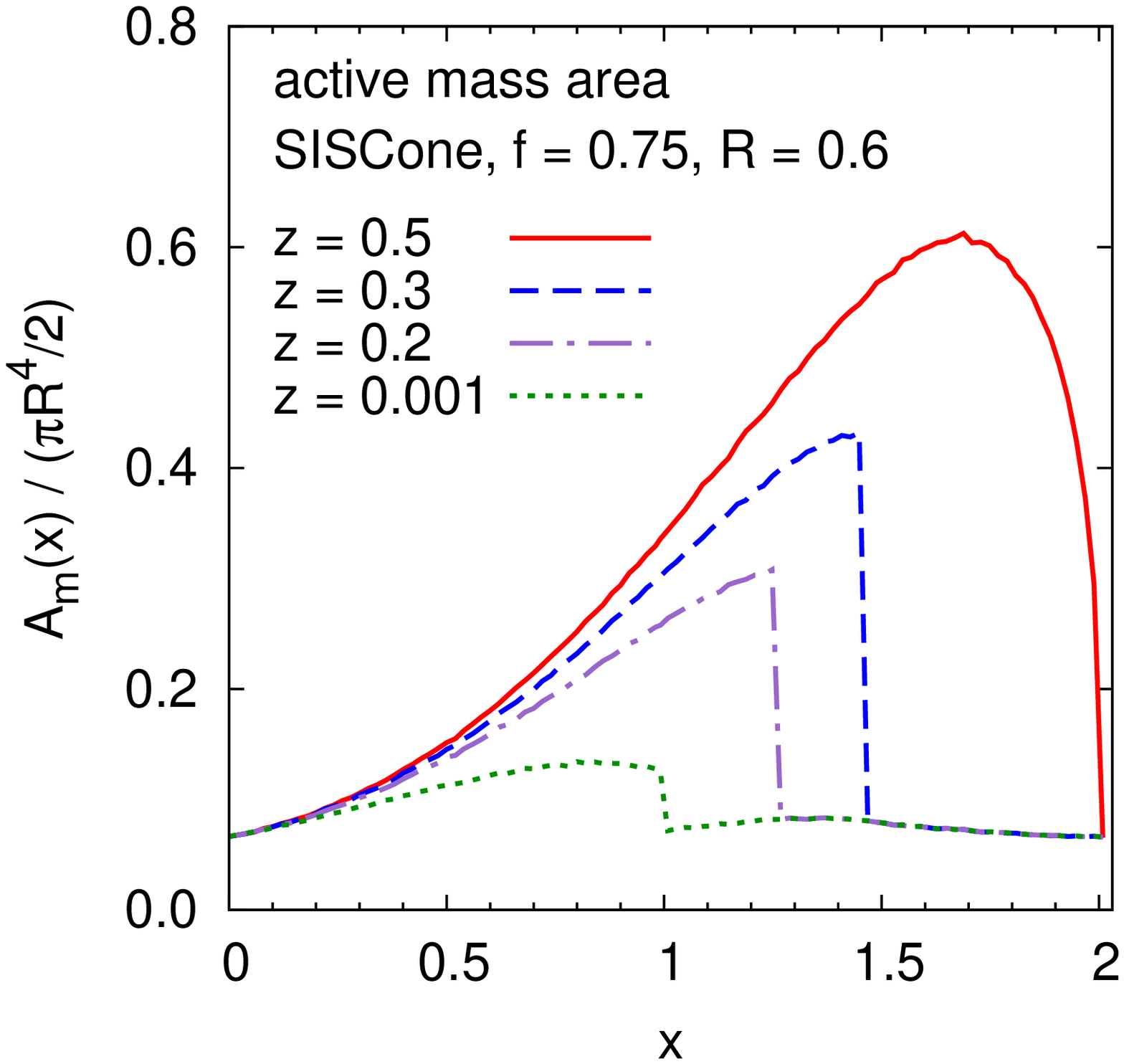}
    \hfill
    \includegraphics[width=0.45\textwidth]
        {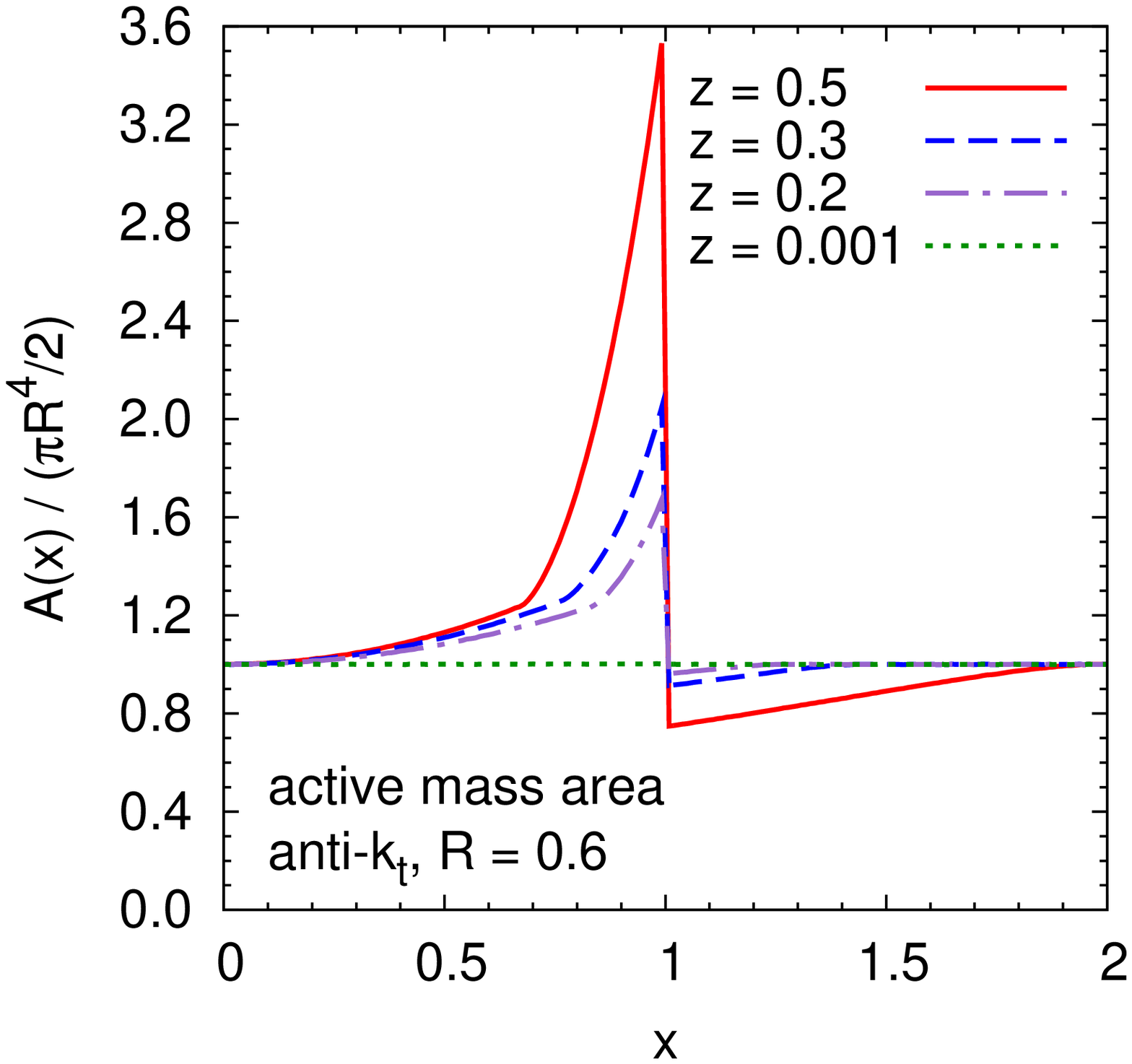}
    \caption{
    Active mass area of the hardest jet for the system of two particles with
    arbitrary ratio of transverse momenta. The results from four jet algorithms
    are shown as functions of the interparticle distance $x$ for several values
    of the asymmetry parameter $z$.
    }
    \label{fig:active-massarea2p_general_allalg}
\end{figure}

The results for active mass area of the hardest jet in a system with two
particles of arbitrary relative hardness are given in
Fig.~\ref{fig:active-massarea2p_general_allalg}. As before, we present the
active mass areas normalised to $\pi R^4/2$ as functions of the separation (in
units of $R$) between the two particles. All curves correspond to numerical
computations with FastJet.
One has to keep in mind that in general, as was the case for 1-particle mass
areas, the 2-particle mass area is a distribution and the curves shown in
Fig.~\ref{fig:active-massarea2p_general_allalg} corresponds to its mean value.

The active mass area from the $k_t$ algorithm does not depend on the asymmetry
parameter~$z$, just as the passive area. For C/A this dependence is weak.
On the other hand, similarly to the case of passive mass areas, the active mass
areas of SISCone and anti-$k_t$ strongly vary depending on whether the two
constituent particles are of comparable hardness or whether their transverse
momenta are significantly different.

The active mass areas from the sequential recombination algorithms are
virtually, for $k_t$ and C/A, or exactly, for anti-$k_t$, identical with their
passive mass area counterparts. 
Regarding only the shape, the situation was quite similar for the 2-particle
areas from $k_t$ and C/A, only that there the normalisation of the active area
was different. Since, as seen from the first two rows of
table~\ref{tab:av-active-massareas}, the active and passive 1-particle areas are
almost identical for $k_t$ and C/A, also the results from
Fig.~\ref{fig:active-massarea2p_general_allalg} and
Figs.~\ref{fig:passive-massarea2p_general_kt+ca} and
\ref{fig:passive-massarea2p_general_sis+akt} coincide to large extent for those
algorithms.
The identity of passive and active mass areas for anti-$k_t$, has the same
origin as the analogous identity for the areas observed in
section~\ref{sec:area-general-2p} and it comes from the fact that the ghosts
cluster among themselves after all clusterings with physical particles have
occurred.
Altogether, for the active mas areas from the sequential recombination
algorithms, one observes the same pattern in the relation of these results to
the active areas as seen earlier for the passive quantities. Specifically, while
the qualitative picture for the active areas and active mass areas is very
similar, quantitatively the effects seen for the latter are much stronger.

The case of SISCone is quite special. Firstly in that its 1-particle active mass
area gets modified very strongly by the presence of the second particle of
comparable hardness. The similar conclusion was drawn already for the passive
mass areas (cf. Fig.~\ref{fig:passive-massarea2p_general_sis+akt}). However, in
absolute terms, the active mass area of a 2-particle SISCone jet remains still
much smaller than both its passive counterpart as well as the active mass areas
from all the other algorithms shown in
Fig.~\ref{fig:active-massarea2p_general_allalg}. This implies that the
sensitivity of the jet mass to soft background should be the lowest for SISCone.
This, in turn, translates into a particularly good mass resolution of this
algorithm seen e.g. in \cite{Butterworth:2008iy}.

The mechanism responsible for the strong relative change of the the 2-particle
jet active mass area with respect to the 1-particle case for SISCone is the same
as discussed in section~\ref{sec:passive-area-2p} for the passive area and
refers to the existence of the third stable cone containing the two physical
particles.
This cone disappears at $x = 1/(1-z)$ and two particles separated by greater
distance form two distinct jets and hence the drop of the active mass area seen
in Fig.~\ref{fig:active-massarea2p_general_allalg} (bottom left).  If the value
of $z$ is sufficiently large and the drop occurs at $x\simeq 2$, the active area
of a 2-particle jet from SISCone starts falling at some $x$, an effect of
split-merge procedure involving the third stable cone (with particles $p_1$ and
$p_2$) and the pure ghost cones. This is seen in
Fig.~\ref{fig:active-massarea2p_general_allalg} (bottom left) for the 2-particle
system with $z=1/2$.  A similar effect was discussed in section~\ref{sec:active-areas-2p} for the active area.

For all algorithms the value of 1-particle passive area from
table~\ref{tab:av-active-massareas}  is recovered at $x=0$. However, at $x=2$
only SISCone and anti-$k_t$ yield $\pi R^4/2$ and a larger value is given by
$k_t$ and C/A. As mentioned in section~\ref{sec:active-areas-2p}, this comes
from the fact that those algorithms build jets starting from formation of
local structures.

A general comment concerning the results of
Fig.~\ref{fig:active-massarea2p_general_allalg} is that sensitivity of the
active mass area to the relative hardness of the constituent particles
(reflected in the value of $z$) is related to the shape of jets produced by a
given algorithm. Namely, the algorithms which depend strongly on $z$ are those
belonging to the class of ``conical algorithms'', i.e. SISCone and anti-$k_t$.
Though, as noticed earlier, in general, their jets are not ideally conical,
still their shapes in the $(y,\phi)$ are usually quite regular.
Conversely, the $k_t$ and C/A algorithms, whose jets are highly irregular in
shape show either none or weak dependence on $z$.

The whole variety of behaviours of the active mass areas observed in
Fig.~\ref{fig:active-massarea2p_general_allalg}, depending on the algorithm,
asymmetry parameter $z$ or the interparticle distance $x$ encourages one to
exploit it on the analysis-by-analysis basis.

\subsubsection{Scaling violation of active mass area of QCD jets}
\label{sec:scaling-violation-active}

\begin{table}[t]
   \centering
\begin{tabular}{c@{\qquad}c@{\qquad}c}
  \toprule
  algorithm & \small $\displaystyle \frac{2 D_{m}}{\pi R^4}$ 
            & \small $\displaystyle \frac{2 S_{m}}{\pi R^4} $  \\[5pt]
  \midrule
   $k_t$      & $1.694$  & $1.415$ \\
   C/A        & $0.387$  & $0.781$ \\
   SISCone    & $0.086$  & $0.101$ \\
   anti-$k_t$ & 0        & 0 \\
 \bottomrule
\end{tabular} 
\caption{%
  Coefficients governing the logarithmic scaling violation of active mass areas
  with transverse momentum of a jet for 2-particle QCD jets, 
  normalised to the passive mass area of a 1-particle jet.
  The numerical results obtained by performing interactions from
  Eqs.~(\ref{eq:delta-active-mass-area}) and (\ref{eq:fluct-active-ma-res})
  using the functions corresponding to those shown in
  Fig.~\ref{fig:active-massarea2p_general_allalg}.
}
\label{tab:scalval-ma-active}
\end{table}

We conclude this section by the study of average leading effect of
perturbative radiation on the active mass areas of QCD jets. In analogy to the
passive mass area, we define
\begin{equation}
  \label{eq:mean-active-mass-area}
  \langle A_{m}\rangle 
  = A_m(0) + \langle \Delta A_m\rangle
  = A_m(\text{1-particle-jet}) + \langle \Delta A_m\rangle\,,
\end{equation}
where $A_m(\text{1-particle-jet})$ depends on jet algorithm as summarised in
table~\ref{tab:av-active-massareas}. The perturbative correction to the
1-particle-jet result can be computed from the formula analogous to
Eq.~(\ref{eq:delta-mean-mass-area}) with $a_m$ replaced by $A_m$ and the upper limit $2R$ removed. The
latter is related to the fact that the active mass area of a 2-particle jet may
in general be different than $A_m(\text{1-particle-jet})$ for $\dist > 2R$, as
noted in the previous subsection. Simple integration gives
\begin{equation}
  \label{eq:delta-active-mass-area}
  \langle \Delta A_{m}\rangle = 
  D_{m} 
  \frac{C_i}{\pi b_0}
   \ln \frac{\alpha_s({Q_0})}{\alpha_s(R p_{t1})}\,,
  \qquad
  \qquad
  D_{m} = \int _{0} \frac{d\theta}{\theta} (A_{m}(\theta) - A_{m}(0))\,,
\end{equation}
and the analogous fixed-coupling result as in section~\ref{sec:scaling-violation-passive}.

As discussed at the beginning of section~\ref{sec:active-mass-area}, the active
mass area comes with an intrinsic fluctuations due to fluctuations of ghosts.
Therefore, the fluctuations of the active mass area of QCD jets can be separated
into two components
\be
\mean{\Sigma^2_m} = \Sigma^2_m(0) + \mean{\Delta\Sigma^2_m}\,,
\ee
where the first one, being just the contribution from one-particle jets, is
given for each of the four algorithms in table~\ref{tab:av-active-massareas}.
The second term comes from 2-particle configurations. It acquires
contributions both from  the change of the mass area caused by the perturbative
radiation (as in the passive case) and from the fluctuations of ghosts used to
determine the mass area of those configurations (absent in the passive case).
The corresponding formulae for active areas was derived in
\cite{Cacciari:2008gn}.  A straightforward, analogous derivation leads to the
following result for logarithmically enhanced, $\order{\alpha_s}$ contribution to
the active mass area
\begin{align}
  \label{eq:fluct-active-ma}
  \langle \Delta \Sigma^2_{m}\rangle &\simeq
  S_{m}^2 \frac{C_1}{\pi b_0}
  \ln \frac{\alpha_s({Q_0})}{\alpha_s(R p_{t1})}
  \;,\\ \qquad
  S_{m}^2 &= \int_0 \frac{d\theta}{\theta} \big[
  (A_{m}(\theta) -  A_{m}(0))^2 + 
  \Sigma^2_{m}(\theta) -  \Sigma^2_{m}(0)
  \big]\, \\
  \label{eq:fluct-active-ma-res}
  &= \int_0 \frac{d\theta}{\theta} 
  (A^2_{m}(\theta) -  A^2_{m}(0))
  -2 A_{m}(0) D_{m}\,.
\end{align}

The results for the coefficients $D_m$ and $S_m$, normalised to $\pi R^4/2$, are
given in table~\ref{tab:scalval-ma-active}. The numbers come from integration of
the numerical results for active mass areas in the case of $k_t$ and C/A
algorithm and the analytic results in the case of SISCone and anti-$k_t$.
One notices that the anomalous dimension and its fluctuations for active mass
areas are very close to those found for passive mass areas with an exception of
SISCone whose active area anomalous dimension, though of similar absolute
magnitude, comes with an opposite sign. Therefore, most of the discussion from
section~\ref{sec:scaling-violation-passive} related to table~\ref{tab:scalval-ma-passive} remains valid also for the results from
table~\ref{tab:scalval-ma-active}. The only qualitative difference of the positive $D_m$ versus
negative $d_m$ for SISCone comes from the fact that for the former the active
mass area of the hardest jet in the 2-particle system never goes below the
1-particle result (see appendix \ref{app:active-formulae-sis}).

\section{Illustration with Monte Carlo events}
\label{sec:mc-study}

\subsection{Mass areas of simulated jets}
\label{sec:mc-study-ma}

Real jets are of course more complex than just the 1- or 2-particle systems
that we studied so far. Nevertheless, we believe that a series of features of
mass areas from those found in the preceding sections  will be present also in
jets measured in the real life. To provide some support to that statement, in
this section, we perform a brief study of mass areas of jets from Monte Carlo
(MC) simulation. Compare to the 1- or 2-particle jets discussed earlier, the
simulated jets will have more accurate modelling of QCD radiation, in particular
that associated with parton shower, as well as hadronization.

We will examine jets from Pythia~6.4~\cite{Sjostrand:2006za} dijets events
with the underlying event switched off. As before, we will be interested in the
hardest jet in an event.  We will not, however,  impose any rapidity or
transverse momentum cuts. Our aim will be to obtain an analog of
Fig.~\ref{fig:active-massarea2p_general_allalg} for the MC jets. 
In the case of two particle system each event was characterized by the the
asymmetry parameter and by the angular distance between the particles, defined
respectively in Eqs.~(\ref{eq:z-def}) and (\ref{eq:x-xJ-def}). 
Those particles were meant as an approximation to two subjets of a realistic
jet. The meaningful subject analysis of real jets is, however, possible only for
some jet algorithms. Moreover, it is not very useful in the region of $x>1$. 
Therefore, for the purpose of this Monte Carlo study, we need to use
slightly more sophisticated strategy. It will be based on the procedure from
\cite{Salam:2007xv, Salam:2009jx} were it was employed to study the reach of jet
algorithms. It involves using a ``reference algorithm'' for which choose C/A
with R=1.2. First, an event is clustered with this algorithm and the hardest jet
is decomposed into its two main subjets $S_1$ and $S_2$. Those subjets are used
to determine $x$ and $z$. Then, the same event is clustered with one of the four
``test algorithms'', $k_t$, C/A, anti-$k_t$ or SISCone with R=0.6. 
Subsequently, one looks for the hardest ``test jet'' which belongs to the same
hemisphere as the hardest jet from the reference algorithm. The mass area of
this jet is assigned to the $(x,z)$ pair determined in the first step.  If the
separation $x$ between the two subjets, $S_1$ and $S_2$, is small, they will
predominantly both end up in the hardest test jet. If, on the other hand, this
separation is large, only one of them, either $S_1$ or $S_2$, will have
significant overlap with the test jet. Each of the two situations should be
reflected in the value of the mass area of the test jet. 

\begin{figure}[t]
    \centering
    \includegraphics[width=0.45\textwidth]
        {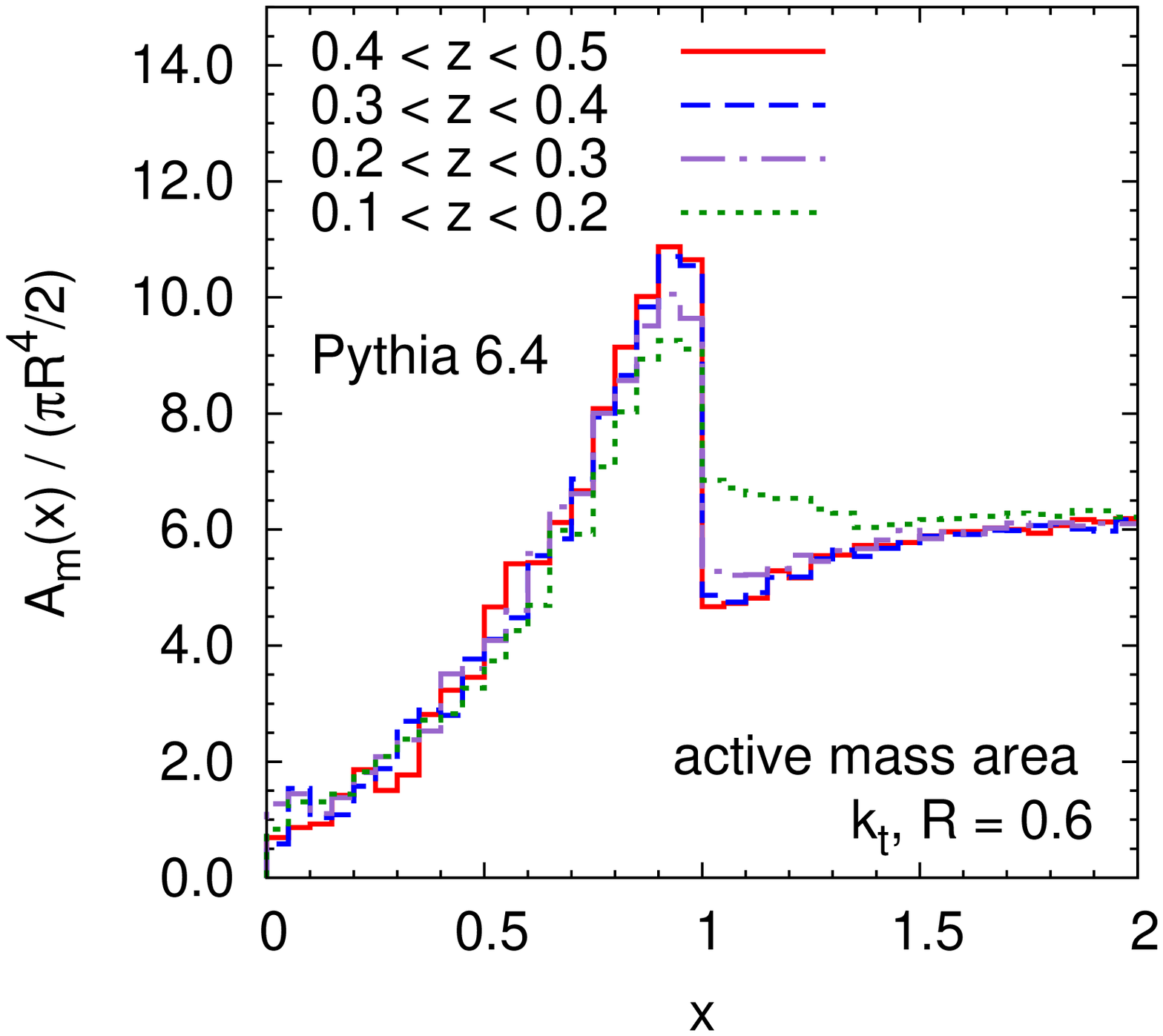}
    \hfill
    \includegraphics[width=0.45\textwidth]
        {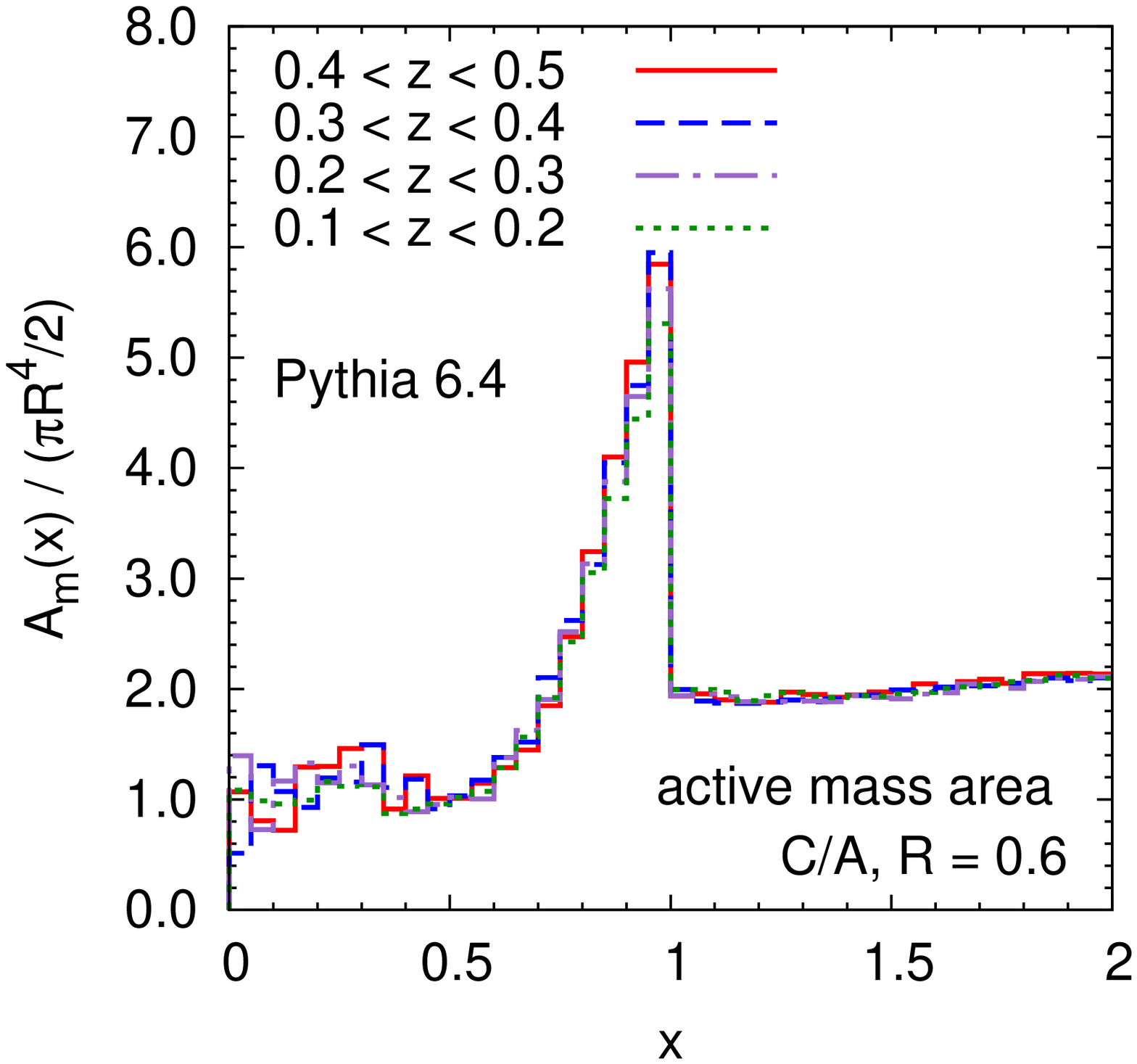}
    \vskip 5pt
    \includegraphics[width=0.45\textwidth]
        {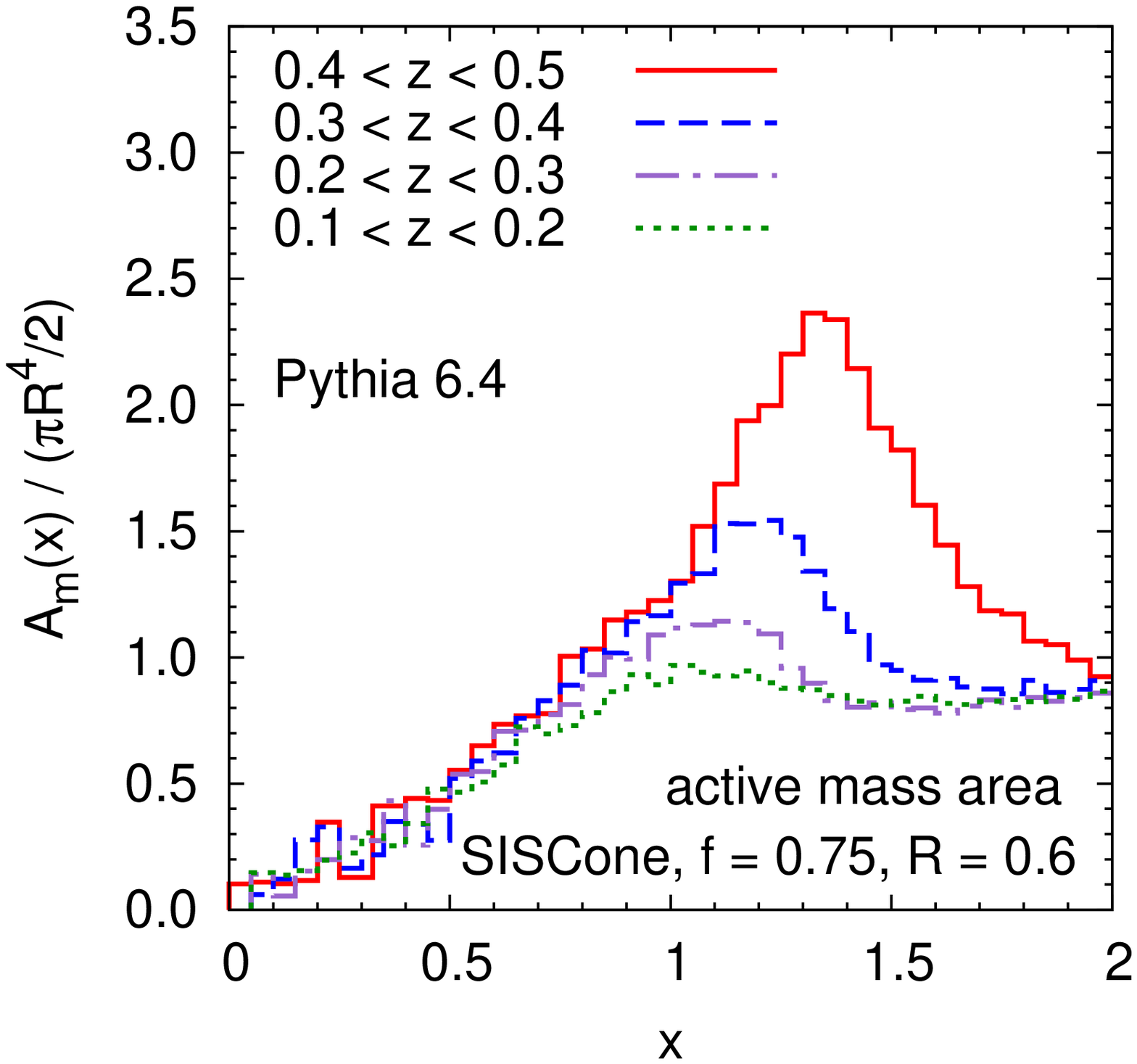}
    \hfill
    \includegraphics[width=0.45\textwidth]
        {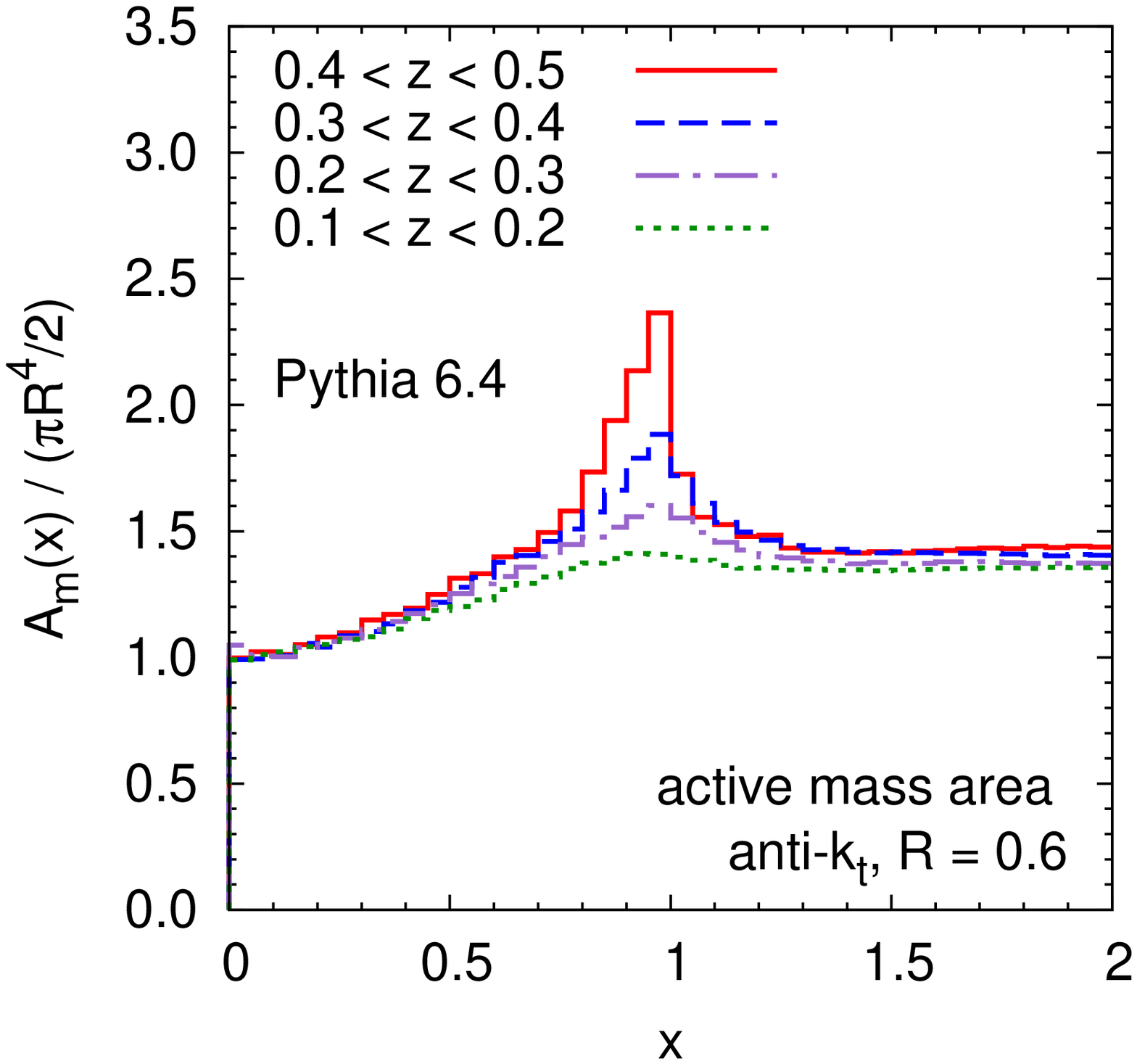}
    \caption{
    Active mass area of the hardest jet from Pythia dijet events with the
    underlying event switched off. The results from four jet algorithms are
    shown as functions of the distance $x$ between two main subjets for several
    bins of the asymmetry parameter $z$. The procedure used to determine $x$ and
    $z$ is described in the main text.
    }
    \label{fig:active-massarea-mc}
\end{figure}

The results obtained after applying the above procedure to the dijet events from
Pythia are shown in Fig.~\ref{fig:active-massarea-mc}, where the mass
areas are presented as functions of $x$ in bins of the asymmetry parameter $z$. 
Assigning correct substructure is difficult for the cases with large asymmetries,
and that is why we do not go below $z=0.1$.
Otherwise, as shown in \cite{Salam:2007xv, Salam:2009jx},  the method works well
with perhaps slightly higher uncertainties for $x\sim 1$ and $z$ being close to
its either lower ($k_t$) or the upper (anti-$k_t$) limit.

The first observation from Fig.~\ref{fig:active-massarea-mc}, is that all four
algorithms give results which are in qualitative agreement with the 2-particle
picture of Fig.~\ref{fig:active-massarea2p_general_allalg}. The general
pattern of the growth of mass area with $x$ and then the drop at some point for
$x\geq1$ is well reproduced. Also the sensitivity to the $z$ value, low for
$k_t$ and C/A, noticeable for anti-$k_t$ and large for SISCone, is consistent
with the 2-particle picture.
There are, however, quantitative differences between
Figs.~\ref{fig:active-massarea2p_general_allalg} and
\ref{fig:active-massarea-mc}. They are clearly related to the extra amount of
perturbative radiation which builds up the structure of physical jets.

Let us begin with the three sequential algorithms, $k_t$, C/A and anti-$k_t$. In
the 2-particle case the active mass areas were in the same ballpark. As seen
from Fig.~\ref{fig:active-massarea-mc}, for MC jets, the three algorithms
exhibit clear hierarchy with the mass areas from $k_t$ being on average
significantly higher than those from C/A, which in turn are much larger than the
mass areas from the \mbox{anti-$k_t$.} 
This can be understood by noticing that the above hierarchy is consistent with
the one found in section~\ref{sec:scaling-violation-active} for the scaling
violation coefficients $D_m$ (table~\ref{tab:scalval-ma-active}). 
The large coefficient for the~$k_t$~algorithm means that even collinear
emissions can lead to a significant increase of mass area. For realistic MC jets
multiple such emissions are provided by parton shower. This also explains
somewhat smaller, but still significant difference in mass areas between the 
2-particle and MC jets from C/A and a very small difference in the case of
anti-$k_t$, whose scaling violation coefficient is identically zero. 
Another quantitative difference is related to the value of the mass area for
$x>1$.  In the 2-particle system this value was close to the passive area of a
1-particle jet. For MC jets, though there is a significant drop above $x=1$,
which we interpret as the two subjets not being merged, the mass area in that
region is not necessarily close to the 1-particle mass area.
The latter is related to the fact that the two widely separated subjets, $S_1$
and $S_2$, with $x>1$ have enough room to develop their own substructure and
hence cannot be approximated by a single particle. The extent to which their
mass area differs from $\pi R^4/2$ is again related to the scaling violation
coefficient and the same hierarchy is observed. We have also checked that the
values of mass areas for $x=2$ are indeed very close to the average mass area of
the hardest jet in the system.

The case of SISCone algorithm is special because of its highly nontrivial
dynamics involving the split-merge procedure and therefore it should be
discussed separately. 
As already mentioned, the general pattern found for MC jets is the same as the
one from 2-particle results.
In particular, we see that the SISCone jets with finite $z$ often have very large mass
areas even for $x>1$, which would point out to the interpretation that their
subjets are likely merged in this region. We note also that this observation is
compatible with the study of reach of the SISCone algorithm from
\cite{Salam:2007xv, Salam:2009jx} and in particular with the discussion therein
related to the $R_{\rm sep}$ parameter. As in the case of $k_t$ and C/A, also
the SISCone jets from Pythia exhibit somewhat larger mass areas than the
2-particle jets. 
Part of the reason is again some sensitivity to additional radiation from
parton shower, though that must be moderate given that fact that the $D_m$
coefficient in table~\ref{tab:scalval-ma-active} is not very big. 
Another important mechanism which leads to larger mass areas of the SISCone MC
jets is related to the split-merge procedure. As discussed in
section~\ref{sec:ama-1p}, the small value of the mass area of 1-particle jet
arises due to the fact that the cone around that particle overlaps with cones of
pure ghost jets and since there is no other particle that they could share such
overlapping cones always split. This must be somewhat different in the realistic
event which is populated with many physical particles. As a consequence, the
number of pure ghost jets is greatly reduced and therefore the above mechanism,
which led to reduction of mass area of the 1-particle jet, is not that efficient
here. 
Similar conclusion can be drawn from the study of the areas of realistic jets
from \cite{Cacciari:2008gn}. To further test this reasoning we varied the 
split-merge parameter $f$  and observed that lower value of this parameter
(corresponding to easier merging of overlapping cones) leads to larger average
value of mass area of the hardest jet and vice versa.

The mass areas of realistic jets from Monte Carlo simulations deserve detailed
study. In this section we gave a brief illustration of what sort of effects one
may expect if one goes beyond the 2-particle approximation of a jet. The main
conclusion from our MC study is that the 2-particle results for $x$ and $z$
dependence, highlighted in Fig.~\ref{fig:active-massarea2p_general_allalg},
together with the study of sensitivity of the algorithms to the perturbative
radiation, allow one to explain most of the features of mass areas of the
simulated jets. Fig.~\ref{fig:active-massarea-mc}  provides additional guidance
for the choice of the jet algorithm which minimizes background contamination.
Consistently with the results of the 2-particle study from preceding sections,
it points at SISCone and anti-$k_t$ disfavouring, in that particular respect,
the $k_t$ and C/A algorithms.

\subsection{Correcting jet mass for pileup contamination}
\label{sec:pileup-cor}

\begin{figure}
    \centering
    \includegraphics[width=0.48\textwidth]{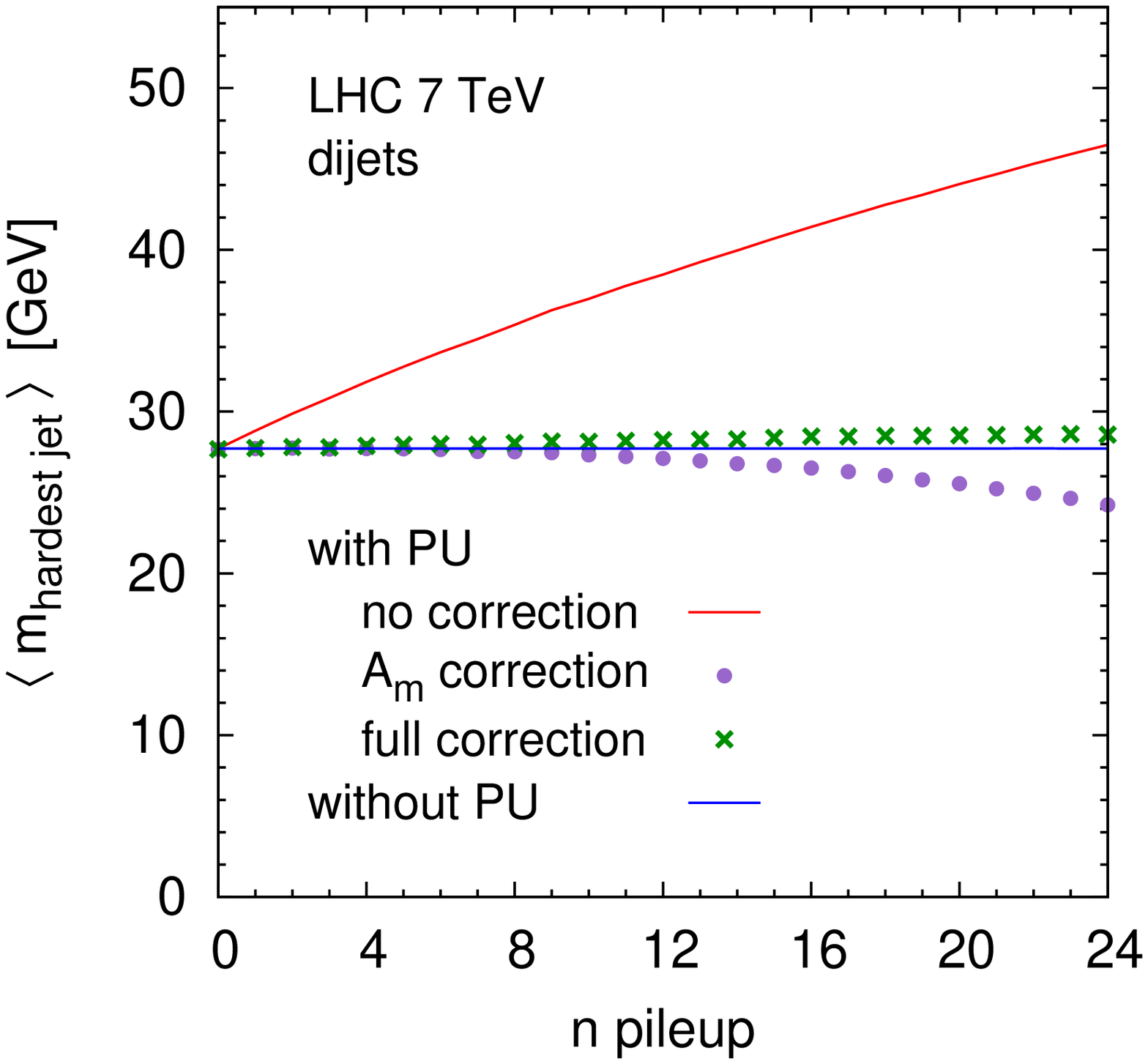}
    \hspace{5pt}
    \includegraphics[width=0.48\textwidth]{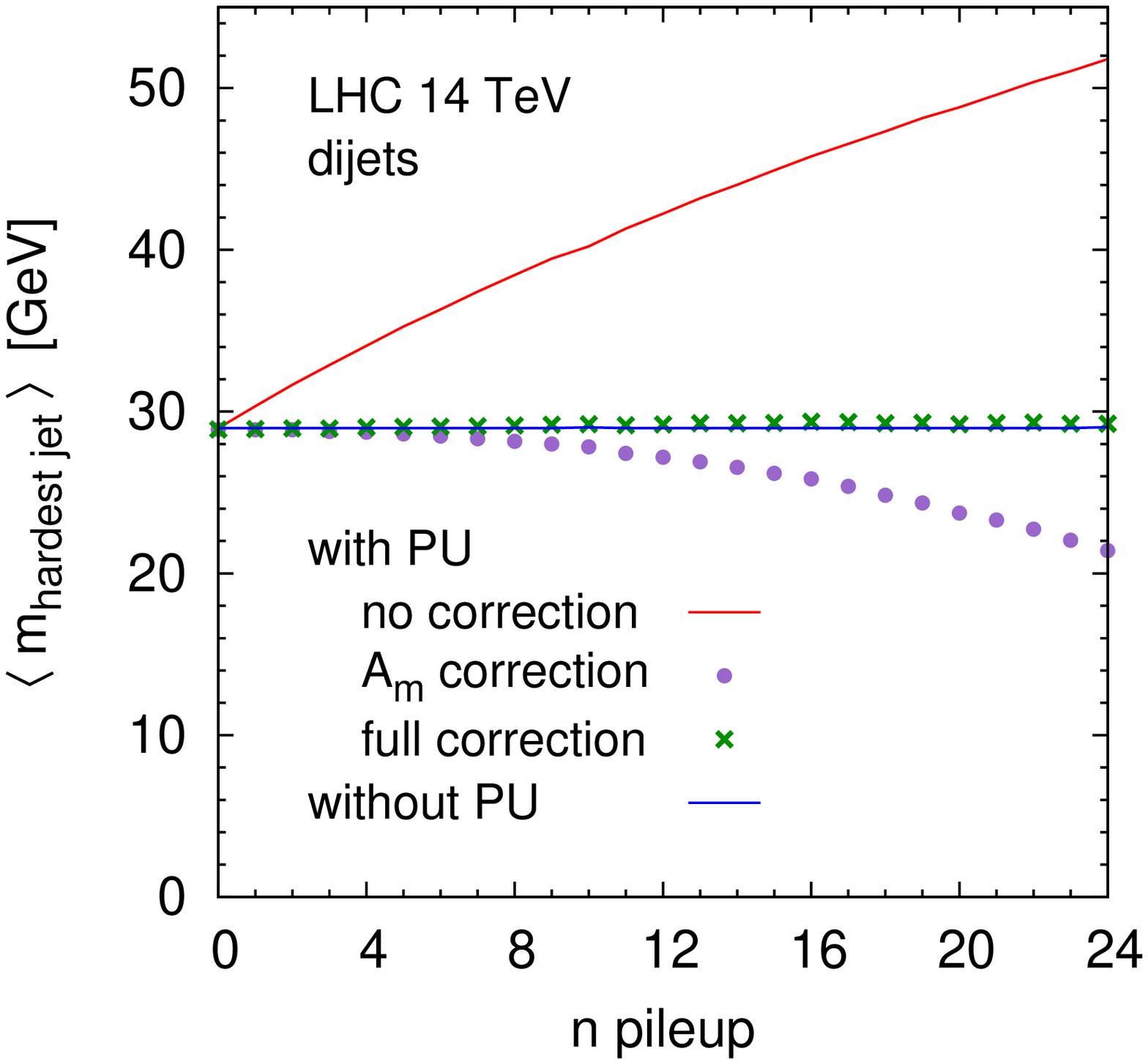}
    \caption{ 
	Average mass of the hardest jet in dijet events at the LHC at $\sqrt{s}
	= 7$~TeV (left) and 14~TeV (right).  The hard event simulated with
	Pythia 6.4 with the underlying switched off. Pileup from Pythia 8, tune
	4C. Massless hadrons.  Jets found with anti-$k_t$, R=0.7. Only events
	with $p_{t, \text{hardest jet}} > 150$ GeV accepted.  
	$\rho$ used to correct for pileup contamination determined on the
	event-by-event basis in the range
	$|y|<4 $ with the area/median method from~\cite{Cacciari:2007fd} taking
	the C/A algorithm with $R=0.5$.
    }
    \label{fig:pileup-cor}
\end{figure}

The definition of the active mass area from Section~\ref{sec:active-mass-area}
suggests its practical application.
Suppose that instead of ghosts we have in our event a dense set of soft
particles distributed fairly uniformly in rapidity and azimuthal angle. Such
particles may be coming, for instance, from pileup (PU). They will normally be
clustered together with genuine jet particles leading to contamination of the
jet. This contamination, in turn, will cause a systematic shift of a mass of
the jet
\be
  m_{J_\PU}^2 = m_{J}^2 + \delta m^2\,,
\ee
where $m_{J_\PU}^2$ is a mass of the jet $J$ from an event with pileup whereas
$m_{J}^2$ is a mass of the same jet from an event without pileup. 
 
Consider an event with the average density of the transverse momentum of PU
particles per unit area equal to $\rho$. Then, in the case of massless hadrons,
the magnitude of the shift of the mass of jet $J_\PU$ due to pileup is given by
\be
  \label{eq:substr-ma}
  \delta m^2 (J_\PU)=  p_{tJ_\PU}\,\rho\, A_m(J_\PU) - \rho^2 A_\mu^2(J_\PU)\,.
\ee
As we see, the leading correction in $\rho$ comes from a term involving active
mass area. It is important to notice that the mass area is computed from
Eq.~(\ref{eq:active-ma-def2}) using  the uncorrected jet $J_\PU$ containing both
genuine jet particles and the contamination from pileup. That means that
$A_m(J_\PU)$ itself acquires a subleading contribution
$\order{\rho/p_{tJ_\PU}}$, which in turn implies that the first term from
Eq.~(\ref{eq:substr-ma}) contains an implicit component $\order{\rho^2}$. 
As long as $\rho$ is not too large compare to $p_{tJ_\PU}$ the presence of PU
particles in the jet does not affect very much the value of $A_m$.  However,
when $\rho$ is comparable with transverse momenta of genuine jet particles, then
the active mass area computed from (\ref{eq:active-ma-def2}) gets a systematic
shift towards larger values. 

Qualitatively, a contribution to $m_{J_\PU}$ proportional to $\rho^2$ is needed
since it accounts for the fact that the change of mass of a jet due to pileup
comes not only from clustering pileup particles with jet particles but also from
clustering pileup particles among themselves. The latter is a subleading effect
in powers of $\rho$. 
Quantitatively, to get the mass correction which is valid also at large $\rho$
one needs a second, negative term in Eq.~(\ref{eq:substr-ma}), which combined
with the $\order{\rho^2}$ component from the first term gives the full
subleading contribution.

We have studied the effects of mass shift due to pileup using dijet events from
Pythia~6.4~\cite{Sjostrand:2006za} combined with pileup from
Pythia~8~\cite{Sjostrand:2007gs}.~\footnote{We would like to thank Gavin Salam
for suggesting using this example to illustrate the procedure of jet mass
correction.}
All hadrons were passed through a simple calorimeter with cells in the
$(y,\phi)$ plane of size $0.1$ and rapidity coverage $|y| < 4.5$. Then, the
calorimeter towers, which correspond to massless 4-vectors, were used as input
to clustering algorithms.  Jets were found with the anti-$k_t$ algorithm with
$R=0.7$ and only events with $p_t$ of the hardest jet greater than 150 GeV were
accepted.  

To correct jet masses for pileup contamination one needs to know the level of
the pileup transverse momentum per unit area, $\rho$, for each event and then
apply the formula~(\ref{eq:substr-ma}). 
To determine $\rho$, we used the area/median method proposed
in~\cite{Cacciari:2007fd} and implemented in the FastJet package~\cite{FastJet}.
The method measures $\rho$ on the event-by-event basis. It starts by adding
ghost particles to an event. Then, all particles (physical and ghosts) are
clustered with an infrared and collinear safe jet algorithm. That leads to a set
of jets, $\{j\}$, with jets ranging from hard to very soft. That set is used to
determine $\rho$, which is defined as a median of the distributions of
$\{p_{tj}/A_j\}$, where $p_{tj}$ is the transverse momentum of the jet $j$ and
$A_j$ is its scalar area.  Using the median is a way to dynamically separate the
soft and hard parts of an event.
The method leaves some freedom in the choice of the jet algorithm, the rapidity
range in which $\rho$ is measured as well as treatment of the hardest jets.
Following suggestions from the literature~\cite{Cacciari:2007fd,
Cacciari:2009dp}, we used the C/A algorithm with $R=0.5$ and the active area
definition. Then, the median was determined taking all jets in the range $|y| <
4$. As shown in~\cite{Cacciari:2007fd, Cacciari:2009dp}, for large rapidity
range, the influence of the two hardest jets on the value of $\rho$ is very
small, therefore we did not remove them from the set of jets used for $\rho$
determination.

One has to remember that $\rho$ characterises pileup in a given event only on
average and that there are always point-to-point fluctuations. Therefore, it may
happen, especially for light jets, that our procedure occasionally leads to
negative values of $m^2_J$. This corresponds to the cases with negative
fluctuation in which the actual contamination from pileup to our jet is locally
smaller than the typical level of PU in that event represented by the value of
$\rho$. In such events, our procedure subtracts too much from a jet. There is,
however, a second class of events with positive point-to-point fluctuation in
the vicinity of a jet and for those events the correction based on global $\rho$
is slightly underestimated. 
The above errors of under or overestimating the mass correction will cancel in
quantities averaged over many events like $\langle m^2_J \rangle$ or $\langle
m_J \rangle$. To make sure that this happens for the latter observable, for
events with $m_{J}^2 < 0$, one needs to set $m_{J} = -\sqrt{-m_{J}^2}$.

In Fig.~\ref{fig:pileup-cor}, we show the average mass of the hardest jet as a
function of the number of pileup, i.e. additional min-bias events accompanying
the production of hard dijet. The plots correspond to the LHC at $\sqrt{s} =
7$~TeV (left) and 14 TeV (right). We see that the average mass of the hardest
jet grows, approximately linearly with $n$. That is easily understood with our
formula~(\ref{eq:substr-ma}) in which the main contribution comes from
the term linear in $\rho$ and it is natural to expect that $\rho$ of pileup will
scale linearly with $n$, which is just the number of alike min-bias events.
We see also that the effect of pileup contamination is strong reaching up to
$70\%$ shift in the mass for the cases with high $n$.  For reference, we show a
horizontal line corresponding to the case without pileup ($n=0$). Then we apply
the ``mass area correction'' using the first term from Eq.~(\ref{eq:substr-ma})
as well as the full correction from that equation involving both mass area and
the $\rho^2$ term. We see that the mass area term dominates the mass shift and
the correction involving this term alone works very well up to fairly high
pileup, $n\sim 12-15$. For larger $n$, however, the second term, subleading in
$\rho$, becomes necessary to get a decent value of corrected jet mass. 
We note that using both terms is equivalent to directly correcting the 4-vector
of a jet with the help of the 4-vector area and then calculating its mass
\cite{GavinCMSTalk}. Our study from this section gives however a significant
additional insight into the structure of these corrections. In particular, we
have gained an understanding which contributions to the mass shift are dominant
and why. 

Overall, the example from this section shows that most of the contamination of
the jet mass due to pileup comes from the term involving mass area. On one hand,
that confirms that the mass area is a robust characteristic of the
susceptibility of the jet mass to contamination. On the other hand it provides a
simple method to correct for that contamination and recover, with excellent
accuracy, the value of the original jet mass. 
More studies of jet mass corrections, in particular a systematic analysis
and optimisation along the lines of~\cite{Cacciari:2007fd, Cacciari:2008gn,
Cacciari:2008gd, Soyez:2010rg}, is left for future work.

\section{Conclusions}
\label{sec:conclusions}

We have proposed a new characteristic of a jet, called \emph{mass  area}, which
is supposed to measure the susceptibility of the jet's mass to soft background
like pileup or underlying event. It is a close relative of the catchment area of
jets introduced in the work of \cite{Cacciari:2008gn, Cacciari:2007fd }.
Two complementary definitions of the mass area were given suitable for two
different limits of the distribution of UE/PU: the \emph{passive mass area},
measuring the sensitivity of mass of a jet to contamination from pointlike
radiation, and the \emph{active mass area}, more appropriate if the soft
background radiation is diffuse and uniform.

We have investigated the properties of the passive and active mass areas for
four jet clustering algorithms, $k_t$, Cambridge/Aachen, anti-$k_t$ and SISCone
by studying systems with one or two particles of arbitrary hardness.
We have also confronted the above results with those obtained with more
realistic jets simulated by Pythia.

As a preparatory step, we have generalised in section~\ref{sec:area-general-2p}
the results for passive and active areas of 2-particle jets to the case where
the two constituent particles have arbitrary transverse momenta. This part of
our study shows that even the ``conical algorithms'', SISCone and anti-$k_t$,
rarely produce jets whose shapes in the $(y,\phi)$ plane are circular. As
discussed in section \ref{sec:area-general-2p} and illustrated e.g. in
Fig.~\ref{fig:2p-general-scheme}, a very simple system of two particles with
comparable hardness leads to the whole variety of jet areas depending on the
algorithm, asymmetry parameter $z$ or the distance between the particles $x$.

The study of mass areas of 1- and 2-particle jets, presented in section
\ref{sec:mass-area}, reveals that similar richness exists also for this
characteristic of a jet.
A general pattern which is seen is that the ``conical algorithms'', SISCone and
anti-$k_t$, exhibit strong dependence on the asymmetry parameter $z$ measuring
how much of the total jet's transverse momentum is taken by the softer particle.
On the contrary, the $k_t$ and C/A algorithms, with jets of highly irregular
areas, show virtually no dependence on~$z$.

The dependence on the distance between the two constituent particles in the
$(y,\phi)$ plane, $xR$, is substantial for all algorithms though its character
varies across them. The results from $k_t$ grow monotonically for $x<1$ whereas
for C/A and anti-$k_t$ they start differing from the 1-particle result for much
larger $x$ in the ballpark of $x>0.5$. Finally, SISCone shows a completely
different $x$-dependence which is the largest for $x>1$, though the active mass
areas changes significantly with $x$ also for $x<1$.

In absolute terms, the active areas and mass areas of 2-particle jets from
SISCone are much smaller than from the other three algorithms.
It is related to the split-merge procedure, which is a part of the SISCone
algorithm, and which results in splittings of stable cones with perturbative
particles overlapping with cones containing only ghosts.
The low active area and mass area of SISCone means that the jets from this
algorithm will be, on average, much less contaminated by a soft and diffuse
background than the jets from $k_t$, C/A or anti-$k_t$. This, in turn, will
result in a very good $p_t$ and mass resolution.

Our study of active mass areas of jets from MC simulation showed the same
pattern of $x$ and $z$ dependence as that found for 1- and 2-particle jets.
This, together with the results from the study of sensitivity of mass areas from
different algorithms to perturbative radiation, was sufficient to account for
most of the features of mass areas of the simulated jets.
We used the simulated jets also to study corrections to jet masses due to
contamination from pileup. We found that most of the systematic shift in mass
caused by soft background comes in the form of a term involving mass area. That
confirms that the mass area indeed captures the essential aspects of the
modification of jet mass and provides simple method to correct for it and
recover its original value. 

As for the the comparison between the areas and the mass areas of jets, we have
seen that, qualitatively, the two characteristics exhibit similar behaviour.
Quantitatively however, the effects observed for mass areas are always
significantly bigger, a fact that we associate with an additional power of angle
in the definitions of the mass areas with respect to the areas.

We envisage several ways in which the concept of mass area, introduced in this
paper, could be used. 
Firstly, as a measure of the susceptibility of the jet's mass to contamination
from soft radiation, it provides a guidance for choosing a given jet algorithm
for a given purpose. For example, to minimise the systematic error in
determination of mass of a jet one may choose the algorithms whose mass area is
smallest.
Secondly, knowing how the mass area depends on the relative hardness of the
subjets may help in designing a discriminating variable which in turn could
allow one to separate the QCD jets from the jets coming from decays of heavy
objects. In this context, one could consider, for instance, using the mass area
as an additional variable entering the Boosted Decision Tree~\cite{Cui:2010km}.
It would be also very interesting to further study the corrections of jet masses
for the contamination from soft background. Possible extensions of the
analysis from section~\ref{sec:pileup-cor} could involve an optimization of jet
definition as well as correcting for contamination from underlying event.
Finally, it would be worth investigating the effects the new procedures of noise
reduction, namely filtering~\cite{Butterworth:2008iy},
pruning~\cite{Ellis:2009me} and trimming~\cite{Krohn:2009th}, have on the mass
area of jets.
We believe that all the above possibilities are worth investigating with jet
events from Monte Carlo simulations. We leave these questions for future work.

\section*{Acknowledgements}

We are indebted to Gavin Salam for suggesting this investigation and for many
useful discussions while the study was carried out. We also thank him and Matteo
Cacciari for careful reading of the manuscript.
In our study we used a set of tools, in particular interfaces to MC generators
and pileup, which were developed by Gavin Salam, Matteo Cacciari and Gregory
Soyez.
The  work was supported in part by the French ANR under contract
ANR-09-BLAN-0060 and by the Groupement d'Int\'er\^et Scientifique ``Consortium
Physique des 2 Infinis''~(P2I).

\appendix
\section{Passive areas and mass areas for general 2-particle system}
\label{app:passive-formulae}

In this appendix, we collect all analytic results for passive areas and mass
areas of the hardest jet in the system with two particles of arbitrary
transverse momenta. 
The calculations  were done within the $E$ recombination scheme and in the limit
of small $R$. The latter corresponds to neglecting the difference between $y$
and $\phi$ dimensions and was used consistently throughout the paper for both
passive and active quantities.
All the analytic results presented here were confirmed  for a range of values of $z$ by numerical study with FastJet. 

As discussed in sections \ref{sec:area-general-2p}, the area and the mass area
of the hardest jet in the system with two particles depends on the interparticle
distance in the $(y,\phi)$ plane. For each algorithm, one distinguishes several
subranges of the range $0<x<2$. In each such subrange the $x$-dependence of the
area and mass area is described by a different function.

Table~\ref{tab:passive-a-ma-2p} summarises the results for passive area and mass
area of the hardest jet from four algorithms normalised to the 1-particle
result, i.e. $\pi R^2$ or $\pi R^4/2$, respectively. The critical $x$~values,
$x_{c1,\dots, 4}$, were defined in Eqs.~(\ref{eq:xc1-solution}),
(\ref{eq:xc2-solution}), (\ref{eq:xc4-approx}) and (\ref{eq:xc3-solution}).

The functions from table~\ref{tab:passive-a-ma-2p}, different for areas
and mass areas are defined below. To make the notation more concise we define
the auxiliary function
{\allowdisplaybreaks
\begin{align}
\label{eq:f-def}
& f(x,x')= \sqrt{\left(1-(x-x')^2\right) \left((x+x')^2-1\right)}\,, \\
\label{eq:xbar-def}
\end{align}
}
and the two following variables
\be
\bar{z}= \frac{z}{1-z}\,, \qquad
\bar{x}= \bar{z} x\,,
\ee
where $x$ is the interparticle distance in units of the jet radius, defined in
Eq.~(\ref{eq:x-xJ-def}), and $z$ is the asymmetry parameter from
Eq.~(\ref{eq:z-def}).

\renewcommand{\tabcolsep}{20pt}
\renewcommand{\arraystretch}{1.4}
\begin{table}[t]
   \centering
   \begin{tabular}{|c|c|c|c|c|} \hline
     &  \multicolumn{4}{c|}{passive area (or mass area) normalised to 1-particle
     result}  \\ \hline
  $x$ range &  $k_t$        & C/A         & SISCone       & anti-$k_t$ \\ \hline
  $[0, x_{c1}]$      & \multirow{4}{*}{$g_X(x)$}     & $p_X(x)$  &
  \multirow{4}{*}{$p_X(x)$}   &  \multirow{2}{*}{$p_X(x)$}   \\
  \cline{1-1}  \cline{3-3} 
  $[x_{c1}, x_{c2}]$ &  & $q_X(x)$    &  &  \\
  \cline{1-1}  \cline{3-3} \cline{5-5}
  $[x_{c2}, x_{c3}]$ & & \multirow{2}{*}{$r_X(x)$} &  & $u_X(x)$  \\
  \cline{1-1}  \cline{5-5} 
  $[x_{c3}, 1]$      & & &   & $v_X(x)$  \\
  \cline{1-1}  \cline{2-5} 
  $[1, x_{c4}]$      & \multirow{2}{*}{$h_X(x)$} &  \multirow{2}{*}{$h_X(x)$}
                     &  $t_X(x)$  & $w_X(x)$  \\
  \cline{1-1}  \cline{4-5}

  $[x_{c4}, 2]$      &  &  &  $h_X(x)$  & 1  \\[5pt]
   \hline
   \end{tabular}
   \caption{
   Passive areas and mass areas of the hardest jet in a system with two
   particles of arbitrary hardness. The critical values of $x$, $x_{c1,\dots,
   4}$, are defined in Eqs.~(\ref{eq:xc1-solution}), (\ref{eq:xc2-solution}),
   (\ref{eq:xc4-approx}) and (\ref{eq:xc3-solution}).
   The functions, corresponding to given range of $x$  are given in
   subsections~\ref{app:passive-area}, for areas, and
   \ref{app:passive-massarea}, for mass areas. The subscript in function
   name, $X=a,ma$, refers to the corresponding quantity. To obtain the area or
   mass area in a given range of $x$, the appropriate function from the table
   should be multiplied by the 1-particle result, $\pi R^2$ or $\pi R^4/2$,
   respectively.
   }
   \label{tab:passive-a-ma-2p}
\end{table}
\renewcommand{\arraystretch}{1}

\subsection{Areas}
\label{app:passive-area}

The passive areas of 2-particle jets are fully specified for the four algorithms
by the following set of functions to be used with table~\ref{tab:passive-a-ma-2p}
{\allowdisplaybreaks
\begin{align}
\label{eq:ga-def}
& g_a(x) =
\frac{1}{\pi}
\Bigg\{
x \sqrt{1-\frac{x^2}{4}} + 2 \left[\pi-\arccos\left(\frac{x}{2}\right)\right]
\Bigg\}\,,\\
\label{eq:ha-def}
& h_a(x) = \frac12 g_a(x)\,,\\
\label{eq:pa-def}
& p_a(x) = 1\,,\\
\label{eq:qa-def}
& q_a(x, x_J)  =  
\frac{1}{\pi}
\Bigg\{
    \pi + \frac{f(x,x_J)}{2}
    + x^2 \left[\pi - \arccos\left(\frac{x^2+x_J^2-1}{2 x x_J}\right)\right]
    + \arctan
      \left(\frac{f(x,x_J)}{x^2-x_J^2-1}\right)
  \Bigg\}\,,\\
\label{eq:ra-def}
& r_a(x, x_J)  =
\frac{1}{\pi}
\Bigg\{
     x^2 \left[\pi + \arcsin\left(\frac{x^2+x_J^2-1}{2 x x_J}\right)
                + \arcsin\left(\frac{x^2+(x-x_J)^2-1}{2 x (x-x_J)}\right)\right]
    +\frac{f(x,x_J)}{2}
    \nonumber \\
  &
    +\frac{f(x,x-x_J)}{2}
    - \arctan \left(\frac{x^2-x_J^2-1} {f(x,x_J)} \right)
    - \arctan \left(\frac{x^2-(x-x_J)^2-1}{f(x,x-x_J)} \right)
\Bigg\}\,, \\
\label{eq:ta-def}
& t_a(x, x_J)  =  
\frac{1}{\pi}
\Bigg\{
    3 \pi - 2 \arccos\left(\frac{x_J}{2}\right)
    - 2 \arccos\left(\frac{x-x_J}{2}\right)
    + \left(x-x_J\right) \sqrt{1-\frac{(x-x_J)^2}{4}}
 \nonumber \\
 &
    + x_J \sqrt{1-\frac{x_J^2}{4}}
\Bigg\}\,, \\
\label{eq:ua-def}
& u_a(x, x_J)  =  
  \frac{1}{\pi}
  \Bigg \{
     \frac{\pi}{2} + 
     \frac{f(x,x-x_J)}{2} +
     x^2 \left[\pi - 
         \arccos\left(\frac{x^2+(x-x_J)^2-1}{2 x (x-x_J)}\right)\right]
         \nonumber \\
  &
      + \arctan
        \left(\frac{1-x^2+(x-x_J)^2}{f(x,x-x_J)}\right)
  \Bigg\}\,, \\
\label{eq:va-def}
& v_a(x, x_J, z,\bar{x})  =  
  \frac{1}{\pi} \Bigg\{
      \frac{f(x, x - x_J)}{2} + \frac{f(x_J, \bar{x})}{2} + 
      x^2 \left[\frac{\pi}{2} + 
      \arcsin\left(\frac{x^2 + (x - x_J)^2 -1}{2 x (x - x_J)}\right)\right] 
      \nonumber \\
&
      + \bar{x}^2 \left[\frac{\pi}{2} + 
      \arcsin\left(\frac{x_J^2 + \bar{x}^2 -1}{2 x_J \bar{x}}\right)\right] - 
      \arctan\left(\frac{2 x x_J - x_J^2-1}{f(x, x - x_J)}\right) - 
      \arctan\left(\frac{\bar{x}^2 - x_J^2 -1 }{f(x_J, \bar{x})}\right)
  \Bigg\}\,, \\
\label{eq:wa-def}
& w_a(x, z,\bar{z})  =  
  \frac{1}{\pi}
  \Bigg\{\pi + 
   \frac{1}{2 (1 - \bar{z}^2)} \sqrt{4 x^2 - (1 + x^2 - \bar{z}^2)^2} - 
      \arcsin\left(\frac{1}{2 x} \sqrt{4 x^2 - (1 + x^2 - \bar{z}^2)^2}\right)
      \nonumber \\
&
      - \frac{x^2 z^2}{(1 - z)^2 (1 - \bar{z}^2)^2}
      \arcsin\left(\frac{1 - \bar{z}^2}{2 x^2 \bar{z}}
      \sqrt{4 x^2 - (1 + x^2 - \bar{z}^2)^2}\right)
  \Bigg\}\,.
\end{align}
}

For the $k_t$ algorithm, the general 2-particle result is identical with that
found in \cite{Cacciari:2008gn} for the case with strongly ordered transverse
momenta. The results from the other three algorithms depend on $z$ via
$x_J$ (see Eq.~(\ref{eq:xJ-Escheme})), in the case of C/A and SISCone, and also 
explicitly in the case of anti-$k_t$. One can also verify that in the limit
$z\to 1/2$, the function $v_a$ reduces to $r_a$ and the function $w_a$ to $h_a$.
Therefore, the areas from C/A and anti-$k_t$ become identical in this limit, as
seen in Figs.~\ref{fig:passive-area2p_general_sis+ca}~(left) and
\ref{fig:passive-area2p_general-akt}. 

\subsection{Mass areas}
\label{app:passive-massarea}

Here we gather all the functions needed, together with
table~\ref{tab:passive-a-ma-2p}, to compute the passive mass areas of the
hardest jet in the 2-particle system for the four algorithms.
The definition of the passive mass area (\ref{eq:passive-ma-def}) introduces
extra $\phi$-dependence via the transverse momentum of a jet, $p_{tJ}$. In line
with all the other calculations presented in this paper, also here, we have
assumed that the jet radius is small. This allows one to approximate the full
jet's transverse momentum by $p_{tJ} \simeq p_{t1}+p_{t2}$.\footnote{ For real
jets, however, there will be some difference depending on whether the two
constituent particles are oriented along the rapidity axis or along the $\phi$
axis. We have performed corresponding numerical study and found that this
difference is indeed negligible except for a very specific case of large ($x
\sim 2$) and symmetric ($z\sim 1/2$) jets from the SISCone algorithm where it
can get up to 20\%.}
The results for the normalised 2-particle mass areas are
{\allowdisplaybreaks
\begin{align}
\label{eq:gma-def}
& g_{ma}(x) =
\frac{1}{\pi}\Bigg\{
\pi  \left(1+x^2\right)+\frac{x}{2} \left(6+x^2\right) \sqrt{1-\frac{x^2}{4}}
+ 2\left(1+x^2\right) \arcsin\left(\frac{x}{2}\right)
\Bigg\}\,, \\
\label{eq:hma-def}
& h_{ma}(x) = 
\frac{1}{\pi}\Bigg\{ \pi
+\frac{x}{6} \left(\frac{x^2}{2}+1\right)\sqrt{1-\frac{x^2}{4}} 
-\arccos\left(\frac{x}{2}\right)\Bigg\}\,, \\
\label{eq:pma-def}
& p_{ma}(x,x_J,z)  = 1 + 2(x - x_J)^2 + 2 z \Big(x_J^2-(x-x_J)^2\Big)\,, \\
\label{eq:qma-def}
& q_{ma}(x, x_J, z)  =
\frac{1}{\pi} \Bigg\{
\frac{\pi}{2} \Big[1 +2 x^2 +3 x^4 -2 x_J(2x-x_J) -2z x (x + x^3 - 2 x_J)\Big]
+ \frac{1}{4 x_J}
\Big[
5 x_J(1+x^2) 
\nonumber \\ & 
+ x_J^3 
- 4x (1 + x_J^2 - x^2) -4 z x (x^2 + x x_J - x_J^2-1) \Big] f(x, x_J) 
+ (3 - 2z) x^4 \arcsin\left(\frac{x^2 + x_J^2-1}{2 x x_J}\right)
\nonumber \\ & 
+ \Big[1 + 2 (x - x_J)^2 - 2 z x (x - 2 x_J)\Big]
\arctan\left(\frac{1 - x^2 + x_J^2}{f(x, x_J)}\right)
\Bigg\}\,, \\
\label{eq:rma-def}
& r_{ma}(x, x_J, z) = 
\frac{1}{\pi} \Bigg\{
\frac{1}{4 x_J} \Big[5 x_J(1+x^2) + x_J^3 - 4 x (1 + x_J^2-x^2) + 
4 z x (1 - x^2 -x x_J + x_J^2)\Big] f(x, x_J)
\nonumber \\  &
+ \frac{1}{4 (x - x_J)} \Big[(x - x_J) (5 + 2 x^2 - 2 x x_J + x_J^2) -
4 z x (1-x^2-x x_J +x_J^2) \Big] f(x, x - x_J) 
\nonumber \\ &
+ \frac{\pi}{2} x^4 (5 + 2 z)
 +x^4 (3 - 2 z)  \arcsin\left(\frac{ x^2 + x_J^2-1}{2 x x_J}\right)
- x^4 (1 + 2 z) \arccos\left(\frac{ x^2 + (x - x_J)^2-1}{2 x (x - x_J)}\right)
\nonumber \\ 
& +(1+2(x-x_J)^2-2 z x (x-2 x_J)) 
\Big[\arctan\left(\frac{1 - x^2 + x_J^2}{f(x, x_J)}\right) + 
     \arctan\left(\frac{1 - 2 x x_J + x_J^2}{f(x, x - x_J)}\right)\Big]
\Bigg\}\,, \\
\label{eq:tma-def}
& t_{ma}(x, x_J, z) = 
\frac{1}{\pi} \Bigg\{
\pi (1 + x^2)+ 
\frac{x_J}{4} \Big[6 + (2 x - x_J)^2 - 4 z x (x - x_J)\Big] \sqrt{4 - x_J^2}
\nonumber \\
& + \frac{x - x_J}{4} \Big[6 + (x - x_J)^2 + 4 z x x_J\Big] \sqrt{4-(x-x_J)^2}
+ 2(1 + (x - x_J)^2 + 2 z x x_J ) \arcsin\left(\frac{x - x_J}{2}\right)
\nonumber \\ 
& + 2\Big[1 + (x - x_J)^2 + x^2 - 2 z x (x-x_J)\Big] \arcsin\left(\frac{x_J}{2}\right)
\Bigg\}\,, \\
\label{eq:uma-def}
& u_{ma}(x, x_J, z) =
\frac{1}{\pi} \Bigg\{
\frac{\pi}{2}\Big((1 + x^2)^2 - 4 x x_J + 2 x_J^2-2 z x (x - x^3 - 2 x_J)\Big) 
\nonumber \\ &
+ \frac{1}{4 (x - x_J)}((x - x_J) (5 + 2 x^2 - 2 x x_J + x_J^2) - 4 z x (1 - x^2 - x x_J + x_J^2)) f(x, x - x_J) 
\nonumber \\ &
- (1 +2 z) x^4 \arcsin\left(\frac{1-x^2-(x-x_J)^2}{2 x(x- x_J)}\right)
+ (1 + 2 (x - x_J)^2 
\nonumber \\ &
- 2 z x (x - 2 x_J)) 
\arctan\left(\frac{1 - 2 x x_J + x_J^2}{f(x, x - x_J)}\right)
\Bigg\}\,, \\
\label{eq:vma-def}
& v_{ma}(x, x_J, z,\bar{x})  = 
\frac{1}{\pi} \Bigg\{
\frac{\pi}{2}\left((-7+2 z)\bar{x}^4+\frac{2x^4(1-2 z- z^2 + 5 z^3)}{(1- z)^4}\right)
- x^4 (2 z
\nonumber \\ &
+ 1) \arccos\left(\frac{x^2 + (x - x_J)^2-1 }{2 x (x - x_J)}\right)
+ \frac{ x^2 \bar{x}^2}{(1 - z)^2} (2 - 6 z + 7 z^2 - 2 z^3)  
\arcsin\left(\frac{\bar{x}^2 + x_J^2-1}{2 \bar{x} x_J}\right) 
\nonumber \\ &
+ (1 + 2 (x - x_J)^2 - 2 x (x - 2 x_J) z) 
\Bigg[\arctan\left(\frac{1 - 2 x x_J + x_J^2}{f(x, x - x_J)}\right) 
+ \arctan\left(\frac{1 - \bar{x}^2 + x_J^2}{f(\bar{x}, x_J)}\right)\Bigg] 
\nonumber \\ &
- \frac{1}{4 (x - x_J)} \Big[x_J (5 + x_J^2) - x (5 + x_J^2 (3 - 4 z) - 4 z) + 
      4 x^2 x_J (1 - z) - x^3 (2 + 4 z)\Big] f(x, x - x_J)
\nonumber \\ &
+ \frac{1}{4 x_J} \Big[x_J (5 + x_J^2) - 4 x (1 + x_J^2) (1 - z) + 
4 \bar{x} x^2 z + \frac{ x^2 x_J (2 - z) (2 - z (5 - 4 z))}{(1 - z)^2}\Big] 
f(\bar{x}, x_J)
\Bigg\}\,, \\
\label{eq:wma-def}
& w_{ma}(x,z,\bar{z}) =
\frac{1}{\pi} \Bigg\{
\pi - \arcsin\left(\frac{1}{2 x} f\left(x, \bar{z}\right)\right) 
+ \frac{(1 - z)^2}{ 4 (1 - 2 z)^3} \Big((1 - 2 z)^2 + x^2 (1 - z)^2 (1 - 2 z 
\nonumber \\ &
+ 6 z^2)\Big) f\left( x, \bar{z}\right)
- \frac{x^4 (1 - z)^4 z^2}{(1 - 2 z)^4} (2 - z (4 - 3 z)) 
\arcsin\left(\frac{1 - 2 z}{2 x^2 (1 - z) z} f\left(x, \bar{z}\right)\right) 
\Bigg\}\,.
\end{align}
}

As for the areas, also here all results except those from the $k_t$
algorithm depend on $z$, both explicitly and via $x_J$. As discussed in
section~\ref{sec:passive-mass-areas-2p}, the $z$-dependence of the mass area of
the $k_t$ jets vanishes due to additional reflection symmetry.
Similarly, for anti-$k_t$, the
function $v_{ma}$ goes to $r_{ma}$ and $w_{ma}$ to $h_{ma}$ in the limit $z\to
1/2$. Hence, the C/A and anti-$k_t$ mass areas become identical as shown also
in  Figs.~\ref{fig:passive-massarea2p_general_kt+ca}~(right) and
\ref{fig:passive-massarea2p_general_sis+akt}~(right). 

\section{Active mass areas from SISCone for strongly ordered 2-particle system}
\label{app:active-formulae-sis}

\begin{figure}[t]
	\centering
 	\includegraphics[width=0.60\textwidth]{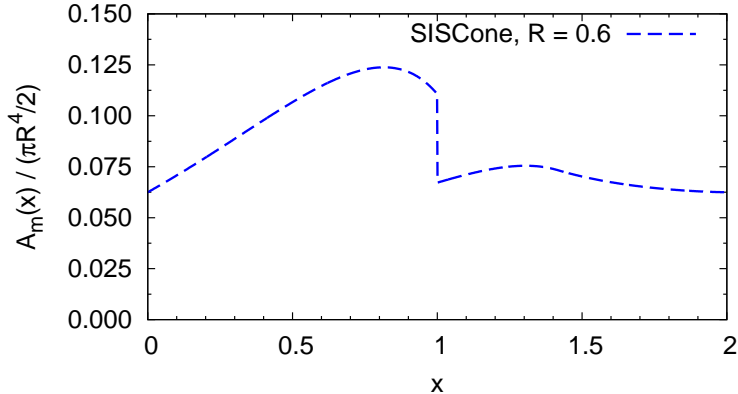}
 	\caption{
	Active mass area of the harder jet from SISCone in a 2-particle system
	with strongly ordered transverse momenta. The curve corresponds to the
	analytic result (\ref{eq:active-massarea-qcd}). At $x=0$ and $x=2$ the
	1-particle result, 1/16, is recovered.
	}
 	\label{fig:active-ma-2p-siscone}
\end{figure}

The active mass area of the hardest jet in the system of two particles with
strongly ordered transverse momenta (e.g. from QCD splitting) is given by
\begin{equation}
\label{eq:active-massarea-qcd}  
A_{m}^{\rm SISCone }(x) = \frac{\pi R^4}{2}\left \{ 
\begin{array}{ll}
  u_1(x) & \quad \text{ for } \quad 0<x<1\\
  u_2(x) & \quad \text{ for } \quad 1<x< \sqrt{2}\\
  u_3(x) & \quad \text{ for } \quad \sqrt{2}<x< 2
\end{array}
\right.\,,
\end{equation}
with the following definitions
{\allowdisplaybreaks
\begin{align}
\label{eq:u1-def}
& u_{1}(x) =
   \frac{1}{ 96 \pi \left(x^2-2\right)^2}
   \bigg\{
   6 \left(2-x^2\right)^2 \left(\pi -\arccos\left(\frac{x}{2}\right)\right)
   +x \sqrt{4-x^2} \left(4 x^6-x^4-39 x^2+48\right)
   \nonumber \\ &
   +3 \sqrt{1-x^2} \left(3
   x^2+2\right) \left(2-x^2\right)^2 \arccos\left(\frac{x}{2-x^2}\right)
   \bigg\}\,, \\
\label{eq:u2-def}
& u_{2}(x) =
  \frac{x}{24 \pi(4-x^2)^{3/2}} \left(10 + 5 x^2 - 6 x^4 + x^6\right) + 
  \frac{1}{16\pi}  \left(\pi - \arccos\left(\frac{x}{2}\right)\right)\,, \\
\label{eq:u3-def}
& u_{3}(x) =
\frac{1}{48 \pi x^3} \bigg\{3 \pi x^3 + 2 \sqrt{4 - x^2} (2 + x^2) - 
   3 x^3 \arccos\left(\frac{x}{2}\right)\bigg\}\,.
\end{align}
}
The corresponding curve in shown in Fig~\ref{fig:active-ma-2p-siscone}.
Eq.~(\ref{eq:active-massarea-qcd}) was used directly, together with the
definitions (\ref{eq:delta-active-mass-area}) and (\ref{eq:fluct-active-ma-res})
to obtain the anomalous dimension and the corresponding fluctuation coefficient
given in table~\ref{tab:scalval-ma-active}.


\end{document}